\documentclass[showpacs,aps,prd,nofootinbib,showkeys,superscriptaddress,twocolumn]{revtex4}
\usepackage{graphicx}
\usepackage{bm}
\usepackage{bbold}
\usepackage{amssymb}
\usepackage{amsmath,latexsym}
\usepackage{amsfonts}
\usepackage[usenames]{color}
\usepackage{subfigure}
\usepackage{subfigure}
\usepackage{slashed}
\usepackage{multirow,array}
\usepackage{mathtools}
\usepackage{mathrsfs}
\usepackage{dsfont}
\usepackage[colorlinks=false,linktocpage=true]{hyperref}
\usepackage{hyperref}
\usepackage[utf8]{inputenc}
\usepackage{aurical}
\usepackage[T1]{fontenc}
\usepackage{multirow}
\newcommand{\be}{\begin{equation}}
\newcommand{\ee}{\end{equation}}
\newcommand{\bea}{\begin{eqnarray}}
\newcommand{\eea}{\end{eqnarray}}
\newcommand{\beal}{\begin{align}}
\newcommand{\enal}{\end{align}}
\newcommand{\bs}{\begin{subequations}}
\newcommand{\es}{\end{subequations}}
\newcommand{\besp}{\begin{split}}
\newcommand{\eesp}{\end{split}}
\newcommand{\mc}[1]{\mathcal{#1}}

\newcommand{\cl}{c_{\ell}}

\newcommand{\nl}{\nonumber \\}
\newcommand{\pd}{\partial}

\begin{document}
\interfootnotelinepenalty=10000
\title{Transasymptotics and hydrodynamization of the Fokker-Planck equation for gluons}
\date{\today}

\author{A.~Behtash}
\affiliation{Department of Physics, North Carolina State University, Raleigh, NC 27695}

\author{S.~Kamata}
\affiliation{
National Center for Nuclear Research, PL-00-681 Warsaw, Poland
}

\author{M.~Martinez}
\affiliation{Department of Physics, North Carolina State University, Raleigh, NC 27695}

\author{T.~Sch{\"a}fer}
\affiliation{Department of Physics, North Carolina State University, Raleigh, NC 27695}

\author{V.~Skokov}
\affiliation{Department of Physics, North Carolina State University, Raleigh, NC 27695}
\affiliation{Riken-BNL Research Center, Brookhaven National Laboratory, Upton, NY 11973, USA}

\begin{abstract}
We investigate the non-linear transport processes and hydrodynamization of a system of gluons undergoing longitudinal boost-invariant expansion. The dynamics is described within the framework of  the Boltzmann equation in the small-angle approximation. The kinetic equations for a suitable set of moments of the one-particle distribution function are derived. By investigating the stability and asymptotic resurgent properties of this dynamical system, we demonstrate, that its solutions exhibit a rather different behavior for large (UV) and small (IR) effective Knudsen numbers. Close to the forward attractor in the IR regime the constitutive relations of each moment can be written as a multiparameter transseries. This resummation scheme allows us to extend the definition of a transport coefficient to the non-equilibrium regime naturally. Each transport coefficients is renormalized  by the non-perturbative contributions of the non-hydrodynamic modes. The Knudsen number dependence of the transport coefficient is governed by the corresponding renormalization group flow equation. An interesting feature of the Yang-Mills plasma in this regime is that it exhibits transient non-Newtonian behavior while hydrodynamizing.  In the UV regime the solution for the moments can be written as a power-law asymptotic series with a finite radius of convergence. We show that radius of convergence of the UV perturbative expansion grows  linearly as a function of the shear viscosity to entropy density ratio. Finally, we compare the universal properties in the pullback and forward attracting regions to
other kinetic models including the relaxation time approximation and the effective kinetic Arnold-Moore-Yaffe (AMY) theory.  
\end{abstract}


\maketitle

\section{Introduction}
\label{sec:intr}

Relativistic fluid dynamics is an effective theory which describes long-wavelength
phenomena. It is widely accepted that its regime of validity 
is
restricted to systems near local thermal equilibrium. However, this traditional
paradigm has recently been challenged by  the overwhelming success of
hydrodynamic models in describing experimental data in high energy nuclear 
collisions~\cite{Chatrchyan:2013nka,Adare:2013piz,Aad:2012gla,Abelev:2012ola,
PHENIX:2018lia,Gale:2013da,Heinz:2013th} as well as cold atom 
systems~\cite{Bluhm:2017rnf,Bluhm:2015bzi,Bluhm:2015raa,Schaefer:2016yzd}.
In these systems the initial state is far from local thermal equilibrium, and 
it is not fully understood how hydrodynamic behavior emerges. 
The search for a kinetic framework that describes far-from-equilibrium 
plasmas has been one of the most important research subjects in high energy 
nuclear collisions and condensed matter 
physics~\cite{Berges:2020fwq,Florkowski:2017olj,Romatschke:2017ejr}. 

An important development in non-equilibrium dynamics was the discovery of emergent
hydrodynamic behavior in far-from-equilibrium conditions which can be understood 
in terms of the mathematical theory of resurgence~\cite{Heller:2015dha}. In this 
work the authors consider an extended hydrodynamic model, the Israel-Stewart 
equation \cite{Israel:1979wp}, and apply it to a strongly coupled plasma undergoing
Bjorken expansion. Subsequently, similar findings were obtained in many other 
transport models. These results show a deep connection between nonlinear relaxation
towards hydrodynamic behavior, also known as \textit{hydrodynamization}, and
transasymptotics and transseries \cite{Aniceto:2018uik,Spalinski:2018mqg,
Spalinski:2017mel,Buchel:2016cbj,Heller:2016rtz,Aniceto:2015mto,Basar:2015ava,
Casalderrey-Solana:2017zyh,Romatschke:2017vte,Blaizot:2020gql}. 

  Since then a very interesting and rich physical picture has emerged: \textit{The
nonlinear relaxation process towards hydrodynamic behavior, also known as
hydrodynamization, is driven by the decay of non-hydrodynamic degrees of freedom. 
Once the non-hydrodynamic modes have died out the system enters into the hydrodynamic
attractor which is entirely determined by the standard asymptotic gradient expansion}.
This new insight might be able to explain why hydrodynamic models work very well 
when applied in extreme experimental scenarios such as ultrarelativistic heavy ion
collisions \cite{Romatschke:2016hle,Weller:2017tsr}. 

Another interesting development in the understanding of far-from-equilibrium 
attractors in relativistic non-equilibrium dynamics is a phase space analysis 
using the language of non-autonomous dynamical systems\footnote{
A non-autonomous dynamical system corresponds to a general set of ODEs for the 
vector $\bold{a}$of the generic form
\begin{equation*}
\frac{d\bold{a}}{d\lambda}=\mc H(\bold{a},\lambda)\,,
\end{equation*}
where $\mc H$ is a function that depends explicitly on both $\bold{a}$ and $\lambda$.
When the affine parameter $\lambda$ does not appear explicitly in the RHS of the
previous expression it is said that the system is autonomous.}
\cite{Behtash:2017wqg,Behtash:2018moe,Behtash:2019txb,Behtash:2019qtk}. This 
particular point of view allows us to characterize in simple terms the behavior of 
the solutions, described as flows, either at early or at late times. For instance, 
a global and local phase space flow analysis led to the conclusion that a large 
class of kinetic models undergoing Bjorken expansion~\cite{Bjorken:1982qr}
hydrodynamize in the long time limit \cite{Behtash:2019txb,Behtash:2019qtk,
Heller:2016rtz,Heller:2020anv}, whereas systems undergoing Gubser flow 
\cite{Gubser:2010ze,Gubser:2010ui} never do~\cite{Behtash:2017wqg,Behtash:2019qtk}.
Moreover, the flow structure in phase space together with the symmetries of the
dynamical system constrains the asymptotic behavior of the solutions of the ODEs. 
For example, for weakly coupled boost invariant systems it was demonstrated 
\cite{Behtash:2019txb} that the solutions of the moments equations admit power 
law series expansions at early times, while at late times linear perturbations 
of the moments decay exponentially~\cite{Behtash:2018moe,Behtash:2019txb,
Heller:2016rtz,Denicol:2016bjh}. This is due to the nature of the early and late-time attractors. 

Attractors are understood as regions in phase space where flows accumulate either 
in the long or short time limit. However, in non-autonomous dynamical systems the 
past and future of the evolution are not the same since time translation invariance is explicitly broken. Each flow solution $\phi\equiv\phi(\phi_0;t,t_0)$ is written 
in terms of its initial value $\phi_0$ and its initial and final times, $t_0$ and 
$t$ respectively. 
In this context, it is important to differentiate the backward 
and forward asymptotic regions of the dynamical system. 
We refer to a forward
attractor as an asymptotic limit of the flows at which solutions converge when
$t\to\infty$ while the initial time $t_0$ is fixed. In contrast, the pullback attractor 
is defined in the limit $t_0\to 0$ while keeping $t$ fixed. It is
important to emphasize that for non-autonomous dynamical systems both limits do not
commute~\cite{caraballo2017applied,kloeden}.

 Most previous work on the convergence properties of kinetic equations in 
relativistic transport theory was based on very simple collision terms, e.g. the relaxation time approximation (RTA). In this work we study the 
resurgent asymptotic properties of the Boltzmann equation for a boost invariant
Yang-Mills plasma governed by the weak coupling collision term in the 
small-angle approximation. In this limit the Boltzmann equation can be 
written as a nonlinear Fokker-Planck equation.

The Fokker-Planck equation (FPE) for gluons is of great interest since it 
captures essential aspects of the early time dynamics of QCD matter produced in 
ultrarelativistic heavy ion collisions~\cite{Mueller:1999pi,Mueller:1999fp}. Due
to the highly nonlinear structure of the collision kernel in the small angle
approximation, the FPE has been solved mostly by numerical means 
\cite{Bjoraker:2000cf,Hong:2010at,Mehtar-Tani:2016bay,Blaizot:2019dut,
Blaizot:2013lga,Blaizot:2014jna,Tanji:2017suk}. Very few analytical solutions 
for rapidly expanding systems are known in the literature \cite{Moore:2004tg}. 
One of the main achieved goals in this work is to fill this gap. 

 Following Grad's method~\cite{Grad} we map the mathematical problem of solving 
the FPE onto finding solutions to the kinetic equations for the Legendre moments 
$c_l$ \cite{Behtash:2017wqg,Behtash:2018moe,Behtash:2019txb,Behtash:2019qtk}. The
moment method turns out to be not only quickly convergent from a numerical point 
of view (see App.~G of Ref.~\cite{Behtash:2019txb}), but it also provides a unique
approach to understand non-trivial aspects of hydrodynamization. New
solutions to the equations of motion of Legendre moments are derived by employing
methods developed in the context of stability analysis of non-autonomous dynamical
systems \cite{caraballo2017applied,kloeden}, as well as techniques from
superasymptotic and hyperasymptotic analysis \cite{boyd:1999,costin2008asymptotics,
costin1995,costin1998}. These tools allow us to analyze the hydrodynamization process in two distinct regimes characterized by the size of dissipative corrections: $Kn\gg1$ (UV/early time) and  $Kn\ll 1$ (IR/late time).

 The solutions of the kinetic equations for the Legendre moments enable us to extend the definition 
of transport coefficients to 
the far-from-equilibrium regime. Within our approach transport coefficients 
depend on the deformation history of the fluid, i.e. its rheology, and their 
values change as a function of the Knudsen number.
Such dependence of the transport
coefficients on the typical gradient size is a  salient  property of for non-Newtonian fluids. 
Our findings provide  further evidence for the 
connection between hydrodynamization and the transient rheological behavior 
of the plasma~\cite{Behtash:2018moe,
Behtash:2019txb}. 

   An interesting aspect of our study is the fact that the evolution of the 
transport coefficients is determined by a RG flow equation where the role of 
the RG scale is played by the Knudsen number. Our work draws inspiration from 
recent arguments that any RG flow can be viewed as a dynamical system
\cite{Gukov:2015qea,Gukov:2016tnp}. We show that, conversely, certain dynamical
systems can be understood as RG flows, provided the existence of the slow invariant
manifold. This idea was first explored in the case of boost invariant plasmas 
governed by a Boltzmann equation in the RTA approximation~\cite{Behtash:2018moe,
Behtash:2019txb,Behtash:2019qtk}. 

 Finally, our study naturally explains the origin of the UV power series expansion considered first in~\cite{Behtash:2019txb} and later studied
in~\cite{Kurkela:2019set,Jaiswal:2019cju,Blaizot:2019scw,Blaizot:2020gql}. The
power series emerges by analyzing the stability behavior of solutions close to 
the UV fixed point. It is shown that the finite radius of convergence of the UV power series expansion depends on the value of the shear viscosity over entropy ratio $\eta/s$. We also outline the intriguing universal properties of the 
FPE related to the pullback and forward attractors, and compare 
them with other kinetic models such as the RTA 
and the full leading order AMY kinetic theory~\cite{Arnold:2002zm}.

The outline of this work is as follows. In the next section we review the 
basic ideas behind the FP equation and introduce the expansion of the distribution
function in terms of its moments whose evolution equations are derived. For
pedagogical purposes we study extensively a truncation of this dynamical system in
Sect.~\ref{sec:l1case} where we present in detail the transasymptotic techniques 
studied in this article. In Sect.~\ref{sec:generalcase} we demonstrate that 
the general solutions of the Fokker-Planck equation are written in terms of 
multiparameter transseries after resumming the fluctuations around the UV and 
IR regimes, respectively. We analyze the universal properties of non-equilibrium
Yang-Mills attractors of different theories in Sect.~\ref{sec:univ}. We conclude 
by giving final remarks. The technical details of our work are presented in the
appendices.

\section{Yang-Mills transport equation in the small angle approximation}
\label{sec:momfokpla}

We consider an interacting gluon plasma described by the Boltzmann equation
in the small angle approximation~\cite{Lifshitz81}. We focus our discussion 
on the case of an expanding system with longitudinal boost invariance 
\cite{Bjorken:1982qr}. The dynamics is invariant under the $ISO(2)\otimes
SO(1,1)\otimes Z_2$ symmetry group. The symmetry becomes manifest in 
Milne coordinates $x^\mu=(\tau,x,y,\varsigma)$, where $\tau=\sqrt{t^2-z^2}$
is the longitudinal proper time, and $\varsigma=\tanh^{-1}(z/t)$ is the 
spatial rapidity. In this coordinate system the metric is simply 
$g_{\mu\nu}=(-1,1,1,\tau^2)$. In Milne coordinates the Fokker-Planck 
equation for the one particle distribution $f(\tau,p_T,p_\varsigma/\tau)
\equiv f_{\bm p}$ is~\cite{Mueller:1999pi,Blaizot:2013lga,Blaizot:2014jna}
\be
\label{eq:FPeq}
 \partial_\tau f_{\bm p} =\mc C[f_{\bm p}]\,. 
\ee
The collision kernel $\mc C[f_{\bm p}(\tau)]$ in the small angle 
approximation takes the form~\cite{Mueller:1999pi,Blaizot:2013lga}
\be
\label{eq:FPcoll}
\besp
\mc C[f_{\bm p}]&=\lambda^2_{YM} \,l_{Cb} \, \nabla_{\bf p} \cdot \left[ {\cal J}\, \nabla_{\bf p} f_{\bf p} + {\cal K} \,\frac{{\bf p}}{p} f_{\bf p} \left( 1+ f_{\bf p} \right) \right]\,, 
\end{split}
\ee
where we introduce the t'Hooft coupling $\lambda_{YM}=g^2 N_c\equiv 
4\pi\alpha_s N_c$. 
The integrals $\mc J$ and 
$\mc K$ are given by
\bs
\label{eq:JKint}
\beal
\label{eq:Jint}
{\cal J} = {\cal J}(\tau) &= \int \frac{d^3 p}{(2 \pi)^3} \, f_{\bf p} \left[ 1 + f_{\bf p} \right], \\
\label{eq:Kint}
 {\cal K} = {\cal K}(\tau) &= 2 \int \frac{d^3 p}{(2 \pi)^3} \, \frac{f_{\bf p}}{p} .
\end{align}
\es
The Coulomb logarithm $l_{Cb}$ in the RHS of Eq.~\eqref{eq:FPcoll} is a 
divergent integral of the form 
\be
\label{eq:lcb}
\besp
l_{Cb}&=\int_{p_{min}}^{p_{max}}\frac{dp}{p}=\log\left(\frac{p_{max}}{p_{min}}\right)\,.
\end{split}
\ee
The IR momentum  divergence originates from $2\to 2$ scattering with small 
momentum transfer. In QCD these divergences are regularized by
static and dynamic screening. The corresponding mass scale is 
on the order of the Debye mass. Near equilibrium, and for particles
obeying Bose-Einstein statistics, we have~\cite{lebellac}
\be
\label{eq:debye}
m_D^2=4\,N_c\,g^2\,\int\,\frac{d^3p}{(2\pi)^3}\,\frac{f_{\bf p}^{eq.}}{p}=2\frac{\zeta(2)\Gamma(2)}{\pi^2}\,\lambda_{YM}T^2\,,
\ee
where $\zeta(n)$ and $\Gamma(n)$ are the Riemann  and Gamma functions
respectively. The UV momentum cutoff is taken to be the mean $p_T^2$, which 
close to equilibrium is 
\be
\label{eq:pt2}
\besp
\langle\,p_T^2\,\rangle&=\frac{\int d^3p\, p_T^2\,f_{\bf p}^{eq.}}{\int d^3p \,f_{\bf p}^{eq.}}
=2\frac{\zeta(5)\Gamma(5)}{\zeta(3)\Gamma(3)}\,T^2\,.
\end{split}
\ee
As a result, near equilibrium the Coulomb logarithm~\eqref{eq:lcb} is 
approximately given by
\be
\label{eq:lcb2}
l_{Cb}=\int_{m_D}^{\sqrt{\langle p_T^2\rangle}}\frac{dp}{p}=\log\left(\frac{\sqrt{\langle p_T^2\rangle}}{m_D}\right)\sim\frac{1}{2}\log\left(\frac{\mc A}{\lambda_{YM}}\right)\,,
\ee
with $\mc A = 72\zeta(5)/\zeta(3)\approx 62.1$. 
This estimate 
gives a Coulomb logarithm which is a constant, independent of the 
energy density of the medium, but logarithmically dependent on the 
coupling constant. This approximation has been used in many 
previous studies~\cite{Blaizot:2013lga,Blaizot:2017lht,Blaizot:2019scw,
Blaizot:2019dut}. 
Replacing $f^{eq.}_{\mathbf p}$ with the general non-equilibrium $f_{\mathbf p}$ in ~\eqref{eq:debye} and~\eqref{eq:pt2}
and 
evaluating the
integrals  numerically at each
time step allowed the authors of Refs.~\cite{Bjoraker:2000cf,Tanji:2017suk}  to  account  for 
the time-dependence of the UV and IR momentum cutoffs. 

 The approximation of a constant Coulomb logarithm gives the 
correct dependence of the shear viscosity on the coupling 
constant in the near-equilibrium, weak coupling limit. However, 
the numerical pre-factor does not agree with calculations based
on the full Hard Thermal Loop (HTL) result~\cite{Baym:1990uj,
Heiselberg:1994vy}. This issue was addressed in~\cite{York:2014wja},
where the authors propose a simple regulator that reproduces the 
leading order HTL result for drag and momentum diffusion in 
soft $2\to 2$ scattering. This is a useful prescription, but it 
does not affect the late-time emergent hydrodynamic behavior. 
We shall comeback to this issue in Sect.~\ref{sec:univ}.

\subsection{Ansatz for the distribution function}
\label{subsec:ans}

One of the most widely used methods to solve the Boltzmann equation was
developed decades ago by Grad~\cite{Grad}. In this approach the problem 
of solving the Boltzmann equation is converted into a set of 
nonlinear PDEs for moments of the one-particle distribution function. This approach
is very useful when analyzing the resurgence properties of the nonlinear 
ODEs for the moments~\cite{Behtash:2017wqg,Behtash:2018moe,Behtash:2019txb,
Behtash:2019qtk}.

In this work we will consider the following ansatz for the distribution 
function in a boost invariant 
system~\cite{Behtash:2018moe,Behtash:2019txb,Behtash:2019qtk} 
\be
\label{eq:ansatzf}
f_{\bm p} = f^{\rm eq}_{\bm p}\,\sum_{\ell=0}^{+\infty} \cl (\tau) P_{2 \ell}(\cos \theta_p), \\
\ee
where $p^\tau=\sqrt{p_x^2+p_y^2+\left(p_\varsigma/\tau\right)^2}$, $\cos \theta_p = p_\zeta/(\tau p^\tau)$, and $P_{2l}$ are the Legendre polynomials. In Eq.~\eqref{eq:ansatzf} the equilibrium distribution function $f^{\rm eq}_{\bm p}$ is 
\be
\label{eq:eqdist}
f^{\rm eq}_{\bm p} =\nu_{eff} \frac{1}{e^{p^\tau/T(\tau)}-1},
\ee
where $T$ is the temperature of the system which is defined below via the 
Landau matching condition and $\nu_{eff}$ are the effective degrees of freedom which for simplicity we set to be $\nu_{eff}\equiv 1$. We implicitly assume that $f_{\bm p}$ is 
independent of spin and color. The ansatz~\eqref{eq:ansatzf} is consistent 
with the restrictions imposed by $ISO(2)\otimes SO(1,1)\otimes Z_2$. In 
particular, the distribution function is invariant under the action of the 
Killing vectors $\phi_i$ of this symmetry group, i.e., $\partial 
f_{\bm p}/\partial \phi_i=0$~\cite{Behtash:2018moe,Behtash:2019txb,
Behtash:2019qtk}. For simplicity we fixed $\mu\equiv 0$ in the equilibrium distribution
function~\eqref{eq:eqdist}. Furthermore, the ansatz~\eqref{eq:ansatzf}
does not carry information about the non-linear relaxation of the transient 
high energy tails of the distribution function which have been studied 
previously within the moment method~\cite{Behtash:2018moe,Behtash:2019txb,
Behtash:2019qtk,Bazow:2015dha,Bazow:2016oky,Blaizot:2013lga,Blaizot:2014jna,
krookwu,bobylev1976fourier}. A more general ansatz which encodes some of 
the information on the high energy tails was discussed recently in
Ref.~\cite{Blaizot:2019dut}. 

The Legendre moments $\cl$ are directly determined by Eq.~\eqref{eq:ansatzf}, 
\be
\label{eq:cl}
\cl =\frac{30(4l+1)}{\pi^2\,T^4}\,\left\langle\,\left(-u\cdot p\right)^2\,P_{2\ell}(\cos \theta_p)\,\right\rangle\,,
\ee
where we denote $\langle\cdots\rangle_X\equiv \int_{\bm p}\cdots f^X_{\bm p}$ 
and the momentum measure is $\int_{\bm p}\equiv \int 
d^2p_Tdp_\varsigma/\left[(2\pi)^3\,\tau\,p^\tau\right]$. 
If the system reaches the thermal equilibrium state, the moments 
$c_l^{eq.}=\delta_{l0}$. 

For the Bjorken flow the energy momentum tensor $T^{\mu\nu}=\langle 
p^\mu\,p^\nu\rangle$ is given by \cite{Molnar:2016vvu,Martinez:2017ibh,
Behtash:2018moe,Behtash:2019txb,Behtash:2019qtk} 
\be
\label{eq:tmn}
T^{\mu\nu}=\epsilon\,u^\mu\,u^\nu\,+\,\mc P_L\,l^\mu\,l^\nu\,+\mc P_T\,\Xi^{\mu\nu}\,,
\ee
where we denote the time-like vector identified with the fluid velocity
$u^\mu=(1,0,0,0)$ (with $u^\mu u_\mu=-1$), the space-like normal vector
pointing along the $\varsigma$ direction is $l^\mu=(0,0,0,1)$ (with $l^\mu
l_\mu=1$) and the projection operator $\Xi^{\mu\nu}=g^{\mu\nu}+u^\mu
u^\nu+l^\mu l^\nu$ which is orthogonal to both $u^\mu$ and $l^\mu$
respectively. The energy density $\epsilon$, longitudinal and transverse
pressures, $\mc P_L$ and $\mc P_T$ respectively, are written in terms of the
angular moments as follows
\bs
\label{eq:tmncomp}
\beal
\label{eq:matchene}
\epsilon&=\left\langle\,\left(-u\cdot p\right)^2\,\right\rangle=\frac{\pi^2}{30}\,c_0\,T^4\,,\\
\mc P_T&=\left\langle\,\frac{1}{2}\Xi^{\mu\nu}p_\mu p_\nu\,\right\rangle=\epsilon\left(\frac{1}{3}-\frac{1}{15}c_1\right)\,,\\
\mc P_L&=\left\langle\,\left(l\cdot p\right)^2\,\right\rangle=\epsilon\left(\frac{1}{3}+\frac{2}{15}c_1\right)\,.
\end{align}
\es
It follows that $\epsilon=2\mc P_T+\mc P_L$ as expected. The Landau matching 
condition for the energy density $\epsilon=\epsilon_{eq.}\equiv \langle\,(-u\cdot p)^2\,\rangle_{eq.}$ implies $c_0\equiv 1$.  For the 
Bjorken case the normalized pressure anisotropy is the ratio of the independent
shear viscous tensor over the energy is written in terms of the Legendre moment
$c_1$, i.e.,
\be
\label{eq:barpi_c1}
\Delta=\frac{\tau^2\pi^{\varsigma\varsigma}}{\epsilon}=\frac{2}{15}c_1\,.
\ee
Moreover, the non-negativity property of the longitudinal and transverse
pressures $\mc P_{L(T)}\geq 0$ implies $-5/2\,\leq\,c_1\,\leq\,5$. This bounds
the basin of attraction from below and above and is satisfied in general only
by the exact kinetic solution to the FP equation~\eqref{eq:FPcoll}.
Nonetheless, it is known that any perturbative approach which aims to find an
approximate solution to the Boltzmann equation does not necessarily obey this
constraint~\cite{Martinez:2009mf}. 

\subsection{Evolution equation of the Legendre moments evolution}
\label{subsec:momfokpla}

 We truncate the expansion in Legendre polynomials at order $L$ and write 
the Legendre moments as a vector ${\bf c}(\tau) \equiv (c_{1}(\tau),\cdots,
c_{L}(\tau))^{\top}$. Following the procedure outlined in 
\cite{Behtash:2018moe,Behtash:2019txb,Behtash:2019qtk} we find that the 
evolution equations for the temperature and the Legendre moments are 
given by the following nonlinear coupled of ODEs
\bs
\label{eq:ODE_tau}
\beal
\label{eq:consqnTtau}
\frac{dT}{d\tau}&=-\frac{T}{3\tau}\left(1+\frac{c_1}{10}\right)\,,\\
\label{eq:cldynsys}
\frac{d {\bf c}}{d \tau} &= {\bf F}(T,{\bf c},\tau)\,, 
\end{align}
\es
with
\begin{align}
{\bf F}(T,{\bf c},\tau)=&-\frac{2}{3}\Big[ \frac{1}{\tau} 
   \left\{ {\frak X}({\bf c}) {\bf c} + \bm{\Gamma} \right\} 
   \nonumber \\
 & \hspace{1cm}\mbox{}     
  + \left\{  \Lambda + {\frak Y}({\bf c}) + {\frak Z}({\bf c})  \right\}
   T\,{\bf c}\Big]
\end{align}
and
\bs
\begin{align}
\label{eq:chimat}
{\frak X}({\bf c}) &= \bar{{\frak X}}- \frac{1}{5} {c}_{1}(\tau){\bf 1}_{L} ,\\
\bar{{\frak X}}   & =\frac{3}{2}
\begin{pmatrix}
  \frak B_{1}   & {\frak A}_{1} &  & &\\
  {\frak C}_{2}  & \frak B_{2} & {\frak A}_{2} & &  \\
   & \ddots & \ddots  & \ddots & &\\
   & &  {\frak C}_{L-1}  & \frak B_{L-1} & {\frak A}_{L-1} \\
   & & &  {\frak C}_{L}  & \frak B_{L}   \\
\end{pmatrix}\, . 
\end{align}
\es
The matrix elements are defined by
\bs
\begin{align}
{\frak A}_{\ell} &= -\frac{ (2 \ell -1) (2 \ell +1)(2 \ell + 2)}{(4 \ell + 3)(4 \ell + 5)},  \\
{\frak B}_l  &=\frac{ 2(14 \ell^2 + 7 \ell - 2 ) }{(4 \ell - 1)(4 \ell + 3)}-\frac{4}{3}\,,  \\
{\frak C}_{\ell} &= \frac{2 \ell (2 \ell - 1)(2 \ell + 2) }{(4 \ell - 3)(4 \ell - 1)}\, .
\end{align}
\es
The Lyapunov exponents and the vector $\Gamma$ are defined by 
\bs
\begin{align}
\label{eq:Lyap_mat}
 \Lambda &= {\rm diag.}(\lambda(1), \cdots, \lambda(L))\,,\\
 \label{eq:gamvec}
\bm{\Gamma} &=(3{\frak C}_{1}/2,0,\cdots, 0)^{\top} =(4,0,\cdots, 0)^{\top}\,.
\end{align}
\es
Finally, we have defined the matrices
\begin{widetext}
\bs
\label{eq:matODE}
\beal
 {\frak Y}({\bf c}) &= \hat{\kappa}
\begin{pmatrix}
  \sum_{m=1}^{2} \Omega_{1m1} c_{m}(\tau)  & \cdots & \sum_{m=n-1}^{1+n}
  \Omega_{1mn} c_{m}(\tau)  &  \cdots & \sum_{m=L-1}^{1+L} \Omega_{1mL}
  c_{m}(\tau) \\
  \vdots  & \ddots & \vdots & \ddots & \vdots  \\
  \sum_{m=\ell-1}^{\ell+1} \Omega_{\ell m1} c_{m}(\tau)  &\cdots &  
  \sum_{m={\rm Max}[|\ell-n|,1]}^{\ell+n} \Omega_{\ell m n} c_{m}(\tau)  &  
  \cdots & \sum_{m=L-\ell}^{\ell+L} \Omega_{\ell mL} c_{m}(\tau) \\
\vdots  & \ddots & \vdots & \ddots & \vdots  \\
  \sum_{m=L-1}^{L+1} \Omega_{Lm1} c_{m}(\tau)  & \cdots & \sum_{m=L-n}^{L+n}
  \Omega_{Lmn} c_{m}(\tau)  &  \cdots & \sum_{m=1}^{2L} \Omega_{LmL} c_{m}(\tau) 
  \\
\end{pmatrix}, \\
\Omega_{\ell m n} &=\frac{\alpha_{m-n+\ell}\,\alpha_{n+m-\ell}\,\alpha_{n-m+\ell}}{\alpha_{n+m+\ell}} \cdot \frac{4 \ell + 1}{2(n+m+\ell)+1}, \qquad \text{with}\,\,\alpha_{\ell}= \frac{(2 \ell - 1)!!}{\ell !}\,,\\
 {\frak Z}({\bf c}) &=  \hat{\kappa}  \sum_{n = 1}^{+\infty} \frac{(2\ell - 1)( \ell + 1) }{3(4 n + 1) } c_{n}(\tau)^2 {\bf 1}_{L}\,.
\end{align}
\es
\end{widetext}

In the previous expressions the parameters $\kappa$, $\theta_0$ and $\hat{\kappa}$ are given by respectively
\be
\label{eq:paramsODEs}
\besp
&\kappa=\pi^2/3-2 \zeta(3)\,,\qquad\theta^{-1}_{0}= \frac{5}{8\,\pi^5}\,\lambda_{YM}^2\,l_{Cb}\,,\\
&\hat{\kappa}=\frac{3\kappa}{2\theta_0}\,,\qquad\lambda(l)=\frac{3}{2\theta_0}\left(\kappa+\frac{l(2l+1)}{3}\pi^2\right)\,.
\end{split}
\ee
The derivation of the equations of motion~\eqref{eq:ODE_tau} is presented in 
App.~\ref{app:eqcl}. The set of ODEs~\eqref{eq:ODE_tau} constitutes a nonlinear
non-autonomous dynamical system~\cite{caraballo2017applied,kloeden} due to the
explicit dependence on the proper time $\tau$ in the RHS of these equations.
The strength of the effect of the collisions enters in the ODEs via the parameter $\theta_0$~\eqref{eq:paramsODEs};  we thus will vary this parameter instead of
the t'Hooft coupling $\lambda_{YM}$ when showing the numerical results.  
The nonlinear nature of the FP 
equation~\eqref{eq:FPeq} is manifest in the mode-mode couplings
among different moments, and in the temperature appearing in the RHS of
Eq.~\eqref{eq:ODE_tau}. Finally, the time-evolution of the energy-momentum
tensor~\eqref{eq:tmn} can be  fully reconstructed from the solutions of the
temperature $T$ and the full set of Legendre moments $c_l$. 

The equations of motion of the Legendre moments~\eqref{eq:cl} 
for the conformal Boltzmann equation within the relaxation time approximation 
(RTA) are given by (see ~\cite{Behtash:2018moe,Behtash:2019txb})
\be
\label{eq:ODE_cl_tau_RTA}
\besp
\frac{d {\bf c}}{d \tau} &= {\bf F}_{RTA}(T,{\bf c},\tau)\,, \\
{\bf F}_{RTA}(T,{\bf c},\tau)&=-\frac{2}{3}\left[ \frac{1}{\tau} \left\{ {\frak X}({\bf c}) {\bf c} + \bm{\Gamma} \right\} + \Lambda_{RTA}\,T\,{\bf c}\right]\,,\\
\Lambda_{RTA}&=\frac{3}{2\,\Theta_0}\,\text{diag.}\,\left(1,1,....1\right)\,.
\end{split}
\ee
Here $\Theta_0$ is a proportionality constant between 
the relaxation time and the shear viscosity over entropy ratio. This constant
can be determined from relativistic kinetic theory methods which for the
conformal RTA approximation gives us $\tau_r=\frac{5}{T}\frac{\eta}{s}$
\cite{Denicol:2010xn,Florkowski:2013lza,Florkowski:2013lya,Denicol:2011fa,Behtash:2019txb,Teaney:2013gca,Chattopadhyay:2014lya,Bhalerao:2013pza,Jaiswal:2013vta,Jaiswal:2013npa,Romatschke:2011qp,Yan:2012jb,Denicol:2012es}, i.e.  $\Theta_0\equiv 5\eta/s$. 

%
%

At linear order the form of the evolution equations for the Legendre moments
for the RTA and FP, Eqs.~\eqref{eq:ODE_cl_tau_RTA} and~\eqref{eq:ODE_tau}
respectively, is very similar. At this order the main difference between 
the two models lies in the specific relaxation scale of the associated mode
$c_l$. For the RTA approximation all modes decay with the same relaxation 
scale $\tau_r$ while in the Fokker-Planck case each $c_l$ has a characteristic
relaxation scale $\lambda(l)^{-1}$ which increases as a function of the $l$ 
(see Eq.~\eqref{eq:paramsODEs}). Therefore for the FPE there is a clear
hierarchy of fast and slow modes\footnote{
For hard spheres and within the kinetic theory framework similar results have 
been found in the past. In those models a hierarchy of scales between the fast 
and soft modes also emerges~\cite{Bazow:2016oky,krookwu,bobylev1976fourier}.}.

The nonlinearities of the FPE are encoded in the mode coupling between 
moments of different order and in the implicit dependence of the temperature 
on $c_1$. These effects drive the system away from equilibrium and
delay the relaxation to the equilibrium state. For instance, for the RTA
Boltzmann equation a new non-hydrodynamic mode survives in the long 
wavelength limit due to a non-linear coupling to the shear viscous
tensor~\cite{Behtash:2019txb}. This result contradicts the common assumption 
of taking the typical relaxation time scale of a mode as a guide for
constructing effective theories in the long time limit.
 
\subsection{Dimensional reduction}
\label{subsec:dimrec}
There is an important simplification of the nonlinear ODEs~\eqref{eq:ODE_tau} 
which makes the asymptotic analysis simpler. It is possible to dimensionally
reduce this dynamical system from $L+1\to L$ by introducing the variable\footnote{
The resurgence analysis can also be studied in terms of the original variable
$\tau$~\cite{Behtash:2019txb}. See the general Fokker-Planck case in
App.~\ref{app:timetrans} and the RTA Boltzmann case is extensively discussed 
in Appendixes A-D in Ref.~\cite{Behtash:2019txb}.}
$w=\tau T(\tau) \in {\mathbb R}^+$. 
Since $w\sim Kn^{-1}$~\cite{Behtash:2017wqg,Behtash:2018moe,Behtash:2019txb,Behtash:2019qtk}, 
the  variable $w$ encodes the strength of the dissipative corrections. 
This fact will be important for re-interpreting the dynamical system of ODEs as RG flows in Sect.~\ref{subsubsec:RGflow}. In terms of $w$
Eqs.~\eqref{eq:ODE_tau} can be written as
\bs
\label{eq:prepODE_c_w}
\beal
\label{eq:T_w_sol}
T(w)&=\mc D (w,w_0)\,T_0\,,\\
\mc D (w,w_0)&=\exp\left[-\frac{1}{2}\int_{w_0}^w\,\frac{dw'}{w'}\,\frac{\left(1+\frac{c_1(w')}{10}\right)}{\left(1-\frac{c_1(w')}{20}\right)}\right]\,,\nonumber\\
& \frac{d {\bf c}}{d w} = {\bf F}({\bf c},w), \\
\label{eq:ODE_cl_w}
{\bf F}({\bf c}, w) &=  - \frac{1}{1-\frac{1}{20}c_{1}(w) } \left[ \frac{1}{w} \left\{ {\frak X}({\bf c}) {\bf c}(w) + \bm{\Gamma} \right\} \right.\nonumber\\
&\left.+ \left\{  \Lambda + {\frak Y}({\bf c}) + {\frak Z}({\bf c})  \right\} {\bf c}(w) \right].
\end{align}
\es
The dynamics of this dimensionally reduced system of ODEs depends only on the
Legendre moments $c_l$~\footnote{
The index $i$ of vectors and matrices runs from $i=1,2,...,L$. If one 
performs a truncation of the non-autonomous dynamical system of
Eq.~\eqref{eq:prepODE_c_w} it is understood that $c_l\equiv 0$ for $l>L$.}
 and does not involve the temperature variable explicitly. It is important to emphasize that 
the dimensionally reduced dynamical system does not preserve the topological
properties of the original nonlinear ODEs~\eqref{eq:ODE_tau}. For instance, 
a new coordinate singularity emerges when $c_1=20$ (see the denominator in 
the integrand of the damping function $D(w,w_0)$ in Eq.~\eqref{eq:T_w_sol}) 
which does not exist in the $\tau$ variable. Furthermore, in the $w$ variable,
one can find a set of bounded solutions at $w_0\rightarrow0$ for a fixed $w$,
the so-called pullback attractor~\cite{Behtash:2019txb,Behtash:2019qtk}; 
whereas the singularity at $\tau=0$ forbids the existence of any bounded 
solutions in the UV. This is of course expected on many grounds but a simple 
explanation is that the system undergoes a topological change under 
$\tau\rightarrow w$ that lifts the $\tau=0$ singularity at the expense of 
introducing a new singularity at $c_{1}=20$ \cite{Behtash:2019txb,
Behtash:2019qtk}. 

\section{Transasymptotic analysis: the truncated l=1 case}
\label{sec:l1case}

  Before presenting the transseries solutions for the nonlinear 
ODEs~\eqref{eq:prepODE_c_w} we first illustrate our techniques by considering
the case where the system is truncated to a single non-hydrodynamic degree 
of freedom. Physically, this degree of freedom corresponds to the viscous
shear tensor. This warm-up exercise illustrates the main aspects 
of the general
asymptotic analysis to be discussed in Sect.~\ref{sec:generalcase}. 

Let us assume that the moments $c_l\equiv 0$ for $l>1$. In terms of the
variable $w$, the truncated $l=1$ case read as
\bs
\label{eq:ODE_w} 
\beal
\label{eq:ODE_c1_w} 
\frac{d c_1}{d w}&= F_1(w,c_1)\,,\\
\label{eq:F1_w}
F_1(w,c_1)&=
-\frac{1}{\left( 1-\frac{1}{20}c_{1} \right)}\left[ \frac{1}{w} \left( 4 +  \frac{5}{7}c_1 - \frac{1}{5}c_1^2    \right) \right.\nonumber\\
&\left.+   \lambda(1)c_1 +  \frac{2 \hat{\kappa}}{7}  c_{1}^2 + \frac{2\hat{\kappa}}{15} c_{1}^3 \right] \,.
\end{align}
\es
%
%
\begin{figure}[t]
\begin{center}
\includegraphics[scale=1.7]{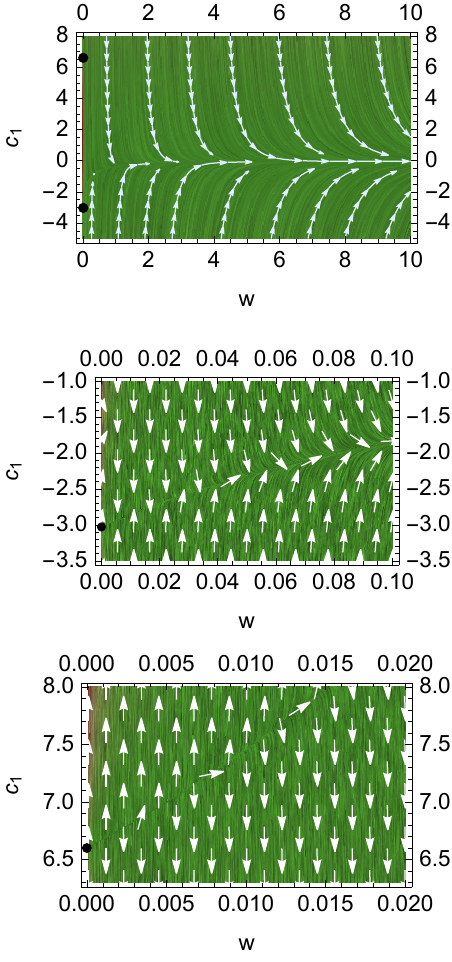}
	\caption{(Color online) Flow diagram of the differential equation for the Legendre moment~\eqref{eq:ODE_c1_w} in the $(c_1,w)$ space. Top, center and bottom panels show the flows close to the IR and two UV fixed points respectively. The black dots showed in each panel correspond to the two UV fixed points of the ODE~\eqref{eq:ODE_c1_w}.} 
	\label{fig:phase}
	\end{center}
\end{figure}
%
%

Before discussing the transasymptotic analysis we would like to comment
on the existence of the pullback and forward attractors of the truncated
ODE~\eqref{eq:ODE_c1_w}. The extension to the full set of nonlinear
ODEs~\eqref{eq:prepODE_c_w} is similar but more difficult to 
visualize~\cite{Behtash:2019txb}. For now, we note that we can identify 
attractors by inspecting the flow diagram of the ODE~\eqref{eq:ODE_c1_w}. 
We will characterize the attractors more fully below. In the top panel of
Fig.~\ref{fig:phase} we observe that in the IR regime all flows converge
asymptotically towards the value $c_1\to 0$ when $w\gg 1$ regardless of 
their origin. This shows that there is only one IR fixed point associated to the forward attractor. 

 In the UV limit, on the other hand, there are two fixed points (black 
dots in Fig.~\ref{fig:phase}) which are located along the $c_1$ axis in the
small $w_0\to 0$ limit ($w_0$ is the initial value of $w$). The UV behavior 
of the flow lines in each fixed point is rather different. For the UV fixed
point located in the positive $c_1$ region one observes that the flow lines 
emerge out of it in all the directions and thus, this fixed point is a source.
In the negative $c_1$ region, we observe that near the UV fixed point the flow lines coming from above or below repel each other so the fixed point is a saddle. However, although the flow lines repel each other, they merge quickly in the vicinity of this fixed point. Thus, the UV fixed point located in the negative $c_1$ region is identified as the pullback attractor. 

 We will see that the behavior of the transseries solution near these fixed points is rather different. We note that that the behavior 
of the flows in the $\tau$ variable differs from the one in $w$. In the 
former case the temperature is a dynamical variable. It diverges in the limit $\tau\to 0$ due to the presence of the singularity at $\tau=0$; therefore there is no meaningful definition of a pullback attractor in the $\tau$ variable (see also~\cite{Behtash:2019qtk,Behtash:2019txb}). 

\subsection{Transseries solution in the IR limit}
\label{subsec:IR_c1}

The leading order term of each Legendre moment in Eqs.~\eqref{eq:prepODE_c_w} 
is $c_l=\mc O(w^{-l})$ $\forall\,l>0$ in the  IR regime~\cite{Behtash:2019txb}. Hence, Eq.~\eqref{eq:ODE_c1_w} admits the following asymptotic expansion 
for $c_1$
\be
\label{eq:asympc_1}
c_1=\sum_{k=0}^\infty\,u^{(0)}_{1,k}\,w^{-k}\,.
\ee
The coefficients $u^{(0)}_{1,k}$ are obtained simply by inserting the IR
expansion into the original ODE~\eqref{eq:ODE_c1_w}. For instance, the 
first three coefficients are given by
\be
\label{eq:g1g2_c1}
\besp
&u^{(0)}_{1,0}=0\,,\qquad
u^{(0)}_{1,1}=-\frac{4}{\lambda(1)}\,,\\
&u^{(0)}_{1,2}=-\frac{8}{7\,\lambda(1)^2}\left(1+4\frac{\hat{\kappa}}{\lambda(1)}\right)\,. 
\end{split}
\ee
Here the reality 
condition of $c_1$ was implicitly taken into account. 
The hydrodynamic gradient expansion of the shear viscous tensor~\cite{Baier:2007ix} provides the following expression for $c_1$~\cite{Behtash:2018moe,Behtash:2019txb}
\be 
\label{eq:c1asympt}
c_1=-\frac{40}{3}\,\frac{1}{w}\,\frac{\eta}{s}\,-\,\frac{80}{9}\,\frac{1}{w^2}\,\frac{T\,(\eta\tau_\pi-\lambda_1)}{s}\,\cdots\,,
\ee
where we use Eq.~\eqref{eq:barpi_c1}. Comparing the previous expression and Eqs.~\eqref{eq:g1g2_c1} one concludes 
\be
\label{eq:asympmatch}
\besp
\frac{\eta}{s}&=\frac{3}{10}\frac{1}{\lambda(1)}\,,\\
\frac{T\,(\eta\tau_\pi-\lambda_1)}{s}&=\frac{9}{70\lambda(1)^2}\left(1+4\frac{\hat{\kappa}}{\lambda(1)}\right)\,.
\end{split}
\ee
with $\lambda(1)$ and $\hat{\kappa}$ given in Eqs.~\eqref{eq:paramsODEs}. 
We will see below that $\lambda(1)$ is related to the rate at which 
fluctuations decay close to the IR fixed point, i.e. the Lyapunov exponent.
The coefficients $u^{(0)}_{1,k}$ that enter the IR expansion
\eqref{eq:asympc_1} are understood as transport coefficients in the IR limit.
Furthermore, the set of the identities~\eqref{eq:asympmatch} turn out be of 
importance when generalizing the concept of a transport coefficient to 
the far-from-equilibrium regime~\cite{Behtash:2018moe,Behtash:2019txb} as we 
shall discuss in Sect.~\ref{subsec:dyn-ren_c1}.

It is known that the IR expansion~\eqref{eq:asympc_1} is divergent since 
its coefficients grow factorially~\cite{Heller:2015dha}. These expansions 
are merely formal expressions and emerge as asymptotic solutions of a 
certain class of differential equations~\cite{costin2008asymptotics}. In 
order to see this let us simply perform a linear perturbation around the
perturbative expansion~\eqref{eq:asympc_1}. Thus, by shifting $c_1\to 
\bar{c}_1+\delta c_1$ with $\bar{c}_1 =\sum_{k=0}^\infty g_k w^{-k}$ while 
keeping terms up to $\mc O(w^{-1})$ one obtains the following linearized 
differential equation for the perturbation $\delta c_1$ 
\be
\label{eq:dc1_linear}
\besp
\frac{d\delta c_1}{dw}&=\frac{\partial F_1}{\partial c_1}\biggr|_{c_1=\bar{c}_1}\,\delta c_1\,,\\
&\approx -\left(\lambda(1)+\frac{1}{w}\left(\frac{18}{35}-\frac{16}{7}\frac{\hat{\kappa}}{\lambda(1)}\right)\right)\delta c_1\,\,,
\end{split}
\ee
whose solution reads 
\be
\label{eq:pertdc1}
\besp
\delta c_1(w)&=\sigma_1\,e^{-S_1 w}\,w^{-b_1}\,,\\
S_1:=\lambda(1)&\,,\, \qquad b_1:=\frac{18}{35}-\frac{16}{7}\frac{\hat{\kappa}}{\lambda(1)}\,,
\end{split}
\ee
where $\sigma_1$ is the integrating constant, $S_1$ is recognized as the 
Lyapunov exponent and $b_1$ is the anomalous dimension. The presence of the
exponential terms is an indication that in order to capture the transient 
behavior of the solutions it is needed to go beyond the perturbative IR 
expansion~\eqref{eq:asympc_1}.

The leading exponential term~\eqref{eq:pertdc1} is the first type of a
large set of exponentials terms that appear when summing over the 
fluctuations around the IR fixed point. In order to include systematically
these type of terms we follow Costin's prescription~\cite{costin1998}. First,
notice that in the IR limit the ODE~\eqref{eq:ODE_c1_w} takes the following 
asymptotic form 
\beal
\label{eq:ODE_prep_c1}
\frac{dc_1}{dw}&\approx-\sum_{n=0}^\infty\,\left(\frac{c_1}{20}\right)^n\,\left[\frac{1}{w} \left( 4 +  \frac{5}{7}c_1 - \frac{1}{5}c_1^2    \right) \right.\nl
&\left.+   \lambda(1)c_1 +  \frac{2 \hat{\kappa}}{7}  c_{1}^2 + \frac{2\hat{\kappa}}{15} c_{1}^3 
\right]\,,\nl
&\approx\,-\left[\lambda(1)\,c_1+\frac{1}{w} \left( 4 +  \frac{32}{35}c_1 \right) \right]\,+\,R(c_1,w)\,,
\end{align}
where $R(c_1,w)$ is a non-linear polynomial function of $c_1$ and $w$. The 
asymptotic limit of the differential equation~\eqref{eq:ODE_prep_c1} coincides 
with the prepared form of the generic class of differential equations studied 
by Costin~\cite{costin1998}. Thus, given the regularity of the solutions at
$w\to\infty$ as well as the non-vanishing value for the Lyapunov exponent 
$S_1$~\eqref{eq:pertdc1}, Eq.~\eqref{eq:ODE_c1_w} has an exact transseries 
solution~\cite{costin1998}
\bs
\label{eq:trans_sol_c1}
\beal
c_1(w)&=\sum_{k=0}^\infty\,\sum_{n=0}^\infty u^{(n)}_{1,k}\,w^{-k}\,[\sigma_1\,\zeta_1(w)]^n\,, 
\\
\label{eq:genIRtransm}
\zeta_1(w)&=\,e^{-S_1 w}\,w^{-b_1}\,.
\end{align}
\es
From the physical point of view, the transseries~\eqref{eq:trans_sol_c1} 
describes deviations from thermal equilibrium due to exponentially damped 
modes multiplied by gradient terms. The exponential damped terms, usually 
called non-hydrodynamic modes, play a role analogous to instantons in 
quantum field theory and quantum mechanics. In order to determine the 
coefficients $u^{(n)}_{1,k}$ we simply equate~\eqref{eq:trans_sol_c1} with
Eq.~\eqref{eq:ODE_c1_w}. As a result we get the following recursive relation
\begin{widetext}
\be
\label{eq:rec_rel_c1}
\besp
& \left( -  m S_1  + \lambda(1)  \right) u_{1,k}^{(m)} + \left( -m b_1 - k + \frac{12}{7} \right) u_{1,k-1}^{(m)} + 4 \delta_{m,0} \delta_{k,1} \\
& - \frac{1}{20} \sum_{m_1,m_2=0}^{m_1+m_2=m} \sum_{k_1,k_2=0}^{k_1+k_2=k}
\left[
\left( -  m_1 S_1 - \frac{40 \hat{\kappa}}{7}  \right) u_{1,k_1}^{(m_1)} + \left( -m_1 b_1 - k_1 + 5 \right) u_{1,k_1-1}^{(m_1)}
  \right] u_{1,k_2}^{(m_2)} \\
& + \frac{2 \hat{\kappa}}{15} \sum_{m_1,m_2,m_3=0}^{m_1+m_2+m_3 = m} \sum_{k_1,k_2,k_3=0}^{k_1+k_2+k_3 = k} u^{(m_1)}_{1,k_1} u^{(m_2)}_{1,k_2} u^{(m_3)}_{1,k_3}  = 0,
\end{split}
\ee
\end{widetext}
where $u_{1,k}^{(m)}=0$ for $k<0$ and we take $u_{1,0}^{(1)}=1$ as the
normalization of the integration constant $\sigma_1$. When  $m=0$ and 
$k=0,1,2$ in the recursive relation~\eqref{eq:rec_rel_c1}  one reproduces 
Eqs.~\eqref{eq:g1g2_c1} as expected.

%
\begin{figure}[t]
\begin{center}
	\includegraphics[scale=.2]{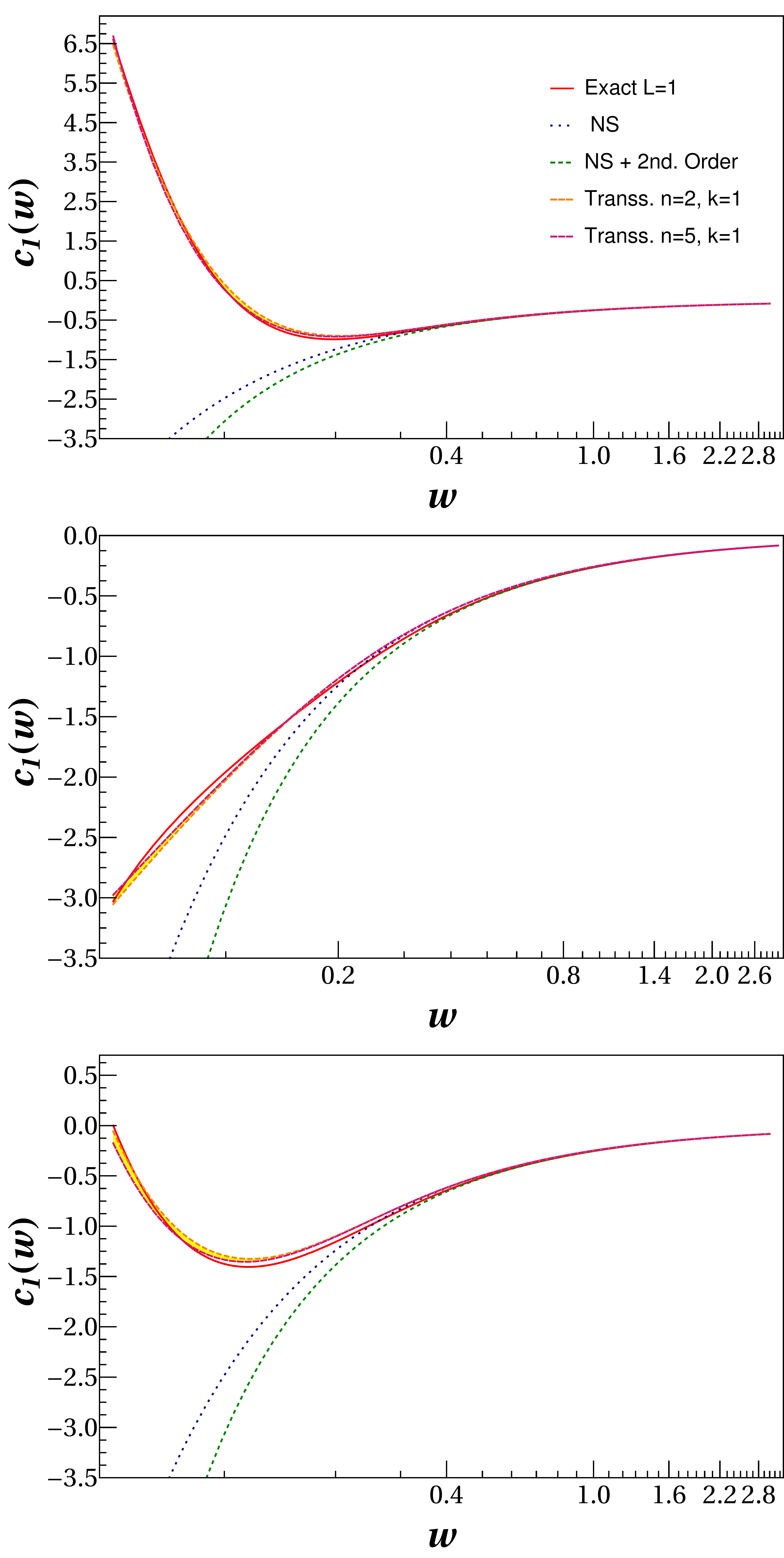}
	\caption{(Color online) Evolution of the non-hydrodynamic mode $c_1$ in terms of $w$ for the exact numerical result (red line), truncated transseries solutions by adding n=2 transmonomials to $k=1$ perturbative order (blue dotted line) and n=15 transmonomials to $k=3$ perturbative order (green dashed line) orders in the IR perturbative expansion respectively, Navier-Stokes (orange dashed line) and NS + 2nd. order (magenta dashed line). See text for further details.} 
	\label{fig:1}
	\end{center}
\end{figure}
%

We conclude this section by presenting some numerical results. In 
Fig.~\ref{fig:1} we show the Legendre mode $c_1$ as a function of $w$ for 
the exact numerical solution (red line), NS (blue dotted line), NS + 2nd order
(green dashed line) and truncated transseries solutions where we added two
(orange dashed line) and five (magenta dashed line) transmonomials to the 
1st IR perturbative order. The initial condition for the numerical solution 
of the ODE~\eqref{eq:ODE_c1_w} was $c_1(w_0=0.03)=\{6.60119,0,-3.02976\}$ 
while fixing $\theta_0=1$~\footnote{The values for $c_1(w_0=0.03)=\{6.60119,-3.02976\}$ correspond to the UV fixed points, see Eq.~\eqref{eq:UVfixedc1}.}. Note that at each order of the resummed theory 
we have to numerically match the transseries parameter $\sigma$. Very 
few cases are known where one can determine $\sigma$ exactly 
\cite{Blaizot:2020gql}. In the present work we have determined $\sigma$ using 
a numerical least-square fit \cite{Behtash:2019txb}. This leads to some
uncertainty, which is not unexpected given the difficulty of matching the 
IR and UV data without performing an all-order resummation. The shaded 
yellow area in Fig.~\ref{fig:1} shows the variation of $\sigma$ between 
each truncated transseries solution. 

 Fig ~\ref{fig:1} shows that the perturbative series at first and second 
orders does not match the numerical results for small values of $w$. On 
the other hand, the truncated transseries~\eqref{eq:trans_sol_c1} where 
only a few transmonomials ($n=2,5$ respectively) were included is very 
close to the exact numerical result. We note that some care has to be 
taken in making such comparisons. Any truncated transseries will
potentially exhibit large deviations from the exact numerical result
in some range of $w$. This is due to the growth of the inherent error 
associated with any type of truncation scheme, see App.~\ref{app:exp_error}.
Any resummed perturbative series with a few transmonomials has a finite 
radius of convergence. The full transseries solution coincides with the 
exact theory only when summing over all the non-perturbative sectors of 
the perturbative series~\cite{costin1998}. 

\subsection{Transasymptotic matching: $l=1$ case}
\label{subsec:dyn-ren_c1}

An interesting property of the IR transseries solution~\eqref{eq:trans_sol_c1}
emerges when rearranging the terms close to the IR fixed point. This
procedure 
 is known as \textit{transasymptotic matching}, and it is a 
well know feature of transseries solutions for a general class of differential
equations~\cite{costin1995,costin1998}. In the small $w$ limit there is a one
to one competition between the exponentially decaying terms entering in the
transseries and the  inverse powers in $w$. However, close to the IR fixed
point the instanton-like contributions are more suppressed than the leading 
IR perturbative terms. It is in this regime where one can indeed reshuffle 
and resum the small exponential terms such that
\be
\label{eq:transmatch}
\besp
c_1(w)&=\left[u^{(1)}_{1,0}\,\sigma_1\,\zeta_1(w)\,+\,u^{(2)}_{1,0}\,[\sigma_1\,\zeta_1(w)]^2\,+\,\cdots\right]\\
&\hspace{-.5cm}+\frac{1}{w}\left[u^{(0)}_{1,1}\,+\,u^{(1)}_{1,1}\,\sigma_1\,\zeta_1(w)\,+u^{(2)}_{1,1}\,[\sigma_1\,\zeta_1(w)]^2\,+\,\cdots\right]\\
&\hspace{-.5cm}+\frac{1}{w^2}\left[u^{(0)}_{1,2}\,+\,u^{(1)}_{1,2}\,\sigma_1\,\zeta_1(w)\,+u^{(2)}_{1,2}\,[\sigma_1\,\zeta_1(w)]^2\,+\,\cdots\right]\,+\cdots\\
&\equiv\sum_{k=0}^{+\infty} G_{1,k}(\sigma_1\zeta_1(w)) w^{-k}\,,
\end{split}
\ee
with $\zeta_1(w)=\,e^{-S_1 w}\,w^{-b_1}$ and the transasymptotic functions $G_{k}(\sigma_1\zeta(w))$ given by
\be
\label{eq:Gfunc}
G_{1,k}(\sigma_1\zeta(w))=\sum_{n=0}^\infty\,u^{(n)}_{1,k}[\sigma_1\,\zeta_1(w)]^n\,.
\ee
The transasymptotic functions $G_{k}(\sigma_1\zeta(w))$ effectively resum the 
full set of instanton-like contributions at a given order $k$ in perturbation
theory. Interestingly this matching procedure is not only valid close to the 
IR but it extends up to the UV, so in this sense it is transasymptotic
\cite{costin2008asymptotics}. Each $G_{k}(\zeta(w))$ is an analytic, Borel 
summable and convergent functions even if one truncates at a given order the IR
perturbative expansion Eq.~\eqref{eq:transmatch}. The full transseries 
solution~\eqref{eq:transmatch} is not Borel summable due to the singularity of 
the original differential equation~\eqref{eq:ODE_c1_w}. This singularity can 
be easily determined by taking the inverse Laplace transformation of the 
ODE~\eqref{eq:ODE_c1_w} (cf. Ref.~\cite{Behtash:2019qtk}). The transasymptotic 
matching procedure coincides with the exact solution of the ODE only when 
summing over all the perturbative and non-perturbative sectors, i.e. 
$k,n\to\infty$ in the upper limit of the sums of 
Eq.~\eqref{eq:transmatch}~\cite{costin2008asymptotics,costin1998,costin1995}.  
In general, a truncation of the perturbative expansion leads to a solution with 
a finite radius of convergence. 

%
\begin{figure}[t]
\begin{center}
	\includegraphics[scale=.22]{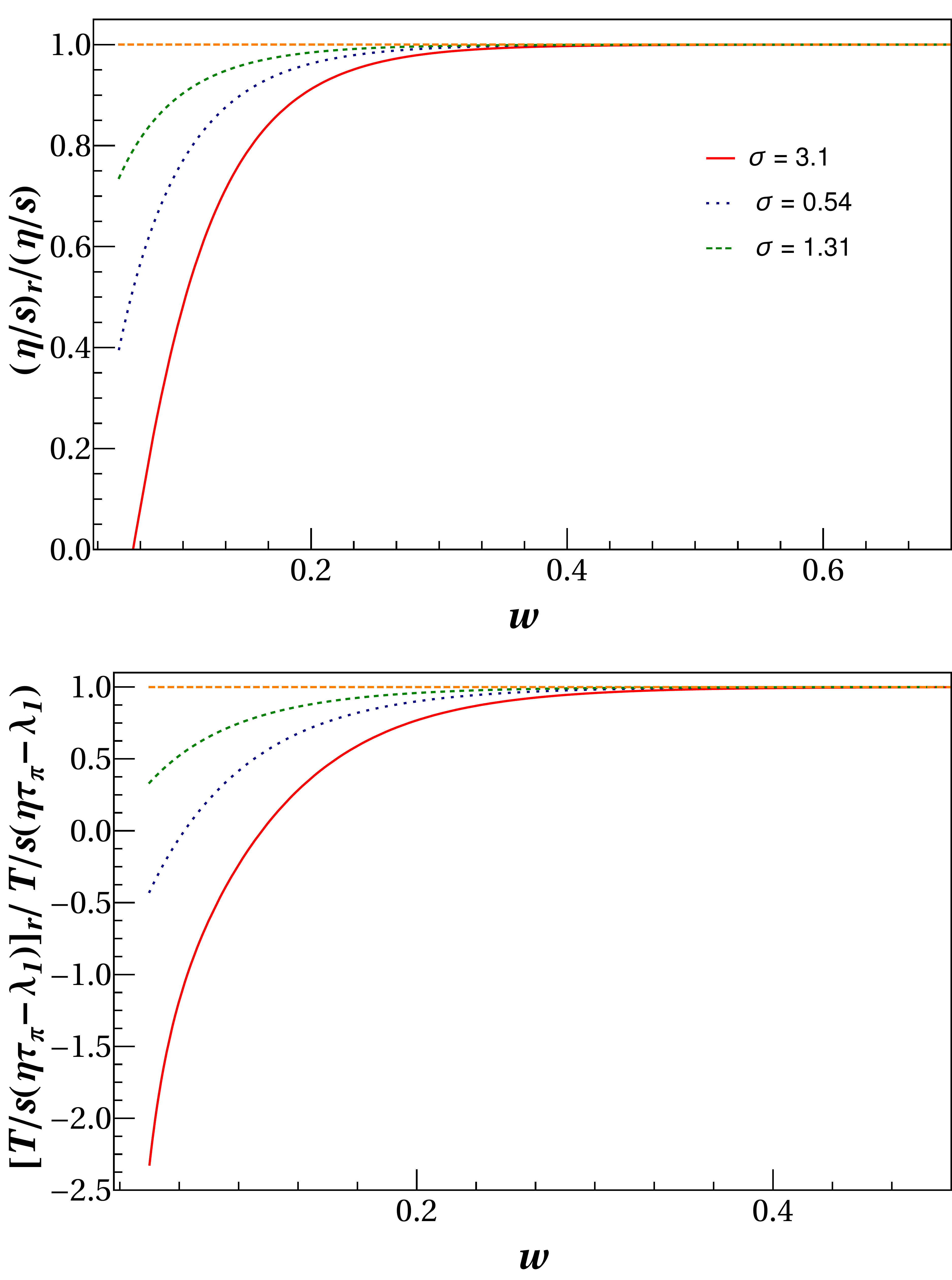}
	\caption{(Color online) Ratio of the dynamically renormalized transport coefficient over its corresponding asymptotic value as a function of $w$ for the $l=1$ case and for different initial conditions. Top and bottom panels correspond to the ratios of $\left(\eta/s\right)_r/\left(\eta/s\right)$ and $\left[(T/s)(\eta\tau_\pi-\lambda_1)\right]_r/\left[(T/s)(\eta\tau_\pi-\lambda_1)\right]$ respectively. See text for further details.} 
	\label{fig:2}
	\end{center}
\end{figure}
%
%

In the large $w$ limit the transasymptotic functions 
$G_{1,k}(\sigma_1\zeta(w))$  is uniquely determine by the coefficients of 
the asymptotic IR expansion, namely 
\be
\label{eq:Glimit}
\lim_{w\to\infty}G_{1,k}(\sigma_1\zeta(w))=u^{(0)}_{1,k}\,.
\ee
If one uses Eqs.~\eqref{eq:g1g2_c1} and~\eqref{eq:asympmatch} for $k=1,2$ in 
the previous expression one finds the following identities which relate the 
transport coefficients determined from linear response theory with the 
asymptotic regime of the function $G_{1,k}(\sigma_1\zeta(w))$ in the IR 
regime~\cite{Behtash:2018moe,Behtash:2019txb}
\be
\label{eq:asympttrans}
\besp
\frac{\eta}{s}&=-\frac{3}{40}\lim_{w\to\infty}G_{1,1}(\sigma_1\zeta(w))\,,\\
\frac{T\,(\eta\tau_\pi-\lambda_1)}{s}&=-\frac{9}{80}\lim_{w\to\infty}G_{1,2}(\sigma_1\zeta(w))\,.
\end{split}
\ee
Thus, the transasymptotic matching procedure automatically allows us to
generalize the concept of a transport coefficient beyond the linear regime by 
promoting the functions $G_{1,k}$ to be an effective transport 
coefficient~\cite{Behtash:2018moe,Behtash:2019txb}. For first and second order 
transport coefficients we have, respectively 
\be
\label{eq:ren_transp}
\besp
\left(\frac{\eta}{s}\right)_r&=-\frac{3}{40}\,G_{1,1}(\sigma_1\zeta(w))\,,\\
\left[\frac{T\,(\eta\tau_\pi-\lambda_1)}{s}\right]_r&=-\frac{9}{80}\,G_{1,2}(\sigma_1\zeta(w))\,.
\end{split}
\ee
As a result, a new physical picture for nonlinear transport emerges: 
\textit{Summing over all the non-perturbative sectors at each order of 
the IR expansion leads to an effective renormalization of the transport
coefficients. Each transport coefficient exhibits a transient non-Newtonian
behavior while relaxing towards its asymptotic value in the IR. This 
transient rheological behavior is described by a dynamical RG flow
equation.} This result extends earlier work in the context of the
Boltzmann equation in the relaxation time approximation
\cite{Behtash:2018moe,Behtash:2019txb}. We emphasize that Eqs.~\eqref{eq:ren_transp} are modified when including higher order moments as we shall explain in Sect.~\ref{subsec:RGtrans}.

 Numerical results for the renormalized transport coefficients in 
Eqs.~\eqref{eq:ren_transp} are shown in Fig.~\ref{fig:2}. In order to 
obtain these results we have to calculate the functions $G_{1,k}$ with
$k=\{1,2\}$ using Eq.~\eqref{eq:Gfunc}. In that expression the transmonomials
are given by Eqs.~\eqref{eq:trans_sol_c1} and the coefficients $u_{1,k}^{(n)}$
are numerically determined by solving order by order the recursive 
relation~\eqref{eq:rec_rel_c1}. Fig.~\ref{fig:2} shows the $w$-dependence 
of the ratios between the dynamically renormalized transport coefficient 
over its asymptotic value. The top panel shows the shear viscosity over 
entropy density ratio, and the bottom panel shows the second order transport
coefficients.

The values of the transasymptotic parameters were determined by fitting
the transseries solution~\eqref{eq:trans_sol_c1} truncated at $k=3$, $n=15$
to  the numerical solution of the ODE for $c_1$~\eqref{eq:ODE_c1_w} 
computed for $\theta_0=1$. The initial condition for $c_1$ was chosen
as $c_1(w_0=0.05)=\{6.60119,0,-3.02976\}$. The corresponding best fit 
transseries parameters are $\sigma=\{3.1,1.31,0.54\}$. Since the determination of the transseries parameter was carried out by using a numerical least-square fit~\cite{Behtash:2019txb,Behtash:2019qtk} it leads to a highly non-linear relation between $\sigma$ and $c_1(w_0)$ so both parameters are not proportional to each other. 

Both panels illustrate that regardless of the initial condition each
renormalized transport coefficient reaches its asymptotic value as it is 
expected from the properties of transasymptotic functions  $G_{1,k}$ in
Eq.~\eqref{eq:asympttrans}. On the other hand, prior to the relaxation 
to the asymptotic values both renormalized transport coefficients feature a
transient phase which depends on the deformation history of the fluid and 
thus, its rheology. During this transient phase the transport coefficients increase monotonically which is expected mainly for two reasons: firstly, at the early stages the expansion rate is larger than the collisional rate such that the effective transport coefficients get reduce compared with their corresponding asymptotic values. On the other hand, as the system hydrodynamizes, entropy production always increases so transport coefficients do the same until their values reached out their asymptotic values determined by linear response. Clearly the transient phase is not universal since it depends on the initial condition. However, one observes that both ratios 
shown in Fig.~\ref{fig:3} saturate their value to the unity around $w\approx 0.4$. These results are consistent with the ones shown in Fig.~\ref{fig:1}. It is important to mention that the effec\-ti\-ve value of the renormalized transport coefficients
change when increasing the number of moments~\cite{Behtash:2019txb,Behtash:2018moe}. We shall comeback to this subtlety issue in Sect.~\ref{subsec:gen_transasymp}.

\subsubsection{Dynamical RG flow equation}
\label{subsubsec:RGflow}

When the transasymptotic matching condition is imposed over the entire domain
of $w$, a set of nonlinear PDEs for the functions $G_{1,k}$ is obtained after 
inserting Eq.~\eqref{eq:transmatch} into the original ODE~\eqref{eq:ODE_c1_w},
\begin{widetext}
\begin{eqnarray}
&& \left( -  S_1  \hat{\zeta} + \Lambda(1)  \right) G_{1,k}  + \left( - \beta_1 \hat{\zeta} - k + \frac{12}{7} \right) G_{1,k-1} + 4 \delta_{k,1} \nl
&&\hspace{-1cm} - \frac{1}{20}  \sum_{k_1,k_2=0}^{k_1+k_2=k}
\left[
\left( -   S_1 \hat{\zeta}  - \frac{40 \hat{\kappa}}{7}  \right) G_{1,k_1} + \left( - \beta_1 \hat{\zeta} - k_1 + 5 \right) G_{1,k_1-1} \right] G_{1,k_2} 
+ \frac{2 \hat{\kappa}}{15} \sum_{k_1,k_2,k_3=0}^{k_1+k_2+k_3 = k} G_{1,k_1} G_{1,k_2} G_{1,k_3}  = 0,
\end{eqnarray}
\end{widetext}
where $\hat{\zeta}=\zeta \pd/\pd {\zeta} = \pd/\pd \log \zeta$. Solving these
PDEs is equivalent to summing over all fluctuation contributions around the 
IR. More importantly, these PDEs can be reinterpreted as RG flow equations for
the transport coefficients given the identities~\eqref{eq:asympttrans}. This
statement can be rigorously proven within the gradient descent  approach to 
the RG flows of quantum field theories in the context of dynamical 
systems, c.f.~\cite{Gukov:2016tnp,Behtash:2018moe,Behtash:2019txb}. Furthermore, a RG flow picture from a global point of view holds for our original dynamical system~\eqref{eq:prepODE_c_w} provided the existence of  
of an effective potential for the ODEs, also known as Lyapunov function. Its mathematical existence is proven in App.~\ref{sec:Lyapfunc}.

Following Refs.~\cite{Behtash:2018moe,Behtash:2019txb} the ODE for $c_1$ in terms of the 
$w$ variable, Eq.~\eqref{eq:ODE_c1_w} can be rewritten as
\begin{widetext}
\bs
\beal
\label{eq:RGc1}
\frac{dc_1}{d\log w}&=\beta_1(c_1,w)\,,\\
\label{eq:beta1}
\beta_1(c_1,w)&=-\frac{1}{1-\frac{c_1}{20}}\left\{4+\frac{5}{7}c_1-\frac{1}{5}c_1^2\,+\,\left[1+\frac{\pi^2}{\kappa}\,+\,\frac{2}{7}\,c_1\,+\,\frac{2}{15}\,c_1^2\right]\,\hat{\kappa}\,w\,c_1\right\}\,\,,
\end{align}
\es
\end{widetext}
where $\hat{\kappa}$ is given in Eq.~\eqref{eq:paramsODEs}. The previous
equation is the RG equation for the moment $c_1$ and the function 
$\beta_1(c_1,w)$ encodes the dependence on the RG time $w$. The variable
$w~\sim Kn^{-1}$ plays the analogous role of the energy scale in QFT since 
it determines the size of deviations from the equilibrium state. 

 We can think of the fugacity $\zeta_1$ as independent of 
$w$~\cite{Costin2001}, so that $G_{1,k}(\sigma_1\zeta_1)$ can be regarded 
as a simple coefficient in the transseries solution of 
$c_1$~\eqref{eq:trans_sol_c1}. Then, the following identity 
follows~\cite{Behtash:2018moe,Behtash:2019txb}
\begin{widetext}
\be
\label{eq:cauchyGk}
\besp
\hat{\zeta}_1 G_{1,k}(\sigma_1\zeta_1)&=-\frac{1}{2\pi i}\oint_{|w|\ll 1}\,dw\,\frac{w^{k-1}}{\left(S_1w+b_1\right)} \left[\sum_{k'=0}^\infty\,k'\,G_{k'}(\sigma_1\zeta_1)w^{-k'}\,+\,\beta_1(c_1(\sigma_1\zeta_1,w),w)
\right]\,,
\end{split}
\ee
\end{widetext}
where we explicitly used Eq.~\eqref{eq:RGc1}~\footnote{
The ordinary derivative respect to $\log w$ was rewritten as
\begin{equation*}
\label{eq:ordder}
\frac{d}{d\log w}=-\left(b_1+S_1 w\right)\hat{\zeta}_1\,+\,\frac{\partial}{\partial\log w}\,.
\end{equation*}
In this loop integration, $\zeta_1$ is temporarily regarded as an independent variable on $w$.}. The Cauchy integral formula in the previous expression simply picks up the $k$th coefficient of the term $\hat{\zeta}_1 c_1$. 

In order to determine the RG equation which describes the transient 
behavior of the transport coefficients we consider an arbitrary observable
${\cal O}={\cal O}(G_{1,k}(\sigma_1\zeta_1(w)))$. Its change along  $w$ results 
in the following RG equation 
\be
\label{eq:RGGkeq}
\besp
\frac{d{\cal O}(G_{1,k}(\sigma_1\zeta_1(w)))}{d\log w}&=
\\&\hspace{-1.2cm}
-\sum_{k=0}^\infty\,\left[\left(b_1+S_1 w
\right)\hat{\zeta}_1 G_{1,k}(\sigma_1\zeta_1)\right]\,\frac{\partial{\cal O}}{\partial G_{1,k}}\,.
\end{split}
\ee
The renormalization of the observable ${\cal O}$ is obtained by solving the
previous equation. In this equation the term $\hat{\zeta}_1 G_{1,k}
(\sigma_1\zeta_1)$ is given by the identity~\eqref{eq:cauchyGk} which 
encodes the dynamics of the original ODE via the beta function 
$\beta_1$~\eqref{eq:beta1}. By setting ${\cal O}=-\frac{3}{40}G_1$ and 
${\cal O}=-\frac{9}{80}G_2$ in Eq.~\eqref{eq:RGGkeq} one determines the 
RG flow equations for the transport coefficients $\eta/s$ and 
$\frac{T\,(\eta\tau_\pi-\lambda_1)}{s}$, respectively. We point out to the reader that the RG flow equation is sensitive to the number of moments involved in any truncation scheme and its more generic form is derived in Sect.~\ref{subsec:gen_transasymp}.

\subsection{Transseries solutions in the UV}
\label{subsec:UV_c1}

In the previous section we showed that perturbations decay exponentially 
in the vicinity of the IR fixed point~\eqref{eq:pertdc1}. For an arbitrarily
chosen ODE linear perturbations do not necessarily decay exponentially in 
the neighborhood of a fixed point. For instance, consider a parameter 
$x\in (0,\infty)$ and a function $y(x)$ which satisfies a well defined 
ODE $d y(x)/dx=\mc L(x,y(x))$. If this ODE has a source fixed point $y_0$ 
in the limit $x\to 0$, i.e. the Lyapunov exponent $\lambda$ close to $y_0$ is
positive in this limit, then the solutions in a vicinity
around it behaves as $y(x)\sim e^{\lambda x}\sim \sum_{n=0}^\infty\left(\lambda\,x\right)^n$ 
in the limit
$x\ll 1$. This approximate solution is a power law series with a 
finite radius of convergence which can be extended by analytic
continuation. 

The stability analysis of the RTA Boltzmann equation indeed showed 
that in the limit when the Knudsen number is large, that is in the 
$w\to 0$ limit, power law series solutions for the Legendre moments
emerge~\cite{Behtash:2019txb}. Similar findings were reported in
\cite{Kurkela:2019set,Jaiswal:2019cju,Blaizot:2019scw}. The power 
series expansion is rather different from non-hydrodynamic modes in 
the gradient expansion~\eqref{eq:asympc_1}. As we shall see in this 
section the main differences can be inferred from the behavior of  
linearized perturbations around the UV and IR fixed points, respectively. 
The case studied here illustrates the importance of the flow structure 
in phase space in controlling the functional form of solutions for 
generic dynamical systems of ODEs~\cite{Behtash:2019qtk}. In the following 
we will explain more carefully the emergence of the power series behavior 
in the case of the Fokker-Planck equation.

\subsubsection{Perturbative power series in the UV}
\label{subsec:UV_pert_pow}

The limit $w\to 0$ of Eq.~\eqref{eq:ODE_c1_w} can be understood by 
changing variables $w\to 1/z$ in Eq.~\eqref{eq:ODE_w}. We obtain
the following differential equation for $c_1(z)$
\begin{widetext}
\bs
\beal
\label{eq:ODE_z}
\frac{dc_1}{dz}&= F_1(c_1,z)  \,,\\
\label{eq:F1_z}
F_1(c_1,z)&=\frac{1}{\left( 1-\frac{1}{20}c_{1} \right)}\left[ \frac{1}{z} \left( 4 +  \frac{5}{7}c_1 - \frac{1}{5}c_1^2    \right) \right.
\left.+ \frac{1}{z^2}\left[  \lambda(1)c_1 +  \frac{2 \hat{\kappa}}{7}  c_{1}^2 + \frac{2\hat{\kappa}}{15} c_{1}^3 \right]\,\right]\,.
\end{align}
\es
\end{widetext}
In the limit $z\to \infty$ the dominant term in the previous expressions is
$\mc O(z^{-1})$. It is straightforward to see that the fixed points correspond
to the roots of the polynomial that multiplies the $\mc O(1/z)$ term
in Eq.~\eqref{eq:F1_z}, 
\be
\label{eq:pol_c1}
4+\frac{5}{7}\bar{c}_1-\frac{1}{5}\bar{c}_1^2\equiv 0 \, . 
\ee
The UV fixed points are given by 
\be
\label{eq:UVfixedc1}
\bar{c}^{\,\pm}_{1}=\left(25\pm  3\sqrt{505}\right)/14\,=
\begin{cases}
6.60119\,\,\,\text{if +}\,,\\
-3.02976\,\,\,\text{if -}\,.
\end{cases}
\ee
In analogy to what we did in the IR limit one can construct a perturbative 
expansion in the limit $z\to\infty$. We consider
\be
\label{eq:uv_c1_pert_z}
c_1(z) =\sum_{k=1}^\infty\, \frac{v_{1,k}}{z^k}\, ,
\ee
where $v_{1,0}=\bar{c}^{\,\pm}_{1}$. A similar power series expansion was 
discussed in the case of kinetic models undergoing Gubser flow in \cite{Behtash:2017wqg,
Behtash:2019qtk}. Inserting the UV series expansion in Eq.~\eqref{eq:uv_c1_pert_z} 
into Eq.~\eqref{eq:ODE_z} we obtain the following recursive relation for the coefficients 
$v_{1,k}$
\begin{widetext}
\be
\label{eq:recrel_pertc1uv}
\besp
&\left(\frac{2}{7}-k \right) v_{1,k-1} - 4 \delta_{k,1} - \lambda(1) v_{1,k-2} + \frac{1}{20}\, \sum_{k_1,k_2=0}^{k_1+k_2=k} \left[  \left( k_1 +3 \right) v_{1,k_1-1}  \right] v_{1,k_2} \\
& - \frac{2 \hat{\kappa}}{7}\,\sum_{k_1,k_2=0}^{k_1+k_2=k-2}  v_{1,k_1}  v_{1,k_2}  - \frac{2 \hat{\kappa}}{15} \sum_{k_1,k_2,k_3=0}^{k_1+k_2+k_3 = k-2}\, v_{1,k_1} v_{1,k_2} v_{1,k_3}  = 0\,.
\end{split}
\ee
\end{widetext}
where $v_{1,k}\equiv 0$ iff $k<0$ and $v_{1,0}=\bar{c}^{\,\pm}_{1}$. 

\subsubsection{Stability analysis and radius of convergence}
\label{subsec:UV_st}

When performing asymptotic expansions it is important to check the 
stability of the perturbative expansion. We address this issue for the UV
power series expansion~\eqref{eq:uv_c1_pert_z} by analyzing the linearized
perturbations around it. Linearizing Eq.~\eqref{eq:F1_z} around the UV 
fixed points, i.e.  $c_1\to\bar{c}_{1}^{\,\pm}+\delta c^\pm_{1}$, one finds 
the following evolution equation for the linearized perturbations in the 
$z\to\infty$ limit
\be
\label{eq:UVpertc1}
\besp
\frac{d\delta c^{\,\pm}_{1}}{dz}&=\,\frac{\partial F_1}{\partial c_1}\biggr|_{c_1=\bar{c}^{\,\pm}_{1}}\,\delta c^\pm_{1}\,+\,\mc O(\delta c^\pm_{1}/z^2)\,\\
&\approx\frac{8\sqrt{505}}{\sqrt{505}\mp 85}\frac{\delta c^\pm_{1}}{\,z}\, .
\end{split}
\ee
The solutions are 
\be
\label{eq:UVpertsol}
\besp
\delta c^\pm_{1}(z)&=\mu_{1}^\pm\,z^{\alpha_1^{\pm}}\,,\\
\alpha_1^{\pm}&=\frac{8\sqrt{505}}{\sqrt{505}\mp 85}\equiv\begin{cases}
&-2.87516\,\,\text{if}\,\,+\,,\\
&1.67278\,\,\text{if}\,\,-
\end{cases}
\end{split}
\ee
where $\mu_1^{\pm}$ is an integration constant associated with the UV fixed
points. The power law behavior of the linearized 
perturbations~\eqref{eq:UVpertsol} is completely different from the behavior
in the IR limit~\eqref{eq:pertdc1}. We note that the exponents $\alpha_1^\pm$
do not depend on the strength of the coupling, but they are uniquely defined 
by the location of the UV fixed points. In the limit $z\to \infty$ the 
solution $\delta c^+_{1}=\mu^+ z^{-2.87516}$~\eqref{eq:UVpertsol} follows 
a power law decay. The linear perturbation $\delta c^-_{1}=\mu^- 
z^{1.67278}$~\eqref{eq:UVpertsol} monotonically increases. This divergence
implies that the associated fixed point only admits a power series of the 
form~\eqref{eq:UVpertc1} which can be extended analytically as it is shown
below. Alternatively one may say that the initial condition $\mu_1^-$ for 
the fluctuation around $\bar{c}_1^-$ vanishes exactly and thus, the only
possible expansion around $c_1^-$ is a power series.  
%
\begin{figure}[t]
\begin{center}
	\includegraphics[scale=.22]{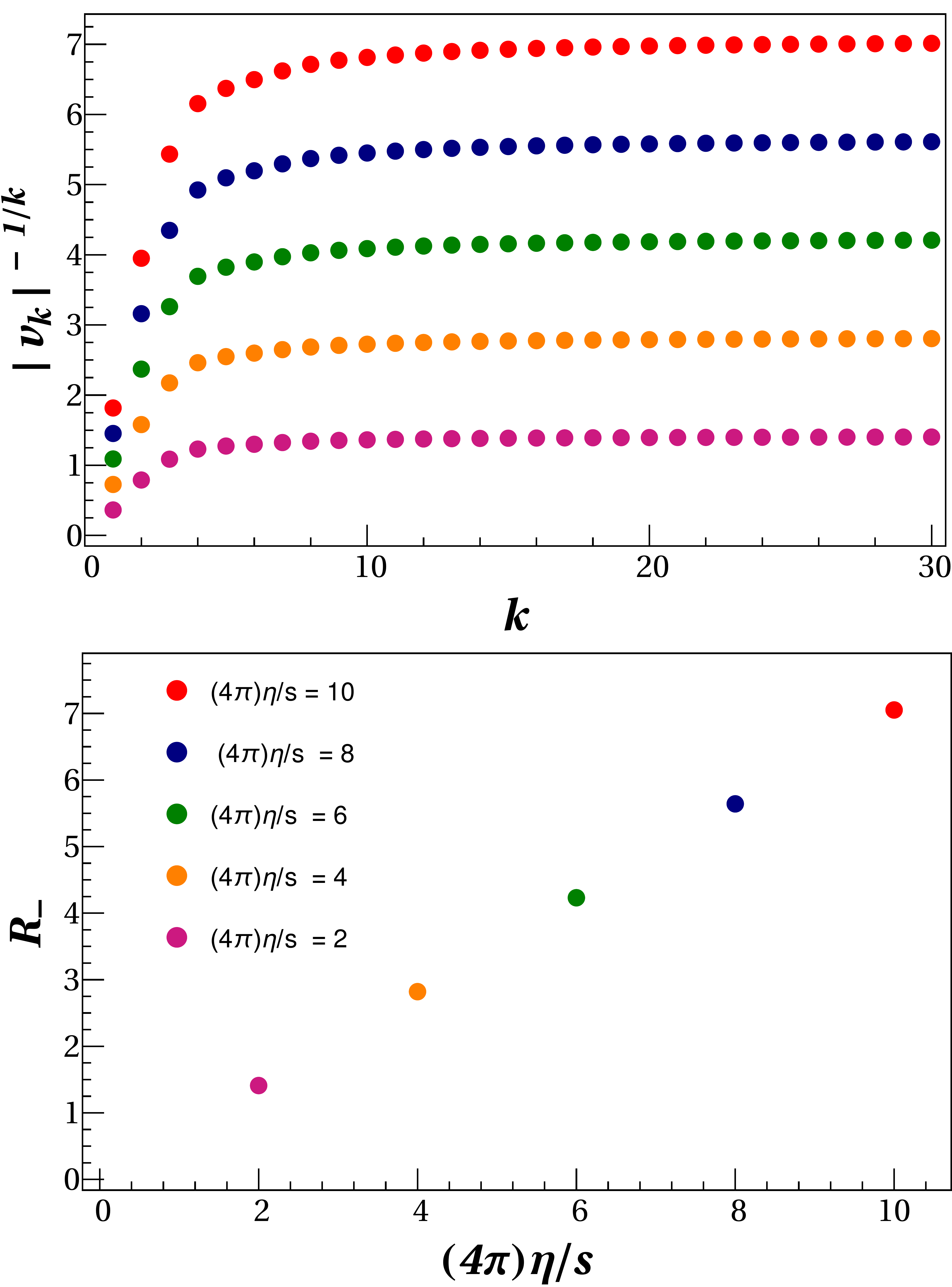}
	\caption{(Color online) \textit{Top panel:} $|v_k|^{-1/k}$ vs $k$ for different values of $(4\pi)\eta/s=\{2,4,6,8,10\}$ for magenta, orange, green, blue and red dots respectively. \textit{Bottom panel:} Numerical value of the radius of convergence vs. $\eta/s$. In this case the UV perturbative series expansion~\eqref{eq:uv_c1_pert_z} is carried out around the UV fixed point $\bar{c}_1^-=\left(25- 3\sqrt{505}\right)/14$.} 
	\label{fig:3}
	\end{center}
\end{figure}
%

The formal power series in the $z$ variable~\eqref{eq:uv_c1_pert_z} has a
finite radius of convergence which can be estimated via the Cauchy-Hadamard 
theorem~\footnote{According to the Cauchy-Hadamard theorem~\cite{lang2010complex},  
the radius of convergence $R$ of the formal power series of a function 
$f(z)$ around the point $a$ (with $b_k,a\in \mathfrak{C}$ )
\begin{equation*}
f(z)=\sum_{k=0}^\infty\,b_k\,(z-a)^k\,,
\end{equation*}
is given by
\begin{equation*}
R=\left[\lim_{n\to\infty}\text{sup.}|b_n|^{1/n}\right]^{-1}\,.
\end{equation*}
}~\cite{lang2010complex}. We proceed to calculate the radius of convergence 
of the power series expansion~\eqref{eq:uv_c1_pert_z} by considering first 
the expansion around $\bar{c}_1^-$. The case around $\bar{c}_1^+$ is analyzed
separately at the end of this section. 
The radius of convergence $R_-$ is calculated by numerically solving the
recursive relation~\eqref{eq:recrel_pertc1uv}. In Fig.~\ref{fig:3} (top panel)
we show the coefficients of the UV expansion $|v_k|^{-1/k}$ vs. the order of
the expansion $k$ for different values of $(4\pi)\eta/s=
\{2,4,6,8,10\}$~\footnote{The asymptotic value of $\eta/s$ and $\theta_0$ 
can be established via Eq.~\eqref{eq:asympmatch} which leads to the following
identity
\begin{equation*}
\frac{\eta}{s}=\frac{\theta_0}{5(\kappa+\pi^2)}\,,
\end{equation*}
where we used explicitly Eqs.~\eqref{eq:paramsODEs}.}. 
The coefficients $|v_k|^{-1/k}$ stabilize for $k\geq 25-30$. This result confirms that the power series~\eqref{eq:UVpertc1} has a finite radius of convergence.
We analyze its dependence  on the value of $\eta/s$ in Fig.~\ref{fig:3} (bottom panel).
This plots shows that $R_-$ depends linearly on the value of the shear over entropy ratio $\eta/s$. The empirical relation between these two quantities extracted from this plot is 
\be
\label{eq:Rvsetas}
R_{-}=0.70515\,\left(4\pi\,\frac{\eta}{s}\right)+b_-\,,
\ee
where the intercept $b_->0$ is very small $\mc {O}(10^{-8})$ and  
consistent with zero.  We verify the extracted numerical value of the radius of convergence by increasing $k$ up to $k_{max}=100$.  We found that the relative difference between the saturated bound $|v_k|^{-1/k}$ for $k=25$ and $k=100$ was only $0.1\%$.

The linear growth of the radius of convergence 
as a function of $\eta/s$ is intuitively understood as follows: If 
$\eta/s$ or, equivalently, the mean free path are large then the rate 
of collisions is small. As a result the Yang-Mills plasma will expand 
freely for a longer period of time. This finding might provide an 
explanation for the partial success of phenomenological models at 
intermediate scales of momentum larger than the typical temperature 
where the expansion is carried out in terms of a small number of  
scatterings~\cite{Heiselberg:1998es,Kurkela:2019kip,Kurkela:2018qeb,
Borghini:2010hy,Romatschke:2018wgi}.  

%
\begin{figure}[t]
\begin{center}
	\includegraphics[scale=.16]{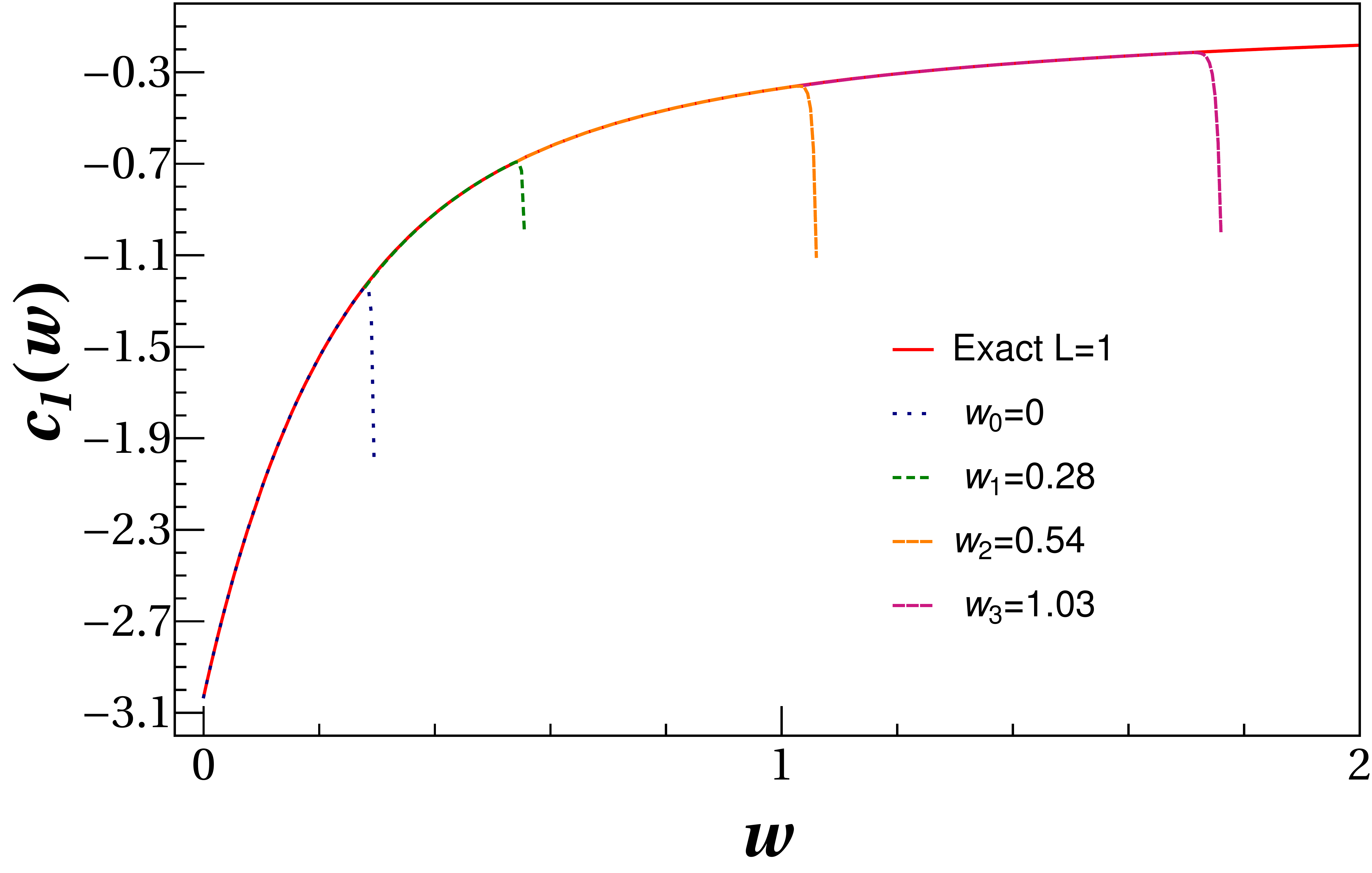}
	\caption{(Color online) Analytical continuation of the perturbative expansion for $c_1(w)$ in the UV around the fixed point $\bar{c}_1^-=\left(25- 3\sqrt{505}\right)/14$. We show the exact 
	solution as we perturbative expansions around the indicated 
	values of $w$.} 
	\label{fig:4}
	\end{center}
\end{figure}
%
%
In general the radius of convergence of a power series can be extended 
by analytical continuation~\cite{lang2010complex}. The idea is to 
consider a power series of the form 
\be
\label{eq:modpowerlaw}
c_1(w)=\sum_{k=0}^\infty\,v_{1,k}\,\left(w-w_0\right)^k\,,
\ee
where both $v_{1,k}$ and $w_0\in\mathfrak{R}$, $v_{1,0}=c_1(w_0)$ and in 
this case $w_0\geq 0$. The coefficients $v_{1,k}$ are determined explicitly 
by plugging Eq.~\eqref{eq:modpowerlaw} into the ODE~\eqref{eq:ODE_c1_w} 
and finding the associated recursive relation analogous to 
Eq.~\eqref{eq:recrel_pertc1uv} which is now evaluated at $w=w_0$. By 
construction the series~\eqref{eq:modpowerlaw} has a finite radius of 
convergence $R_0$ so the series will diverge outside of $|w-w_0|<R_0$. One can 
thus expands again the function at another point $w_1$ located within a 
distance $R_0-\delta$ (with $\delta$>0) centered at $w_0$ and determine the 
new coefficients of the power series. Successively applying this 
approach one is extending the original power series significantly  beyond 
its original radius of convergence $R_0$. We apply this method to our 
solution by initializing the expansion at  $w_0=0$ around $\bar{c}_1^-$  
and $\theta_0=1$ (i.e., extremely low values of $\eta/s\to 0$). The 
agreement between the exact numerical result and the analytical
continuation of the perturbative expansion in the UV is truly remarkable,
as shown in Fig.~\ref{fig:4}. 

\subsubsection{Resummation of the fluctuations around $\bar{c}_1^+$}
\label{subsub:newress}

The power law decay of linearized perturbations $\delta c_1^+$
in Eq.~\eqref{eq:UVpertc1}, together with the UV stability analysis,
strongly suggests a resummation scheme around the UV fixed point
$\bar{c}_1^+$. The form of the equation in the $z$ variable, see
Eq.~\eqref{eq:ODE_z}, resembles the corresponding results obtained
in the case of the RTA Boltzmann equation for Bjorken and Gubser 
flows~\cite{Behtash:2019txb,Behtash:2019qtk}. Following these approaches 
we consider the following UV transseries ansatz
\be
\label{eq:UVtrans}
\besp
c_{1}&=\sum_{k=0}^\infty\,\sum_{n=0}^\infty\,\,v_{1,k}^{(n)}\,
\varphi_{k}^{n}\,,
\end{split}
\ee
where the UV transmonomial is
\be
\label{eq:UVtransm}
\varphi_{k}^n:=z^{-k}\,[\xi_{1}(z)]^n\,,\qquad\xi_{1}(z)=\mu_1\,z^{\alpha_1}\,.
\ee
with $\alpha_1=-2.86516$. We note that, unlike the IR transseries where 
fluctuations decay exponentially close to the fixed point, the transmonomials
$\xi_{1}(z)$~\eqref{eq:UVtransm} are non-negligible close to the UV fixed 
point. For instance, the first fluctuation term $z^{\alpha_1}\approx
z^{-2.87516}$ is comparable to terms of order $\mc O(z^{-3})$. This 
implies that the complete solution near $\bar{c}_1^+$ requires a double
summation, over both the perturbative exponent $k$  as well as the 
transmonomials~\eqref{eq:UVtransm}~\cite{Behtash:2019txb,Behtash:2019qtk}. 
We also observe that the approximate UV solution~\eqref{eq:UVtrans} has 
a finite radius of convergence even after resumming the fluctuations around
$\bar{c}_1^+$~\cite{Behtash:2019qtk}. Finally, in the UV regime it is not
possible to implement IR transasymptotic matching (see 
Sect.~\ref{subsec:dyn-ren_c1}) since the solution~\eqref{eq:UVtrans} is 
not valid for all values of $z$ given the finite radius of convergence. 

 The coefficients $v_{1,k}^{(n)}$ in Eq.~\eqref{eq:UVtrans} are determined 
by inserting this ansatz into Eq.~\eqref{eq:ODE_z}. We obtain the recursion
relation
\begin{widetext}
\be
\label{eq:recrel_c1uv}
\besp
&\left( n \alpha_1 - k + \frac{2}{7} \right) v_{1,k-1}^{(n)} - 4 \delta_{n,0} \delta_{k,1} - \lambda(1) v^{(n)}_{1,k-2} - \frac{1}{20} \sum_{n_1,n_2=0}^{n_1+n_2=m}\, \sum_{k_1,k_2=0}^{k_1+k_2=k} \left[  \left( n_1 \alpha_1 - k_1 -3 \right) v_{1,k_1-1}^{(n_1)}  \right] v_{1,k_2}^{(n_2)} \\
& - \frac{2 \hat{\kappa}}{7} \sum_{n_1,n_2=0}^{n_1+n_2=n}\, \sum_{k_1,k_2=0}^{k_1+k_2=k-2}  v_{1,k_1}^{(n_1)}  v_{1,k_2}^{(n_2)}  - \frac{2 \hat{\kappa}}{15} \sum_{n_1,n_2,n_3=0}^{n_1+n_2+n_3 = n}\, \sum_{k_1,k_2,k_3=0}^{k_1+k_2+k_3 = k-2}\, v^{(n_1)}_{1,k_1} v^{(n_2)}_{1,k_2} v^{(n_3)}_{1,k_3}  = 0\,.
\end{split}
\ee
\end{widetext}
For the UV transseries~\eqref{eq:UVtrans} the coefficients 
$v_{1,0}^{(0)}=\bar{c}_{1}^{\,+}$ and the normalization of the integration 
constant is $v_{1,0}^{(1)}=1$. 

\section{Transasymptotic analysis: the general case}
\label{sec:generalcase}

In this section we generalize the transasymptotic analysis outlined in the
previous section to the case when we consider the full vector of Legendre
moments $\bar{c}=\left(c_1,\cdots,c_L\right)^T$ with $L\geq 1$. We briefly
outline how to generate IR and UV transseries solutions for the non-autonomous
dynamical system~\eqref{eq:ODE_w}. The techniques presented in the section 
were extensively discussed in~\cite{Behtash:2018moe,Behtash:2019txb,
Behtash:2019qtk}, and we will focus on issues specific to the 
Fokker-Planck equation. More rigorous mathematical aspects of the techniques outlined in this section can also be found in Refs.~\cite{costin1998,costin2008asymptotics}.

\subsection{Transseries solutions in the IR}
\label{subsec:IR_gen}

We seek to find multiparameter transseries solutions to the ODEs
\eqref{eq:prepODE_c_w} by following the generic procedure developed by 
Costin~\cite{costin1998,costin2008asymptotics}. In the large $w$ limit 
and for the nonlinear ODEs~\eqref{eq:prepODE_c_w} the Legendre moments admit 
IR expansions of the form~\cite{Behtash:2018moe,Behtash:2019txb}
\be
\label{eq:IRgenpert}
\bar{c}_l=\sum_{l=0}^\infty\,\frac{u_{l,k}}{w^k}\,.
\ee
By mathematical induction one can show that by inserting the expansion
\eqref{eq:IRgenpert} into Eq.~\eqref{eq:prepODE_c_w}, and considering 
only the linear regime of the ODE, then the leading large $w$ behavior 
of the moments $c_l$ is $c_l\sim\mc O(w^{-l})$ \cite{Behtash:2018moe,
Behtash:2019txb}. These results can be used to linearize
Eq.~\eqref{eq:prepODE_c_w} by expanding around the IR 
expansion~\eqref{eq:IRgenpert}, i.e. ${\bf c}\to \bar{\bf c}+\delta {\bf c}$ 
where $\bar{\bf c}=(\bar{c}_1,\cdots \bar{c}_L)$. Here, each $\bar{c}_l$ 
given by Eq.~\eqref{eq:IRgenpert}. By keeping only the leading terms 
$\mc O(1/w)$ we get the following linearized equation for the IR 
perturbation $\delta {\bf c}$
\bs
\label{eq:dcl_ODE_w_}
\beal
\frac{d\delta {\bf c}}{dw}&=\sum_{l=1}^L\frac{\partial {\bf F}({\bf c},w)}{\partial{c}_l}\biggr|_{{\bf c}_l=\bar{{\bf c}}_l}\cdot\delta {\bf c}_l\,,\nl
&\approx -\left[\Lambda+\frac{1}{w}\left(\bar{\frak{X}}-\frac{8}{\Lambda(1)}\bar{\frak{Y}}+\frac{\Lambda}{5\lambda(1)}-\frac{\bold{\Gamma}}{20}\right)
\right]\,\delta {\bf c}\,\nl
&=-\left[\Lambda+\frac{1}{w}\,\frak W\right]\,\delta {\bf c}\,,\\
\label{eq:anomat}
\frak W&=\bar{\frak{X}}-\frac{8}{\lambda(1)}\bar{\frak{Y}}+\frac{\Lambda}{5\lambda(1)}-\frac{\bold{\Gamma}}{20}\,,
\end{align}
\es
where 
\be
\label{eq:ubarmat}
\bar{\frak Y} = \hat{\kappa} 
\begin{pmatrix}
  \Omega_{111}   & \Omega_{112} &  &  &    \\
  \Omega_{211} & \Omega _{212} & \Omega_{213} & & \\
     & \ddots & \ddots & \ddots & \\
  & &  \Omega_{L-11L-2} &  \Omega_{L-11L-1} & \Omega_{L-11L}  \\
  & &  & \Omega_{L1L-1} & \Omega_{L1L}  \\
\end{pmatrix}\,. 
\ee
We can check that the truncated $l=1$ linear equation~\eqref{eq:dc1_linear} 
is a special case of Eq.~\eqref{eq:dcl_ODE_w_}. In order to solve the 
linearized equation~\eqref{eq:dcl_ODE_w_} we perform a transformation in 
terms of the pseudomodes $\delta \tilde{\bf c}$~\cite{costin2008asymptotics}
\be
\label{eq:pseudodcl}
\delta \tilde{\bf c} = \tilde{V}(w) \delta {\bf c}, \qquad \tilde{V}(w) := {\bf 1}_{L} + \frac{V}{w}\,,
\ee
where $V$ is a constant matrix.
Thus, the equation of the pseudomodes $\delta \tilde{\bf c}$ is 
\be
\label{eq:ODE_pseudodcl}
\besp
\frac{d \delta \tilde{\bf c}}{d w} 
&= - \tilde{V}(w) \left[ \Lambda + \frac{1}{w} \frak W  \right] \tilde{V}(w)^{-1} \delta \tilde{\bf c}(\hat{\tau})  + O( \delta \tilde{\bf c}/w^2) \\
&= -  \left[ \Lambda + \frac{1}{w} \left( \frak W + [V,\Lambda]  \right)   \right]  \delta \tilde{\bf c}(w)\,, 
\end{split}
\ee
where the commutator $[A,B]:=AB-BA$ was introduced. One can take ${\rm
diag.}(V)=(0,\cdots,0)$, and the other components can be chosen in such a 
way that one can diagonalize the matrix $\frak{W}$ as follows
\be
\frak W + [V,\Lambda] \ : \mapsto \ \tilde{\frak W} := {\rm diag.}({\frak w}_1,\cdots,{\frak w}_L).
\ee
Therefore, the solution of Eq.(\ref{eq:ODE_pseudodcl}) is
\be
\label{eq:dcl_sol_lin}
\delta \tilde{c}_{\ell}(w) = \sigma_{\ell} \frac{e^{- \lambda(\ell) w}}{w^{{\frak w}_{\ell}}} \quad \Rightarrow \quad \delta {c}_{\ell}(w) = \sigma_{\ell} \frac{e^{- \lambda(\ell) w}}{w^{{\frak w}_{\ell}}} \,.
\ee
The Lyapunov exponents are the diagonal components of the matrix
$\hat{\Lambda}$~\eqref{eq:Lyap_mat}. These exponents govern the rate
at which each mode relaxes towards its equilibrium value, 
$w_l^*=[\lambda(l)]^{-1}$. We observe that for larger value of $l$ 
relaxation is faster, and that there is a clear hierarchy of scales 
$w^*_1\,>\,w^*_2\,>\cdots> w^*_L$. For the FP equation the slowest 
non-hydro mode is $c_1$ which is proportional to the normalized shear 
viscous tensor component.
A common misunderstanding in the literature is to assume that the existence 
of this hierarchy of scales determines immediately the full set of the 
slowest degrees of freedom of the physical system. Recent studies
\cite{Behtash:2019txb,Bazow:2016oky}  have shown that this assumption is incorrect since nonlinear mode-to-mode coupling among moments plays a relevant role close 
to the forward attractor and thus, the determination of the slow invariant
manifold of the dynamical system is not uniquely determined by the mere 
existence of a hierarchy of scales \cite{Behtash:2019txb,Behtash:2019qtk}.
in Eq.~\eqref{eq:barpi_c1}. The eigenvalue ${\frak w}_{\ell}\in {\mathbb R}^+$ 
of the matrix $\hat{\frak W}$ is the anomalous dimension of the linear
perturbation $\delta c_l$. Their values ${\frak w}_{\ell}$ depends explicitly 
of the coupling constant as well as the truncation order $L$
\cite{Behtash:2018moe,Behtash:2019txb,Behtash:2019qtk}; only in the limit
$L\to\infty$ do their values coincide with the exact underlying microscopic 
theory. We note that ${\frak w}_{\ell}$ is not a universal constant. Its value
depends on the kinetic model, and is sensitive to nonlinearities encoded in
mode-to-mode couplings.

The general transseries solutions of the nonlinear coupled 
ODEs~\eqref{eq:prepODE_c_w} are constructed by rewriting these equations 
in terms of the pseudomodes basis, i.e., 
\be
\label{eq:pseudo_basis}
\tilde{\bf c}=\tilde{V}(w){\bf c}
\ee
where $\tilde{V}(w)$ is given in Eq.~\eqref{eq:pseudodcl}. Under this 
transformation the equation of the pseudomodes $\tilde{\bf c}$ is
\begin{widetext}
\bs
\begin{align}
\label{eq:pseudo_ODE}
\frac{d\tilde{\bf c}}{dw}&=\tilde{\bf F}(\tilde{\bf c},w)\,,\\
&\hspace{-1cm}\tilde{\bf F}(\tilde{\bf c},w)=-\frac{1}{\left(1-\frac{c_1}{20}\right)}\,\left[\frac{1}{w}\left(\tilde{\frak X}({\bf c})\,\tilde{\bf c}\,+\,\tilde{\bm{\Gamma}}(w)\right)\,\right.\left. +\,\left(\tilde{\Lambda}+\tilde{\frak Y}({\bf c})+\tilde{\frak Z}({\bf c})\right)\,\tilde{\bf c}
\right]\,
-\,\frac{1}{w^2}V\tilde{V}^{-1}(w)\tilde{\bf c}\,.\nl
\end{align}
\es
\end{widetext}
where $c_1(w)=\sum_{k=1}^L\tilde{V}^{-1}_{1k}(w)\tilde{c}_k(w)$. We have denoted
matrices in the pseudomode basis by $\tilde{M}(w)\equiv \tilde{V}(w)M\tilde{V}^{-1}(w)$
and the vector $\tilde{\bm{\Gamma}}(w)\equiv\tilde{V}(w)\bm{\Gamma}$. Having 
determined the linearized pseudomodes~\eqref{eq:dcl_sol_lin} we can write
the solutions of the pseudomodes as multiparameter transseries
\bs
\label{eq:multi_pseudosols}
\beal
\label{eq:multsol_pseudo}
& \tilde{c}_{\ell}(w) = \sum_{{\bf n} \in {\mathbb N}_0^L} \sum_{k=0}^{+\infty} \,\tilde{u}^{(\bf n)}_{\ell,k} \Phi^{\bf n}_{k},\\
\label{eq:transmon}
& \Phi^{\bf n}_{k}:= \left( \prod_{j=1}^{L} \sigma_j^{n_{j}}\zeta_j^{n_{j}} \right) w^{-k}, \qquad\text{with}\,\,\zeta_{j}:= \frac{e^{-S_j w}}{w^{\beta_{j}}}\,,
\end{align}
\es
where ${\bf n}=(n_1,n_2,\cdots,n_L)$ is a vector where each component $n_i$ 
is a non-negative integer which labels the non-perturbative sectors of the
pseudomodes, $\sigma_{i}\in {\mathbb C}$ is the integration constant,  and 
$L \in {\mathbb N}$ is the truncation order. The term $\zeta_{j}$ in
Eq.~\eqref{eq:transmon} must match the linearized solutions
\eqref{eq:dcl_sol_lin} and thus, it determines the IR data
\be
S_{j} = \lambda(j), \qquad
\beta_{j} = {\frak w}_{j} .
\ee
The set of ODEs for the pseudomodes~\eqref{eq:pseudo_ODE} 
satisfy the asymptotic and regular conditions needed in  Costin's prescription~\cite{costin1998} and 
thus, it justifies mathematically why the exact solutions are indeed multiparameter 
transseries. Now, given the solutions for the pseudomodes~\eqref{eq:multi_pseudosols} one obtains the ones corresponding to the Legendre moments $c_l$ as follows 
\bs
\beal
\label{eq:multsol_cl}
&c_{\ell}(w) = \sum_{\ell'=1}^L\,\tilde{V}_{\ell \ell'}^{-1}(w)\tilde{c}_{\ell'}(w)=\sum_{{\bf n} \in {\mathbb N}_0^L} \sum_{k=0}^{+\infty} u^{(\bf n)}_{\ell,k} \Phi^{\bf n}_{k}\,,\\
&\text{with}\,\, u^{(\bf n)}_{\ell,k}=\sum_{\ell'=1}^L \sum_{k^\prime=0}^k\, (-1)^{k^\prime} [V^{k^\prime}]_{\ell \ell'} \, \tilde{u}^{(\bf n)}_{\ell',k-k^{\prime}}.
\end{align}
\es
The reality condition on the distribution function implies that the Legendre
moments $c_l$ are real. Although the solutions for the pseudomodes
$\tilde{c}_l$~\eqref{eq:multsol_pseudo} are complex the reconstructed 
transseries solutions of $c_l$~\eqref{eq:multsol_cl} are real since 
the following conditions are satisfied $\forall j\geq 1$~\cite{Behtash:2019qtk} 
\begin{enumerate}
\item $\lambda(j)\in {\mathbb R}^+$ (with $\lambda_j$ given by 
Eq.~\eqref{eq:Lyap_mat}) and $\lambda(j) \ne \lambda(j^\prime)$ for any $j \ne 
j^{\prime}$.
\item If $\beta_{j}\,\in {\mathbb R}^+$ which implies $\sigma_{j}\in {\mathbb 
R}^+$.
\item If $\beta_{j}\,\in {\mathbb C}$ then there should be a complex conjugate 
pair $\beta_{k}= \beta^*_{j}$ ($j\neq k$) which implies $\sigma_k=\sigma_j^*$.  
\end{enumerate}
Finally the coefficients $\tilde{u}^{(\bf n)}_{\ell,k}$ are determined by 
inserting the transseries solution of the pseudomodes~\eqref{eq:multsol_pseudo}
into its corresponding Eq.~\eqref{eq:pseudo_ODE}. As a result one gets the 
following recursive relation 
\begin{widetext}
\be
\label{eq:pseudo_recur}
\besp
&  -  \left( {\bf m} \cdot {\bf S} \right) \, \tilde{\bf u}_{k}^{({\bf m})} - \left( {\bf m} \cdot \bm{\beta} + k \right) \tilde{\bf u}_{k-1}^{({\bf m})} + \frac{1}{20} \sum_{{\bf m}_1,{\bf m}_2={\bf 0}}^{{\bf m}_1+{\bf m}_2={\bf m}} \sum_{k_1,k_2=0}^{k_1+k_2=k}\left[  \left( {\bf m}_1 \cdot {\bf S} \right) \, \tilde{\bf u}_{k_1}^{({\bf m}_1)} + \left( {\bf m}_1 \cdot \bm{\beta} + k_1\right) \tilde{\bf u}_{k_1-1}^{({\bf m}_1)}  \right] {u}_{1,k_2}^{({\bf m}_2)} \\
& + \left[ \left\{ \tilde{\Lambda} + \tilde{\frak Y}({\bf u}) + \tilde{\frak Z}({\bf u}) \right\}  \star \tilde{\bf u} \right]^{({\bf m})}_{k}+ \left[ \tilde{\frak X}({\bf u}) \star \tilde{\bf u} + \tilde{\bm{\Gamma}} \right]^{({\bf m})}_{k-1} +  V  {\bf u}^{({\bf m})}_{k-2} = 0,
\end{split}
\ee
\end{widetext}
where $\tilde{\bf u}^{({\bf m})}_{k}$ and ${\bf u}^{({\bf m})}_{k}$ are 
coefficients of $\tilde{\bf c}$ and ${\bf c}$, respectively,
$[\bullet]^{({\bf m})}_{k}$ denotes the coefficient for the basis $\Phi^{\bf m}_{k}$, which can be projected out by the loop integrations as
\be
\besp
& a^{(\bf m)}_{k} =  \oint_{|w| \ll 1} \frac{d w }{2 \pi i} w^{k-1} \left[  \prod_{\ell=1}^{L} \oint_{|\zeta_\ell| \ll 1} \frac{d \zeta_\ell}{2 \pi i} \frac{\zeta_\ell^{-m_\ell-1}}{\sigma_\ell^{m_\ell}} \right]  A, \\
& \mbox{with} \ A=\sum_{{\bf m} \in {\mathbb N}_0^{L}} \sum_{k=0}^{\infty} a^{({\bf m})}_{k}  \Phi^{{\bf m}}_k, 
\end{split}
\ee
and $\star$ denotes the convolution product summing over ${\bf m}$ and $k$, equipping the usual matrix-vector product as well, defined as
\be
\besp
&   [a \star b]^{({\bf m})}_k = \sum_{{\bf m}^\prime \ge {\bf 0} }^{{\bf m}} \sum_{k^\prime=0}^{k}   a_{k^\prime}^{({\bf m}^\prime)} b^{({\bf m}- {\bf m}^\prime)}_{k-k^\prime}, \\
& \mbox{with} \ A=\sum_{{\bf m} \in {\mathbb N}_0^{L}} \sum_{k=0}^{\infty} a^{({\bf m})}_{k}  \Phi^{{\bf m}}_k, \quad {B}=\sum_{{\bf m} \in {\mathbb N}_0^{L}} \sum_{k=0}^{\infty} b^{({\bf m})}_{k} \Phi^{{\bf m}}_k. 
\end{split}
\ee
Here, we replaced the ${\bf c}$-dependence in the matricies with ${\bf u}$, and the multiplication between ${\bf c}$ (and also $w$) should be replaced with the convolution product.
Notice that $\tilde{\bm{\Gamma}}^{({\bf m})}_{k}$ is non-zero when ${\bf m}={\bf 0}$ and $k=0,1$.

\subsection{Transsasymptotic matching: general case}
\label{subsec:gen_transasymp}

In this section we present the transasymptotic matching condition for the multiparameter transseries solutions of the dynamical system~\eqref{eq:prepODE_c_w}. The rigorous mathematical demonstration of the generalized transasymptotic matching condition can be found in Refs~\cite{costin2008asymptotics,costin1998}. In this general case is more convenient to use the pseudomode basis defined in Eq.~\eqref{eq:pseudo_basis}. Following the same arguments outlined in Sect.~\ref{subsec:dyn-ren_c1} the solutions of the pseudomodes $\tilde{c}_i$ (with $i=1,\cdots,L$) can be written as
\bs
\label{eq:gen_transasym}
\beal
\label{eq:transcond_gen}
\tilde{c}_i(w)&=\sum_{k\geq 0}^\infty\,\tilde{G}_{i,k}\left(\boldsymbol{\sigma\,\zeta}(w)\right)\,w^{-k}\,,\\
\label{eq:gen_transfunc}
\text{with}\,\,\tilde{G}_{i,k}&=\sum_{{\bf n} \in {\mathbb N}_0^L}\,\tilde{u}^{(\boldsymbol{n})}_{i,k}\left( \prod_{j=1}^{L} \sigma_j^{n_{j}}\zeta_j^{n_{j}} \right).
\end{align}
\es
The transasymptotic matching condition~\eqref{eq:transcond_gen} can be directly obtained from~(\eqref{eq:pseudo_recur}) by taking the summation of ${\bf m}$ after putting $\prod_{\ell=1}^{L} \sigma^{m_\ell} \zeta^{m_\ell}$ in the each terms, and it leads to the following 1st-order PDE for the functions $\tilde{\bf G}_{k}\equiv (\tilde{G}_{0,k},\tilde{G}_{1,k},\cdots,\tilde{G}_{L,k})$
\begin{widetext}
%
%
%
\be
\label{eq:PDE_gentrans}
\besp
&  -  (  {\bf S} \cdot \hat{\bm{\zeta}} ) \, \tilde{\bf G}_{k} - (  \bm{\beta} \cdot \hat{\bm{\zeta}} + k ) \, \tilde{\bf G}_{k-1}
+ \frac{1}{20} \left[ {G}_{1} \star ({\bf S} \cdot \hat{\bm{\zeta}}) \,  \tilde{\bf G} \right]_k + \frac{1}{20} \left[ {G}_{1} \star (\bm{\beta} \cdot \hat{\bm{\zeta}}) \,  \tilde{\bf G} \right]_{k-1} + \frac{1}{20} \sum_{k_1,k_2=0}^{k_1+k_2=k} k_1  \tilde{\bf G}_{k_1-1} G_{1,k_2}\\ 
& + \left[ \left\{ \tilde{\Lambda} + \tilde{\frak Y}({\bf G}) + \tilde{\frak Z}({\bf G}) \right\}  \star \tilde{\bf G} \right]_{k}+ \left[ \tilde{\frak X}({\bf G}) \star \tilde{\bf G} + \tilde{\bm{\Gamma}}  \right]_{k-1}+  V  {\bf G}_{k-2} = 0,
\end{split}
\ee
\end{widetext}
where $\left[ \bullet \right]_k$ denotes the coefficient for the basis $w^{-k}$, and $\star$ denotes the convolution product summing only over $k$.
Notice that $\tilde{\bm{\Gamma}}_k = \tilde{\bm{\Gamma}}^{({\bf 0})}_k$.
The functions $\tilde{G}_{i,k}$ depend only on $\sigma_i\zeta_i$. The solution of this equation gives us automatically the time evolution of any operator $\mathcal{O}(\tilde{c}_i)$ by including all the transmonomials. If one increases the number of moments it is not a trivial task to solve Eq.~\eqref{eq:PDE_gentrans}. In principle, Eq.~\eqref{eq:PDE_gentrans} is a nonlinear PDE which can be solved provided a well defined initial condition as well as a correct choice for the eigenbasis (see Ref.~\cite{costin2008asymptotics} for technical details). To the best of our knowledge, the only case known in the literature where the transasymptotic matching condition has been solved explicitly is the $l=1$ case for the RTA Boltzmann equation~\cite{Behtash:2019txb,Behtash:2018moe}.
However, one can always reconstruct the functions $G_{i,k}$ from their definition~\eqref{eq:gen_transfunc} provided previous knowledge of the coefficients $\tilde{u}^{(\boldsymbol{n})}_{i,k}$ which are determined via the recursion relation~\eqref{eq:pseudo_recur}. The information of the IR data is encoded in the integration constants $\sigma_i$ which in principle can be determined from the UV data. 

\subsubsection{Dynamical RG flow equation: general case}
\label{subsec:first_second}

In this section we generalize the procedure of the RG flow equation when the number of moments $c_l$ is $l\geq 1$. Consider an observable $\mc O\equiv\mathcal{O}(\boldsymbol{c}(w))$ with $\boldsymbol{c}=(c_1,\cdots,c_L)$. Starting from the original ODE for the moments $c_l$~\eqref{eq:prepODE_c_w} and using Eq.~\eqref{eq:pseudo_ODE}, one finds that the change of $\mc O$ along the RG time $w$ is
%
\be
\label{eq:generRGeqn}
\besp
\frac{d\mc O(\boldsymbol{c}(w))}{d\log w}&
=-\sum_{i=1}^L\,\left[\sum_{k=0}^\infty\,\left(\tilde{\bf b}+w\boldsymbol{S}\right)\cdot \hat{\bf \zeta}\tilde{G}_{i,k}\,+\,k\,\tilde{G}_{i,k}\right]\,\frac{\partial \mc O}{\partial\tilde{c}_i}\,,\\
&=\sum_{i=1}^L\,\tilde{\bf \beta}_i\cdot \frac{\partial \mc O}{\partial\tilde{c}_i},
\end{split}
\ee
%
where $\tilde{\bf F}=\tilde{V}{\bf F}$ being the matrix $\tilde{V}$ the inverse of $\tilde{V}^{-1}$ in Eq.~\eqref{eq:pseudo_basis} (see also Eq.~\eqref{eq:pseudo_ODE}) and $\tilde{\bf \beta}=w\tilde{\bf F}$. In the previous equation we use that $\frac{d}{d\log w}=\left(\tilde{\bf b}+w\boldsymbol{S}\right)\cdot \hat{\bm{\zeta}}\,+\,\frac{\partial}{\partial\log w}$ and we denote $\hat{\zeta}_i=\partial/\partial \log \zeta_i$. The first line of Eq.~\eqref{eq:generRGeqn} encodes the scaling behavior of the observable $\mc O$ while the second one encodes the dynamics of the nonlinear ODEs through $\beta$. By solving this RG equation one summs over all the non-perturbative contributions of the multiparameter transseries and thus, determines the renormalization of $\mc O$. 

We are seeking for a RG flow equation of the transport coefficients so it is more convenient to consider that $\mc O=\mc O(G_{i,k})$. As a result, the transport coefficients depend intrinsically only on ${\bf \zeta}(w)$. Thus, for this case the RG flow equation of $\mc O$ reads as follows
\be
\label{eq:generRG_2}
\frac{d\mc O(G_{i,k}(w))}{d\log w}=-\sum_{i=1}^L\,\left[\sum_{k=0}^\infty\,\left(\tilde{\bf b}+w\boldsymbol{S}\right)\cdot \hat{\bm{\zeta}}\tilde{G}_{j,k}\right]\cdot \frac{\partial \mc O}{\partial\tilde{G}_{j,k}}\,,
\ee
 At this level the connection with the beta function $\tilde{\boldsymbol{ \beta}}$ is highly non-trivial and thus, one must consider particular cases like the $l=1$ case studied in Sect.~\ref{subsec:dyn-ren_c1} (see also Refs.~\cite{Behtash:2019qtk,Behtash:2018moe} for further examples). 

\subsubsection{Renormalization of first and second order transport coefficients}
\label{subsec:RGtrans}

From the results derived in the previous section we have the tools to study qualitatively the effect of adding more moments on the renormalized transport coefficients. In general, it is proven that the asymptotic leading order behavior of the moments $c_l\sim \mc O(w^{-l})$. The perturbative Chapman-Enskog expansion up to second order~\cite{Teaney:2013gca} leads to the following asymptotic series expansion of the moments $c_{1}$ and $c_{2}$ as follows~\cite{Blaizot:2017lht,Blaizot:2017ucy,Behtash:2019txb}
\bs
\label{eq:asympc1c2}
\beal
c_{1} &= -\frac{40}{3}\frac{1}{w}\frac{\eta}{s}-\frac{80}{9}\frac{1}{w^2}\frac{T(\eta\tau_\pi-\lambda_1)}{s}+ \mathcal{O}(1/w^3)\,,\\
c_{2} &= \frac{80}{9}\frac{T(\lambda_1+\eta\tau_\pi)}{s}\frac{1}{w^2} + \mathcal{O}(1/w^3)\,.
\end{align}
\es
If one now we implement the transasymptotic matching condition~\eqref{eq:transcond_gen} together with the asymptotic expressions of the moments we conclude
\bs
\label{eq:fs_limit_gen}
\beal
\frac{\eta}{s}&=-\frac{3}{40}\lim_{w\to\infty} G_{1,1}(\boldsymbol{\sigma\zeta}(w))\,,\\
\frac{T}{s}\eta\tau_\pi &= -\frac{9}{160}\lim_{w\to\infty}\left( G_{1,2}(\boldsymbol{\sigma\zeta}(w))-G_{2,2}(\boldsymbol{\sigma\zeta}(w))\right)\,,\\
\frac{T}{s}\lambda_1&=\frac{9}{160}\lim_{w\to\infty}\left( G_{1,2}(\boldsymbol{\sigma\zeta}(w))+G_{2,2}(\boldsymbol{\sigma\zeta}(w))\right)\,.
\end{align}
\es
The previous expressions generalize the results derived for the $l=1$ case, Eqs.~\eqref{eq:ren_transp}. It is then  straightforward to implement the newly developed concept of non-equilibrium transport coefficients when adding more moments as follows
\bs
\label{eq:ren_transp_general}
\beal
\left(\frac{\eta}{s}\right)_r &= -\frac{3}{40}G_{1,1}(\boldsymbol{\sigma\zeta}(w))\,,\\
\left(\frac{T}{s}\eta\tau_\pi\right)_r&= -\frac{9}{160}\left( G_{1,2}(\boldsymbol{\sigma\zeta}(w))-G_{2,2}(\boldsymbol{\sigma\zeta}(w))\right) \,,\\
\left(\frac{T}{s}\lambda_1\right)_r&= \frac{9}{160}\left( G_{1,2}(\boldsymbol{\sigma\zeta}(w))+G_{2,2}(\boldsymbol{\sigma\zeta}(w))\right)\,.
\end{align}
\es
The RG evolution equations of the first and second order transport coefficients are automatically determined by using the previous expressions and replacing $\mc O\to \mc O^r_i$ into Eq.~\eqref{eq:generRG_2} where $\mc O^r_i=\left(\left(\eta/s\right)_r,\left((T/s)\eta\tau_\pi\right)_r, \left((T/s)\lambda_1\right)_r\right)$ . 
These equations also illustrate that the inclusion of more moments indeed affects the values of the renormalized transport coefficients. One can also determine the renormalized transport coefficient by solving the recursive relation~\eqref{eq:pseudo_recur} and then reconstructing the multiparameter transseries as it was done in Fig.~\ref{fig:2}. However, the anomalous dimension $\beta_i$, the coefficients $u_{l,k}^{(\boldsymbol{n})}$ and the integration constants $\sigma_i$ (see Eq. (31)) become very sensitive to the number of moments added~\cite{Behtash:2019txb,Behtash:2019qtk}. Furthermore, the number of integration constants $\sigma$'s increase so the uncertainty to compare the IR and UV data becomes more cumbersome~\cite{Behtash:2019txb,Behtash:2019qtk}. The only terms that are insensitive to the changes in the number of dynamical variables are the Lyapunov exponents $\lambda(l)$. In general, any truncation scheme would lead to a multiparameter transseries with a finite radius of convergence (see App. D in Ref.~\cite{Behtash:2019txb} and Sect. 3.3. in Ref.~\cite{Behtash:2019qtk}). A possible way to circumvent this problem was proposed recently~\cite{Kamata:2020mka} where the ratio between the inverse Reynolds number and the Knudsen number was determined numerically.

\subsection{Transseries solutions in the UV}
\label{subsec:UV_gen}

In this section we derive the UV transseries solutions by generalizing
the results of Sect.~\ref{subsec:UV_pert_pow}. We show that the linear
perturbations of the Legendre modes $c_l$ around the UV fixed point, 
which is a source, follow a power-law decay. We begin by discussing 
some features of the stability of the dynamical system of 
ODEs~\eqref{eq:ODE_cl_w} in the UV regime.

Consider the $w\to 0$ limit of Eq.~\eqref{eq:ODE_cl_w} by first 
changing the variable $w=1/z$. We get
\bs
\label{eq:ODE_cl_z}
\beal
\frac{d{\bf c}}{dz}&={\bf F}({\bf c},z)\,,\\
{\bf F}({\bf c},z)&=\frac{1}{\left(1-\frac{c_1}{20}\right)}\,\left[\frac{1}{z}\left(\frak{X}({\bf c}) \bf{c}+{\bf \Gamma}\right)\,\right.\nl
&\left. +\,\frac{1}{z^2}\left(\hat{\Lambda} + {\frak Y}({\bf c}) + {\frak Z}({\bf c})\right){\bf c}
\right]\,.
\end{align}
\es
In the $z\to \infty$ limit the dominant terms of the previous equation
are $\mc O(1/z)$. In this limit one can determine the fixed points by 
solving the following equation 
\be
\label{eq:fix_point_z}
\frak{X}({\bf c}){\bf c}+{\bf \Gamma}=0\,.
\ee
The solutions to this equation provides a set of vectors $\bar{\bf 
c}=(\bar{c}_1,...,\bar{c}_L)$ which determine the UV fixed points of 
the dynamical system. The solutions to Eq.~\eqref{eq:fix_point_z}
depend on the truncation order $l_{max}$, so these are not necessarily
real and we need to impose the reality conditions. Furthermore, the
original FPE admits two UV fixed points for the moment $c_1$, which
have the property that the transverse and longitudinal pressures 
are minimized, $c_1=5$ and $c_1=-5/2$ respectively. We call these 
configurations \textit{maximally prolate} and \textit{maximally oblate} 
respectively. The two UV fixed points act as bounds for the basin of 
attraction from above and below along the $c_1$ ray in the infinite
dimensional space of the moments $c_l$. Thus, any truncation scheme 
of the distribution function can be considered a good approximate 
solution of the Boltzmann equation iff the UV stability properties are 
reproduced approximately. In this case, the truncated dynamical system 
of ODEs should have at least two UV fixed points for the moment $c_1$. 

%
\begin{figure}[t]
\begin{center}
	\includegraphics[scale=.17]{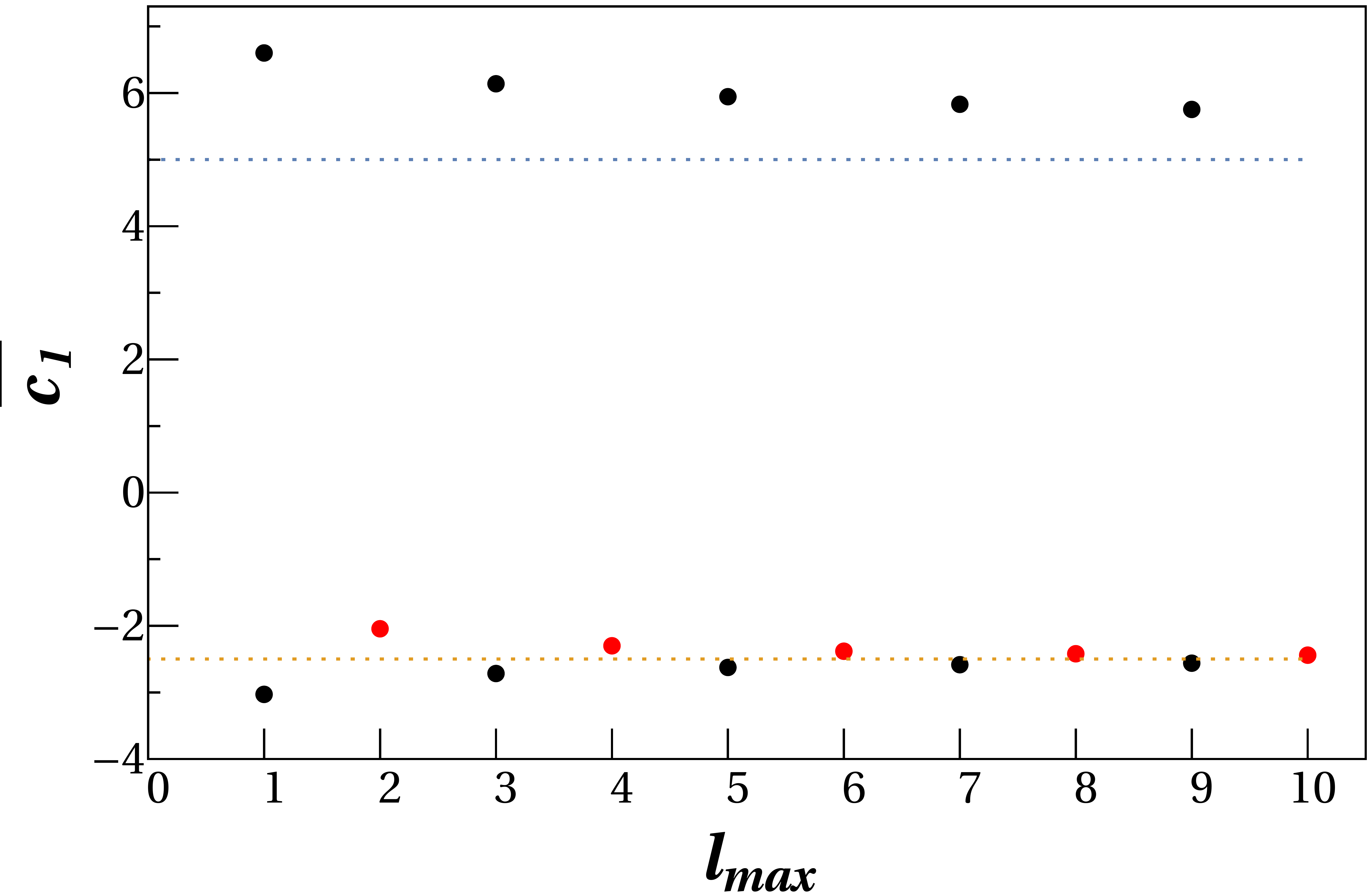}
	\caption{(Color online) Distribution of the UV fixed points for the moment $\bar{c}_1$ as a function of the truncation order $l_{max}$. Black and red dots correspond to $l_{max}$ being either an odd or an even number respectively. Blue ($\bar{c}_1=5$)and brown ($\bar{c}_1=-5/2$) dashed lines correspond to the UV fixed points of the original FPE.} 
	\label{fig:Uv}
	\end{center}
\end{figure}
%
%
In Fig.~\ref{fig:Uv} we plot the real solutions to 
Eq.~\eqref{eq:fix_point_z} for the moment $\bar{c}_{1}$ at the fixed 
points as a function of the truncation order $l_{max}\in[1,10]$. We 
find that if $l_{max}$ is odd then the real solutions for $\bar{c}_1$
come in pairs. As $l_{max}$ increases (for odd values) the UV fixed
points approach the expectation for the original FPE, $\bar{c}_1=
\{5,-5/2\}$. On the other hand, there is only one real solution for  
$\bar{c}_1$ if $l_{max}$ is even. This observations also hold for 
the UV fixed points of higher order moments $\bar{c}_l$. Thus, the
standard idea of obtaining more accurate approximate solution to the 
Boltzmann equation by adding more moments to the distribution function 
is not necessarily correct. A good truncation scheme must reproduce
the flow structure of the original Boltzmann equation in the IR and 
UV regimes. Similar findings were reported for the RTA Boltzmann
equation~\cite{Behtash:2019txb,Behtash:2019qtk}. In the rest of this
section we will consider the case that $l_{max}$ is odd. 

In the limit $z\to\infty$ we linearized Eq.~\eqref{eq:ODE_cl_z} 
around a particular UV real fixed point $\bar{\bf c}$, namely 
${\bf c}\to \bar{\bf c}+ \delta{\bf c}$, which lead to the 
following equation for the linear perturbations
%
\be
\label{eq:linODE_dcl_z}
\besp
\frac{d\delta {\bf c}}{dz}&=\sum_{k=1}^L\frac{\partial {\bf F}({\bf c},w)}{\partial c_k}\biggr|_{ c_k=\bar{c}_k}\,\delta c_k\,\\
&=\frac{1}{1-\frac{\bar{c}_1}{20}}\,\left[\bar{\frak{X}}
  -\frac{\bar{c}_1}{5}\, {\bf 1}_L -\frac{1}{5} 
  \begin{pmatrix}
    \bar{c}_1 & 0 &\cdots & 0 \\
    \bar{c}_2 & 0 &\cdots & 0 \\
    \vdots & \vdots &\ddots & \vdots \\
    \bar{c}_L & 0 &\cdots & 0 \\
  \end{pmatrix}
  \right]\left(\frac{\delta {\bf c}}{z}\right)\,\\
&\equiv \, \frak{T}(\bar{\bf c})\,\cdot\left(\frac{\delta {\bf c}}{z}\right)\,\,\,,
\end{split}
\ee
%
where the matrix 
$\frak{X}$ is defined in
Eq.~\eqref{eq:chimat}. Upon inspection we find that 
Eq.~\eqref{eq:UVpertc1} is a particular case of this linearized 
equation for the perturbations. We introduce the linear 
pseudomodes $\delta \tilde{\bf c}=\frak{U}\delta {\bf c}$ so 
that $\frak{U}$ diagonalizes the matrix $\frak{T}(\bar{\bf c})$, 
\be
\label{eq:eigenT}
\besp
\frak{T}(\bar{\bf c})&\to \frak{U}\frak{T}(\bar{\bf c})\frak{U}^{-1}\equiv\tilde{\frak{T}}(\bar{\bf c})\\ \text{with}\,\, &\tilde{\frak{T}}(\bar{\bf c})=\,\text{diag.}\,(\hat{\frak{t}}_1,\cdots\hat{\frak{t}}_L)\,.
\end{split}
\ee
This procedure gives the equation for the linearized pseudomodes
\be
\label{eq:linODE_psdcl_z}
\frac{d\delta\tilde{\bf c}}{dz}=\frac{1}{z}\,\tilde{\frak{T}}(\bar{\bf c})\cdot \delta\tilde{\bf c}\,. 
\ee
Its solution is 
\bs
\label{eq:genpertcl}
\beal
\delta\tilde{c}_i&=\mu_i z^{\hat{\frak{t}}_i}\,,\\
\Rightarrow \qquad\delta { c}_i&=\frak{U}^{-1}_{ij}\delta\tilde{c}_j=\frak{U}^{-1}_{ij}\mu_j w^{-\hat{\frak{t}}_j}\,.
\end{align}
\es
This result shows that close to the UV fixed points the linearized
perturbations have a power-law behavior which is opposite to their IR
counterparts~\eqref{eq:dcl_sol_lin}. In general the eigenvalue
$\hat{\frak{t}}_i$, the integration constant $\mu_i$ as well as the
matrix element $\frak{U}_{ij}$ are complex numbers. The reality 
condition of the linearized modes is satisfied since the coefficients 
of the polynomial eigenvalue problem of the matrix $\frak{T}$ are 
real and thus, their complex eigenvectors and eigenvalues come in
complex conjugate pairs. When summing over the complete eigenbases of 
pseudomodes the linearized modes are real as expected. Moreover, what 
happens in the vicinity of any UV fixed point $\bar{\bf c}$ along the 
ray $\bar{c}_i$ is encoded by the real part of the eigenvalues of the 
matrix $\frak{T}$, namely $\frak{R}(\hat{\frak{t}}_i)$. In the limit 
$z\to\infty$ the linearized perturbation $\delta\tilde{c}_i$ decays 
rapidly as a power law if $\frak{R}(\hat{\frak{t}}_i)<0$ while 
$\delta\tilde{c}_i$ increases when $\frak{R}(\hat{\frak{t}}_i)>0$. 

In order to generate approximate UV solutions in the limit $z\to\infty$
limit it is convenient to rewrite the equations of motion in terms of 
the pseudomode basis $\tilde{\bf c}_i=\frak{U}_{ij}{\bf c}_j$ (with 
$\frak{U}$ given by Eq.~\eqref{eq:eigenT}). In terms of the pseudomodes 
the nonlinear ODEs~\eqref{eq:ODE_cl_z} read 
\bs 
\label{eq:ODEpseudo_z}
\beal
\frac{d\tilde{\bf c}}{dz}&=\tilde{\bf F}(\tilde{\bf c},z)\,,\\
\tilde{\bf F}(\tilde{\bf c},z)&=\frac{1}{\left(1-\frac{c_1}{20}\right)}\,\left[\frac{1}{z}\left(\tilde{\frak{X}}({\bf c})\,\tilde{\bf c}+\tilde{\bm{\Gamma}}\right)\,\right.\nl
&\left. +\,\frac{1}{z^2}\left(\tilde{\Lambda} + \tilde{\frak Y}({\bf c}) + \tilde{\frak Z}({\bf c})\right)\tilde{\bf c}
\right]\,.
\end{align}
\es
In the previous equation it must be understood that 
$c_1=\sum_{j=1}^L\,\frak{U}^{-1}_{1j}\tilde{c}_j$. The transseries 
ansatz solution is built up based on the solutions of the linearized 
perturbations~\eqref{eq:genpertcl}. These solutions can be taken as the 
transmonomials of the transseries which carry out the information about 
the non-perturbative contributions. Therefore, in the limit $z\to
\infty$ the transseries ansatz is given 
by~\cite{Behtash:2019txb,Behtash:2019qtk}
\bs
\label{eq:transsol_pseudo}
\beal
\tilde{c}_l(z)&=\sum_{|{\bf m}|\geq 0}^\infty\,\sum_{k=0}^\infty\,
\tilde{v}^{({\bf m})}_{l,k}\boldsymbol{\varphi}^{{\bf m}}_{k}\,,\\
\boldsymbol{\varphi}^{{\bf m}}_{k}&:=\frac{1}{z^k}\,\left(\prod_{j=1}^L\,\xi_j^{m_j}\right)\,,\qquad\text{with}\,\,\xi_j=\mu_j\,z^{\hat{t}_j}\,,
\end{align}
\es
where ${\bf m}\in {\mathbb N}^L_{0}$, $\mu_{i}\in {\mathbb C}$ is the 
integration constant and $L \in {\mathbb N}$ is the truncation order. 
The UV data is determined by matching the transmonomial $\xi_j$ with 
the linearized solutions~\eqref{eq:genpertcl} and thus, the anomalous
dimensions $\alpha_j$ entering in the previous expression are the set 
of eigenvalues $\hat{\bf t}$ of the linearization matrix 
$\hat{\frak{T}}$, Eq.~\eqref{eq:eigenT}, evaluated at the UV fixed 
points $\bar{\bf c}$. A similar transseries ansatz was proposed for 
the RTA Boltzmann equation in systems undergoing
Bjorken~\cite{Behtash:2019txb} and Gubser flow~\cite{Behtash:2019qtk}.
The solutions of the Legendre moments in terms of the pseudomodes 
are given by
\begin{align}
\label{eq:genUVsol}
 c_l(z)=\sum_{l'=1}^{L}\frak{U}^{-1}_{ll'}\tilde{c}_{l'}&=\sum_{|m|=0}^\infty\,\sum_{k=0}^\infty\,v_{l,k}^{({\bf m})}\boldsymbol{\varphi}^{\bf m}_k\,,\\
 v_{l,k}^{({\bf m})}&=\sum_{r=1}^L\,\frak{U}^{-1}_{lr}\tilde{v}_{r,k}^{({\bf m})}\,.
\end{align}
The coefficients $\tilde{v}^{({\bf m})}_{l,k}$ entering in the transseries
ansatz are determined entirely by inserting 
Eq.~\eqref{eq:transsol_pseudo} into Eq.~\eqref{eq:ODEpseudo_z} 
which leads to the following recursive relation
\begin{widetext}
\be
\label{eq:pseudo_recur_2}
\besp
 \left({\bf m} \cdot {\bf \hat{t}}  -k+1 \right) \tilde{\bf v}_{k-1}^{(m)} - \frac{1}{20} \sum_{{\bf m}_1,{\bf m}_2={\bf 0}}^{{\bf m}_1+{\bf m}_2={\bf m}} \sum_{k_1,k_2=0}^{k_1+k_2=k}&\left[ \left({\bf m} \cdot {\bf \hat{t}}-k+1 \right) \tilde{\bf v}_{k_1-1}^{({\bf m}_1)}  \right] {v}_{1,k_2}^{({\bf m}_2)} \\
&+ \left[ \left\{ \tilde{\Lambda} + \tilde{\frak Y}({\bf u}) + \tilde{\frak Z}({\bf u}) \right\} \star \tilde{\bf u} \right]^{({\bf m})}_{k-2}+ \left[  \tilde{\frak X}({\bf u})\star \tilde{\bf u} + \tilde{\bm{\Gamma}} \right]^{({\bf m})}_{k-1}= 0,
\end{split}
\ee
\end{widetext}
where $\star$ denotes the convolution product. The normalization condition is chosen to be $\tilde{v}^{({\bf n})}_{i,0}=1$ when $n_{j} = \delta_{j,i}$. 
and the coefficients $\tilde{v}^{({\bf 0})}_{i,0}=\frak{U}_{ij}\bar{c}_j$ 
since one expands around a given UV fixed point ${\bf \bar{c}}$. In 
Eq.~\eqref{eq:pseudo_recur_2} $\tilde{\bf v}^{({\bf m})}_{k}$ and ${\bf 
v}^{({\bf m})}_{k}$ are coefficients of $\tilde{\bf c}$ and ${\bf c}$ 
respectively, and $[\bullet]^{({\bf m})}_{k}$ denotes a coefficient 
proportional to $\varphi^{\bf m}_{k}$.

 We emphasize that the transseries ansatz~\eqref{eq:transsol_pseudo} 
is a convergent series with a finite radius of convergence. The rate
of convergence depends on the anomalous dimension 
$\alpha_i\equiv\frak{R}(\hat{t}_i)$ governing linearized perturbations
$\delta \tilde{c}_i$ ~\eqref{eq:genpertcl}. The lessons learned from 
the $l=1$ truncation studied in Sect.~\ref{subsec:UV_c1} (see also 
Refs.~\cite{Behtash:2019txb,Behtash:2019qtk}) show that the power law 
transseries solutions~\eqref{eq:genUVsol} can be constructed iff 
$\frak{R}(\hat{t}_i)<0$. When $\frak{R}(\hat{t}_i)>0$ the best option 
is to perform an analytical continuation while cancelling the
transmonomial contributions in Eq.~\eqref{eq:transsol_pseudo} by 
setting the integration constant $\mu_i\equiv 0$. 

\section{Universal aspects of attractors for different kinetic models}
\label{sec:univ}

In the previous sections we analyzed the behavior of the solutions 
of the FPE in both the UV and IR limits. In this section we study
universal aspects of the attractors in the FPE as well as the RTA 
Boltzmann equation and the AMY kinetic theory~\cite{Arnold:2002zm}. 

Following the approach of Refs.~\cite{Giacalone:2019ldn,Kamata:2020mka}
we consider a distribution function which is squeezed along the beam
direction at early time~\footnote{
For the RTA Boltzmann equation the form of the initial distribution
function does not play a major role. However different processes of
Yang-Mills plasmas, i.e. elastic and inellastic interactions, affect
the parametrization of the momentum
distributions~\cite{Schlichting:2019abc,Dusling:2009df}. In our approach
these effects do not play a role at the level of the moments since the 
free streaming expansion dominates at early times over the collision 
rate of the FPE as discussed in Sect.~\ref{subsec:UV_gen}.
}. In this situation the initial longitudinal pressure vanishes 
exactly and thus, this configuration determines the pullback 
attractor of the distribution function at early 
times~\cite{Behtash:2018moe,Behtash:2019qtk}. Under this condition 
the initial phase space distribution can be modeled 
as~\cite{Kamata:2020mka}
\begin{eqnarray}
\label{eq:inidistfunc}
f_{0}(\tau_0,p_T,p_\varsigma) =  (2\pi)^3~\delta(p_{\varsigma}) \frac{dN_{0}}{d\varsigma d^2\bold{p_T} d^2\bold{x_T}}\;. \nonumber \\
\end{eqnarray}
The normalization constant is chosen such that the initial energy
density per unit rapidity per transverse area is 
constant
\begin{eqnarray}
\frac{dE_{0}}{d\varsigma d^2\bold{x_T}}=\lim_{\tau_0\to0} \tau_0 e(\tau_0) = \left( \tau e \right)_0 = const\;.
\end{eqnarray}
The initial distribution function determines the initial conditions 
for the Legendre moments
\be
\label{eq:cl0}
\besp
\cl&=\left(4l+1\right)\,(-1)^l\frac{\Gamma\left(l+1/2\right)}{\Gamma\left(l+1\right)}\,.
\end{split}
\ee
Previous studies~\cite{Baier:2007ix,Giacalone:2019ldn} have shown 
that universal behavior of the numerical solutions can be by analyzed
in terms of observables that are less sensitive to the initial
conditions. Following Refs.~\cite{Giacalone:2019ldn,Kamata:2020mka} 
we study the following observable
\be
\label{eq:normene}
\besp
\mc E&=\frac{\tau^{4/3}\,\epsilon(\tau)}{\left(\epsilon\,\tau^{4/3}\right)_{hydro}}\,,\\
\left(\epsilon\,\tau^{4/3}\right)_{hydro}&\equiv\lim_{\tau\to\infty}\tau^{4/3}\epsilon(\tau)\,.
\end{split}
\ee
%
%
\begin{figure*}[t]
\begin{center}
\includegraphics[scale=.27]{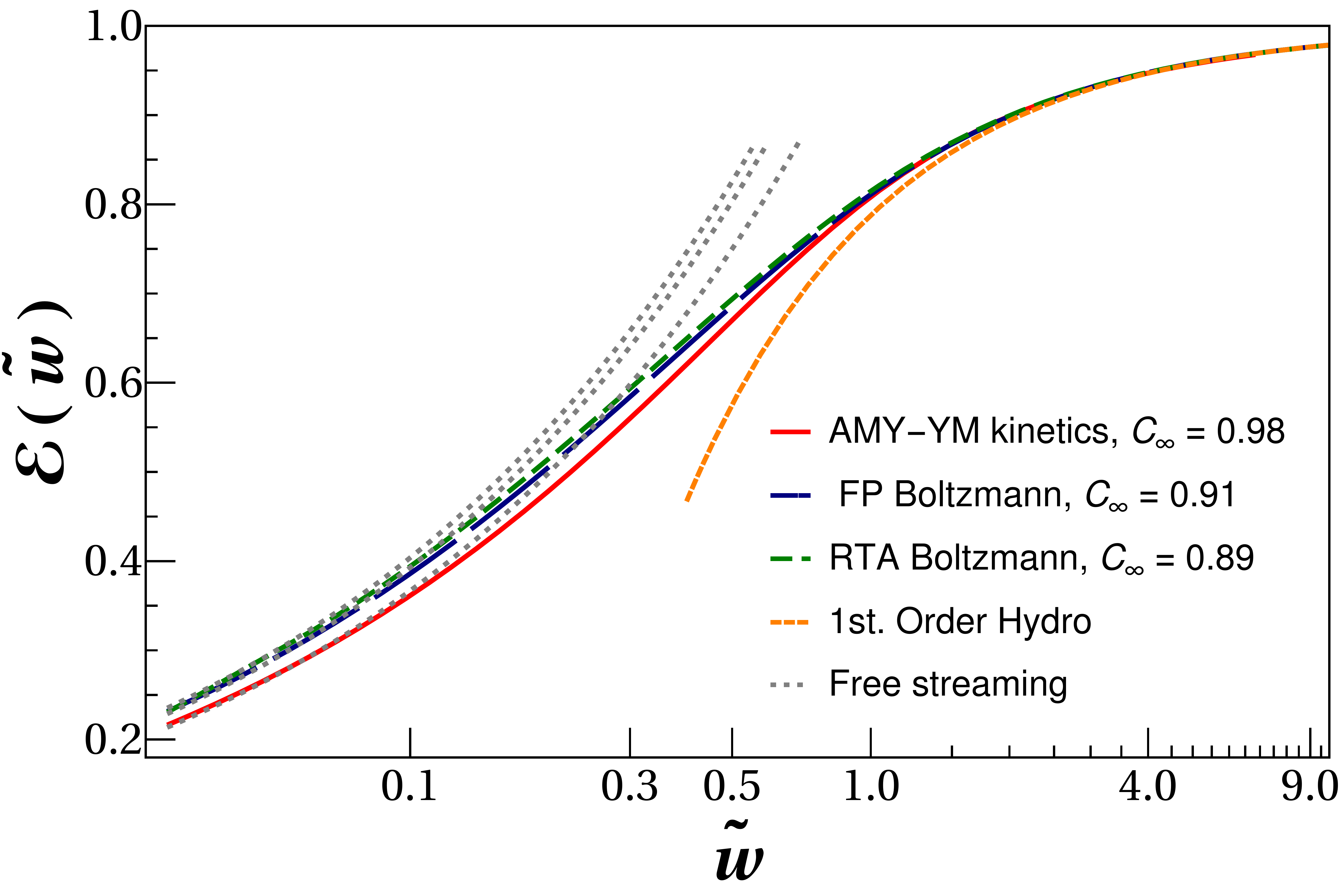}
	\caption{(Color online) Evolution of the normalized energy density~\eqref{eq:normene} vs. $\tilde{w}=\tau T(\tau)/[(4\pi)\eta/s]$. We show the numerical results obtained for the FPE (dashed blue line), YM-AMY kinetics (red line) and RTA Boltzmann (dashed green line). In addition we present the universal hydrodynamic behavior (dashed orange line) and early-time free streaming (dotted gray line).} 
	\label{fig:univatt}
	\end{center}
\end{figure*}
%
%
We will analyze this observable as a function of the variable 
$\tilde{w}=\tau\,T(\tau)/\left[(4\pi)\eta/s\right]$. In terms of 
this variable the results are insensitive to the strength of the
coupling. Eq.~\eqref{eq:normene} has two interesting limits in the 
forward and pullback attracting regions. The former is described by 
a few terms of the non-hydrodynamic expansion~\cite{Baier:2007ix}, 
while the latter is determined by expanding around the UV fixed point 
where the longitudinal pressure vanishes~\cite{Giacalone:2019ldn}, 
\be
\label{eq:normene_limit}
\mc E(\tilde{w})=\begin{cases}
C^{-1}_\infty\,\tilde{w}^{4/9}\,\,(\text{pullback attractor})\,,\\
1-\frac{2}{3\pi}\frac{1}{\tilde{w}}\,\,(\text{forward attractor})\,.
\end{cases}
\ee
where the constant $C_\infty$ is determined from Eq.~(51) in
Ref.~\cite{Kamata:2020mka}. Alternatively one can determine 
this constant by fitting to the numerical data. We verified that 
both methods lead to the same approximate value for this constant,
see also \cite{Giacalone:2019ldn,Kamata:2020mka}. 

We numerically solved the evolution equations for the Legendre moments,
Eqs.~\eqref{eq:ODE_tau} and~\eqref{eq:ODE_cl_tau_RTA} respectively, 
together with the conservation law~\eqref{eq:consqnTtau}. The solutions 
for the Legendre moments $c_l$ were obtained by truncating the 
expansion at $l<l_{max}$, setting $c_{l}\equiv 0$ for $l>l_{max}$. 
This method converges rapidly and no sizable deviations are observed 
for $l_{max}\gtrsim 35$ for both the RTA Boltzmann and FPE equations.
In the case of the Yang-Mills AMY theory we used publicly available 
numerical results~\footnote{
The numerical data obtained in \cite{Kurkela:2018wud,Kurkela:2018vqr,
Giacalone:2019ldn} are available at 
https://pub.uni-bielefeld.de/record/2939684. We thank to 
S.~Schlichting for directing us to this site.}
\cite{Kurkela:2018wud,Kurkela:2018vqr,Giacalone:2019ldn}.

In Fig.~\ref{fig:univatt} we show numerical solutions of $\mc E$ 
as a function of $\tilde{w}$ for the RTA Boltzmann equation~(dashed
green line), FPE~(dashed blue line) and AMY kinetic theory~(red line). 
The general behavior of the different kinetic models is quite similar.
They all exhibit a smooth transition from early-time free streaming
to universal late-time hydrodynamic behavior. Deep in the UV regime 
the expansion (dashed gray lines) dominates over collisions regardless 
of the underlying collision kernel, and the pullback attractor is 
determined by free streaming expansion. A more surprising fact is 
that the interactions modifies the expansion for similar values of 
$\tilde{w}\approx 0.08$, i.e. at a large Knudsen number $Kn\approx$ 
12.5, independent of the underlying microscopic theory. We note that 
the existence of a pullback attractor is manifest only in the $w$ 
variable. In the original proper time variable $\tau$ the limit 
$\tau_0\to 0$ is not well defined because the ODEs have an essential 
singularity at this point~\cite{Behtash:2019txb,Behtash:2019qtk}.    

In the UV regime the main difference among the kinetic models 
depicted in Fig.~\ref{fig:univatt} is the values of the constant 
$C_\infty=\{0.98,0.91,0.89\}$ for the AMY, FPE, and RTA models, 
respectively. These values allows us to quantify the difference 
among the predictions of RTA and FPE respect the YM-AMY model. 
The former differs by $\sim 10\%$ while the latter $\sim 8\%$. 
It is somewhat surprising that a relatively simple approximation 
such as RTA captures many aspects of the non-equilibrium dynamics 
compared with YM-AMY kinetics. This observation was also noticed 
previously in~\cite{Giacalone:2019ldn,Kamata:2020mka}.

In the IR regime we observe in Fig.~\ref{fig:univatt} that all 
kinetic models reach the non-hydrodynamic behavior and thus, the 
forward attractor is encoded by the late-time non-hydrodynamic 
expansion. All the kinetic models reach the asymptotic hydrodynamic 
gradient expansion around $\tilde{w}\approx 2$, and thus $Kn\approx 
0.5$ so the deviations from equilibrium are quite large. This 
finding implies that one can reach out the non-hydrodynamic behavior 
without having local thermal equilibrium.  

 Given that we find forward attractors for a number of different 
kinetic equations, one may ask how general this feature is. In 
App.~\ref{sec:invman} we discuss a sufficient set of mathematical 
conditions which ensure its existence for weakly coupled boost 
invariant systems with highly non-linear collision kernels beyond 
those studied in this work. 

\section{Discussion and final remarks}
\label{sec:concl}
 
   In this work we studied the hydrodynamization of a boost invariant system
of gluons described by kinetic equation derived from QCD in the small angle,
diffusive, approximation. We demonstrate that the physics of the Fokker-Planck
equation can be recast in terms of a set of non-linear ODEs for the Legendre 
moments of the one-particle distribution function. We show that these kinetic 
equations admit transseries solutions in both the UV and IR regimes. These  
findings extend previous results obtained in the context of the Boltzmann 
equation in the relaxation time approximation
\cite{Behtash:2018moe,Behtash:2019txb,Behtash:2019qtk}. 

 We employ techniques from the theory of nonlinear dynamical systems. Applying
these methods to the kinetic equations we investigate the stability properties 
of linearized perturbations around the IR and UV fixed points. We analyze the 
the emergence of transseries solutions and show that their functional form is 
a rather natural consequence of the stability properties of the nonlinear ODEs 
around the UV and IR fixed points, respectively. 

In the IR regime we prove that the solutions of the moments equations are
multiparameter transseries. The associated transmonomials are built up using 
the behavior of linearized perturbations around the IR fixed point. These 
perturbations involve the product of an integration constant, an exponentially
decaying term and a power law term. The exponentially decaying term determines 
the rate at which linearized perturbations decay near the IR fixed point. This
decay time is controlled by the product of the Lyapunov exponent and the inverse
Knudsen number. These terms play an analogous role to `instanton-like' contributions
in QFT. The IR transseries effectively resums non-perturbative dissipative
contributions that are absent in the usual perturbative gradient expansion. 
These results strongly suggest that a consistent formulation of non-hydrodynamic
theories in far-from-equilibrium regimes requires the inclusion on non-perturbative
physics. 

In the IR regime the constitutive relations of each Legendre moment $c_l$ 
contain the non-perturbative information encoded in the multiparameter transseries.
The result suggests a non-perturbative dynamic renormalization scheme which 
goes beyond standard linear response theory. In this theoretical framework, 
transport coefficients can be defined in far-from-equilibrium regimes when 
the Knudsen number is large. At each order of the pertubative IR expansion, 
transport coefficients are dynamically renormalized by non-perturbative 
corrections in the transseries, a method known as ``transasymptotic matching''
in the resurgence literature. This approach has a nice interpretation in terms
of an RG flow. 

From the physical standpoint of view, the renormalized transport coefficients 
depend on the rheology of the fluid. They describe the relaxation of dissipative
coefficients to their values dictated by the linear response approach. Thus, the 
fluid experiences transient non-Newtonian behavior prior to hydrodynamization. 

On the mathematical side, we explored the relation between dynamical systems 
and RG flows. Indeed, it is known that the field theoretical RG flows may be 
discussed in the framework of autonomous dynamical systems \cite{Gukov:2015qea,
Gukov:2016tnp}. Our work shows that this link can be extended to non-autonomous
systems that hydrodynamize, where time plays the role of scale parameter. In 
other words, a non-autonomous dynamical system can be considered as an RG flow 
equation provided there is a slow invariant manifold, i.e. when the long-time
physics is dominated by slow degrees of freedom~\cite{Behtash:2019qtk}. In the 
case studied here and in~\cite{Behtash:2018moe,Behtash:2019txb,Behtash:2019qtk}
the invariant manifold is shown to exist if the IR perturbations go to zero in 
the long-time limit, a fact that points to the existence of a forward attractor. 
In particular, there are bounded solutions at $w\rightarrow\infty$ or $\tau 
\rightarrow\infty$ for a fixed initial time that approach the equilibrium point. 
For expanding systems such as Gubser flow~\cite{Behtash:2017wqg,Behtash:2019qtk}, 
the system is completely perturbative so that full hydrodynamization does not
occur. Therefore, there is no slow invariant manifold, which in turn explains 
why the RG-flow paradigm for the transport coefficients is no longer available. 

In the UV regime the nonlinear ODEs for the moments admit transseries solutions 
with a rather different behavior compared to their counterparts in the IR limit. 
These solutions are power series with a finite radius of convergence. The radius 
of convergence $R$ grows linearly with the shear viscosity over entropy ratio
$\eta/s$. The linear relation between $R$ and $\eta/s$ provides a rather simple
explanation of previous numerical findings~\cite{Martinez:2017ibh,Bazow:2015cha,
Denicol:2014tha,Denicol:2014xca,Tinti:2015xra,Denicol:2014mca,Florkowski:2014bba,
Florkowski:2014sfa,Florkowski:2013lya,Florkowski:2013lza} where it was noticed 
that initial conditions close to the maximally oblate UV fixed point converge 
slowly to the forward non-hydrodynamic attractor than the ones located in the 
vicinity of the maximally prolate UV fixed point. The latter is a saddle point 
for any viable truncation (e.g. if the truncation bound for $l$ is odd) and it 
is found that there are a set of solutions for the Boltzmann equation that 
remain bounded near this point. In contrast, the maximally oblate UV fixed point 
is a source from which flow lines are streaming away in $w_0\rightarrow0$ 
for a fixed $w$, meaning that there are no attracting regions to probe 
in the past of the dynamical system around this fixed point. 

 We show that the validity of the power series expansion can be extended by 
analytic continuation. As an alternative, we also introduce a new resummation 
scheme which accounts for the non-perturbative physics of the power-law behavior 
of the fluctuations around the UV fixed points. We note that the UV stability 
analysis also unveils a non-trivial aspect of truncating the moment expansion
in relativistic kinetic theory: In order to preserve the UV structure of the 
original kinetic equation for Bjorken flow we have to include an odd number of
Legendre moments $L=2n+1$ moments in the $n\to\infty$ limit. 

We discussed the general properties of the IR and UV regimes of the 
Fokker-Planck equation in comparison to the RTA Boltzmann and AMY kinetic
equations. The pullback attracting region is entirely determined by the free 
streaming limit where the longitudinal extent of the distribution function is
negligible. The attracting region in the long time limit is corresponds to
hydrodynamization and can be characterized by a few terms in the gradient 
expansion. There are, however, minor between different models. We find that 
the RTA and AMY differ by at most 10\% while the latter disagrees by up to
8\% with the FPE result. 

The techniques presented in this work are not necessarily restricted to
far-from-equilibrium systems undergoing longitudinal boost invariant 
expansion. Indeed, these methods can be applied to describe non-equilibrium 
dynamics in other physical systems of relevance such as cold atoms systems, 
condensed matter and cosmology. 

\acknowledgments

We would like to thank M.~Spalin\'ski, J.~Jankowski, S.~Schlichting and 
A.~Mazeliauskas for useful discussions. M.~M.~and T.~S.~are supported in 
part by the US Department of Energy Grant No.~DE-FG02-03ER41260. M.~M., T.~S., 
and V.~S.~are partially supported by the BEST (Beam Energy Scan Theory) 
DOE Topical Collaboration. S.~K.~is supported by the Polish National 
Science Centre grant 2018/29/B/ST2/02457. A.~B.~was partially supported 
by the DOE grant DE-SC0013036 and the National Science Foundation under 
Grant No. NSF PHY-1125915. V.~S.~acknowledges support by the DOE Office 
of Nuclear Physics through Grant No.~DE-SC0020081. V.~S.~thanks the 
ExtreMe Matter Institute EMMI (GSI Helmholtzzentrum für Schwerionenforschung, 
Darmstadt, Germany) for partial support and hospitality. 

\appendix
\begin{widetext}
\section{Evolution equations for the Legendre moments $c_{l}$}
\label{app:eqcl}
In this appendix we briefly describe the main elements to derive the evolution equation for the Legendre moments $c_l$. Some important identities of the Legendre polynomials were used and explicitly listed at the end of this section, see~\ref{subsec:leg}. The equation for the temperature is obtained from the conservation law which in our case reads as 
\be
\label{eq:ODE_T}
\frac{d\epsilon}{d\tau}+\frac{1}{\tau}\left(\epsilon+p_L\right)=0\qquad\Rightarrow\qquad
\frac{d T(\tau)}{d \tau}  = - \frac{T(\tau)}{3 \tau} \left( 1 + \frac{1}{10} c_{1}(\tau) \right) \,,
\ee
where we used explicitly the matching condition for the energy~\eqref{eq:matchene} with $c_0\equiv 1$. In order to obtain the equations for the Legendre moments $c_l$ we multiply first both sides of the FPE~\eqref{eq:FPeq} by $\int_p (-u\cdot p)^2\,P_{2l}(\cos\theta_{\bf p})$. As a result we get 
\be
\label{eq:int_FP}
\besp
\int_{\bf p}\,\left(-u\cdot p\right)^2\,P_{2l}(\cos\theta_p)\,\partial_\tau f_{\bf p}\,
&\,=\lambda^2 \,l_{Cb} \,\int_{\bf p}\,\left(-u\cdot p\right)^2\,P_{2l}(\cos\theta_p)\,\left\{
 \nabla_{\bf p} \cdot \left[ {\cal J}(\tau)\, \nabla_{\bf p} f_{\bf p} + {\cal K}(\tau) \,\frac{{\bf p}}{p} f_{\bf p} \left( 1+ f_{\bf p} \right) \right] 
\right\}
\end{split}
\ee
Now we equate in the previous expression the ansatz~\eqref{eq:ansatzf} into the previous expression. In the LHS of Eq.~\eqref{eq:int_FP} one simply gets 
\be
\label{eq:LHS_FP}
\besp
\int_{\bf p} \left(-u\cdot p\right)^2 P_{2 \ell}(\cos \theta_p) \frac{\pd f(\tau,{\bf p})}{\pd \tau}&=
\int_{\bf p} \left(-u\cdot p\right)^2 P_{2 \ell}(\cos \theta_p) \quad
\left[\sum_{\ell'=0}^{+\infty} \left( \frac{dc_{\ell'}(\tau)}{d \tau} P_{2 \ell'}(\cos \theta_p) f_{\rm eq}(\tau,{\bf p}) \right. \right. \\
 & \left.\left. + c_{\ell'}(\tau) \frac{\pd P_{2 \ell'}(\cos \theta_p)}{\pd \tau} f_{\rm eq}(\tau,{\bf p})    + c_{\ell'}(\tau) P_{2 \ell'}(\cos \theta_p) \frac{\pd f_{\rm eq}(\tau,{\bf p})}{\pd \tau} \right)\right] \,,\\
&= \frac{\epsilon(\tau)}{ (4 \ell + 1)} \left[ \frac{d c_{\ell}(\tau)}{d \tau} 
+ \frac{1}{\tau}  \left\{ {\frak U}_l\, c_{\ell+1}(\tau) + \left({\frak B}_l-\frac{2}{15}c_1\right) c_{\ell}(\tau) + {\frak C}_l\,c_{\ell-1}(\tau)   \right\}   \right],
\end{split}
\ee
where we used explicitly the conservation law for the temperature~\eqref{eq:ODE_T}. In the previous expression the coefficients  ${\frak U}_l$, ${\frak B}_l$ and ${\frak C}_l$ are respectively
\be
{\frak A}_{\ell} = -\frac{ (2 \ell -1) (2 \ell +1)(2 \ell + 2)}{(4 \ell + 3)(4 \ell + 5)}\,,  \qquad
{\frak B}_l  =\frac{ 2(14 \ell^2 + 7 \ell - 2 ) }{(4 \ell - 1)(4 \ell + 3)}-\frac{4}{3}\,,  \qquad
{\frak C}_{\ell} = \frac{2 \ell (2 \ell - 1)(2 \ell + 2) }{(4 \ell - 3)(4 \ell - 1)}\, .
\ee

The calculation of the momentum integrals in the RHS of Eq.~\eqref{eq:int_FP} simplifies if one replaces the ansatz~\eqref{eq:ansatzf} in the definition of the integrals $\mc J$ and $\mc K$, Eqs.~\eqref{eq:JKint}, i.e., 
\bs
\label{eq:JK_ans}
\beal
{\cal J} (\tau) &= \int \frac{d^3 p}{(2 \pi)^3} \, f(\tau,{\bf p}) \left[1 + f(\tau,{\bf p}) \right] = \frac{T(\tau)^3}{2 \pi^2}  \left[ 2 \zeta(3) + \sum_{n = 0}^{+\infty} \frac{\pi^2 - 6 \zeta(3)}{3(4 n + 1)} c_{n}(\tau)^2 \right], \\
{\cal K} (\tau) &= 2 \int \frac{d^3 p}{(2 \pi)^3} \, \frac{f(\tau,{\bf p})}{p} \, = \, \frac{T(\tau)^2}{6} \,,
\end{align}
\es
where $\zeta(n)$ is the Riemann zeta function and the Landau matching condition for energy density~\eqref{eq:matchene} was explicitly used. 

In order to perform the momentum integrals in the RHS of Eq.~\eqref{eq:int_FP} we change the variable
$p_z=p_\varsigma/\tau$. This results into changing the momentum measure  $\frac{d^2{\bf p}_Tdp_\varsigma}{\tau}\to d^2{\bf p}_Tdp_z$ as well as the comoving energy $p^\tau\to p=\sqrt{{\bf p_T}^2+p_z^2}$. Furthermore, this change of variable allows us to write the spatial components of the momentum as $p^i=p\,(\cos\phi_p\,\sin\theta_p\,,\,\sin\phi_p\,\sin\theta_p\,,\,\cos\theta_p)$. For instance, one of the integrals in the RHS of Eq.~\eqref{eq:int_FP} gives us
\be
\label{eq:momint1}
\besp
\int_{\bf p}\,(-u\cdot p)^2\,P_{2l}(\cos\theta_p)\,\nabla_{\bf p}^2 f(\tau,{\bf p}) &= \int\,\frac{d^3p}{(2\pi)^3}\,p\,P_{2l}(\cos\theta_p)\,\\
&\times\left[ \frac{\pd^2}{\pd p^2} + \frac{2}{p} \frac{\pd}{\pd p} + \frac{1}{p^2}  \frac{\pd}{\pd \cos \theta_p}  \sin^2 \theta_p \frac{\pd}{\pd \cos \theta_p} \right] f(\tau,p\sin\theta_{\bf p},p\cos\theta_{\bf p})\\
&= -\frac{T(\tau)^2}{6(4 \ell + 1)}  \left[ \ell (2\ell + 1) -1\right] c_{\ell}(\tau)\,.
\end{split}
\ee
In the previous expression we make explicit use of the Bjorken constraints over the total distribution function, i.e., $f_{\bf p}=f(\tau,|{\bf p}_T|,p_\varsigma/\tau)\to  f(\tau,p\sin\theta_{\bf p},p\cos\theta_{\bf p})$ after the aforementioned change of variable is carried out. 

The remaining momentum integral in the RHS of Eq.~\eqref{eq:int_FP} reads as
\be
\label{eq:momint2}
\besp 
\int_{\bf p}\,(-u\cdot p)^2\,P_{2l}(\cos\theta_p)\,\nabla_{\bf p} \cdot \frac{{\bf p}}{p} f(\tau,{\bf p}) \left( 1+ f(\tau,{\bf p}) \right) &=-\frac{T(\tau)^3}{2 \pi^2(4 \ell + 1)} \\
&\times \left[ 2 \zeta(3) c_{\ell}(\tau)  +\left( \frac{\pi^2}{3} - 2 \zeta(3) \right) \sum_{m,n=0}^{|m-n|=\ell} \Omega_{\ell m n} c_{m}(\tau) c_{n}(\tau)  \right]
\end{split}
\ee
where
\be
\Omega_{\ell m n} =\frac{\alpha_{m-n+\ell}\,\alpha_{n+m-\ell}\,\alpha_{n-m+\ell}}{\alpha_{n+m+\ell}} \cdot \frac{4 \ell + 1}{2(n+m+\ell)+1}, \qquad \text{with}\,\,\alpha_{\ell}= \frac{(2 \ell - 1)!!}{\ell !}\,.
\ee
By equating Eqs.~\eqref{eq:momint1} and~\eqref{eq:momint2} together with Eqs.~\eqref{eq:JK_ans} into the RHS in Eq.~\eqref{eq:int_FP} we finally get 
\be
\label{eq:RHS_FP}
\besp
\int_{\bf p}\,(-u\cdot p)^2\,P_{2l}(\cos\theta_p)\,\mc C[f]&=-\frac{ T(\tau)\,\epsilon(\tau)}{(4 \ell + 1) \theta_0}  \left( 2 \zeta(3) + \sum_{n = 0}^{+\infty} \frac{\pi^2 - 6 \zeta(3)}{3(4 n + 1)} c_{n}(\tau)^2 \right)  \left( \ell (2\ell + 1) -1\right) c_{\ell}(\tau) \\
& - \frac{T(\tau)\,\epsilon(\tau)}{(4 \ell + 1) \theta_0} 
 \left[  2 \zeta(3) c_{\ell}(\tau)  + \left( \frac{\pi^2}{3} - 2 \zeta(3) \right) \sum_{m,n=0}^{|m-n|=\ell} \Omega_{\ell m n} c_{m}(\tau) c_{n}(\tau)  \right],
\end{split}
\ee
with $\theta_0^{-1}=\frac{5}{8\,\pi^5}\,\lambda_{YM}^2\,l_{Cb}$. Thus, by putting together the LHS~\eqref{eq:LHS_FP} and RHS~\eqref{eq:RHS_FP} of Eq.~\eqref{eq:int_FP} we get the following evolution equations for the Legendre moments $c_l$
\be
\label{eq:ODE_c} 
\besp
\frac{d c_{\ell}(\tau)}{d \tau}   & = -\frac{1}{\tau}  \left[ {\frak A}_{\ell} c_{\ell+1}(\tau) + \left( \frak B_{\ell} - \frac{2}{15} c_{1}(\tau) \right)  c_{\ell}(\tau) + {\frak C}_{\ell} c_{\ell-1}(\tau)   \right]  \nl
& - \frac{ T(\tau)}{\theta_0} \left[  \left\{ \kappa + \frac{\pi^2  \ell (2\ell + 1) }{3}  \right\}  c_{\ell}(\tau)
+  \kappa \sum_{m,n=1}^{|m-n| \le \ell} \Omega_{\ell m n} c_{m}(\tau) c_{n}(\tau)
  + \kappa   
  \sum_{n = 1}^{+\infty} \frac{(2\ell - 1)(\ell + 1) }{3(4 n + 1) } c_{n}(\tau)^2   c_{\ell}(\tau)   \right].
\end{split}
\ee
After some redefinitions of the variables and writing them in a matrix form one gets Eqs.~\eqref{eq:ODE_tau}.

\subsection{Some useful identities of the Legendre polynomials}
\label{subsec:leg}
In the previous section we used explicitly the following identities of the Legendre polynomials
\bs
\beal
& P_{n}(x) = 2^{n} \sum_{k=0}^{n}
\begin{pmatrix}
  n \\
  k
\end{pmatrix}
\begin{pmatrix}
  \frac{n+k-1}{2} \\
  n
\end{pmatrix} x^{k}
= \frac{1}{2^{n}} \sum_{k=0}^{n}
\begin{pmatrix}
  n \\
  k
\end{pmatrix}^2
(x-1)^{n-k}(x+1)^{k},\\
&P_{n}(x) P_{m}(x) 
= \sum_{k=0}^{{\rm Min}[n,m]} \frac{\alpha_{m-k}\alpha_k \alpha_{n-k}}{\alpha_{n+m-k}} \cdot \frac{2(n+m-2k)+1}{2(n+m-k)+1} \cdot P_{n+m-2k}(x). \qquad \text{with}\,\,\alpha_{k}= \frac{(2k-1)!!}{k!} \,,\\
 &\int^{+1}_{-1}  d x \ P_{2 n}(x) P_{2 m}(x) = \frac{2}{4 n+1} \delta_{n,m}, \\
 &\int^{+1}_{-1}  d x \ P_{2 n}(x) P_{2 m}(x)P_{2 \ell}(x)=\frac{\alpha_{m-n+\ell}\alpha_{n+m-\ell}\alpha_{n-m+\ell}}{\alpha_{n+m+\ell}} \cdot \frac{2}{2(n+m+\ell)+1}\,,\\
 &x^2 \, P_{2 \ell}(x) = \frac{2 \ell (2 \ell-1)}{(4 \ell-1)(4 \ell + 1)} P_{2 \ell-2}(x) 
 + \frac{8 \ell^2 + 4 \ell - 1}{(4 \ell-1)(4 \ell + 3)} P_{2 \ell}(x) + \frac{(2 \ell +1)(2 \ell + 2)}{(4 \ell+1)(4 \ell + 3)} P_{2 \ell+2}(x)\,,\\
&  \frac{\pd }{\pd \cos \theta_p} \sin^2 \theta_p \frac{\pd }{\pd \cos \theta_p} P_{2 \ell}(\cos \theta_p) = -2 \ell (2 \ell + 1) P_{2 \ell}(\cos \theta_p).
\end{align}
\es
%

\section{On the inherent exponential error}
\label{app:exp_error}

We conclude this section by estimating the error made when truncating the transseries solutions based on super/hyper asymptotics~\cite{boyd:1999}. In Sect.~\ref{subsec:IR_c1} we observe a good agreement between the numerical results and the IR transseries solutions (see also Refs.~\cite{Behtash:2018moe,Behtash:2019txb,Behtash:2019qtk}). Naively, one might think that a better agreement between the transseries and the numerical results can be obtained when adding more transmonomials and/or higher orders in the transseries. Nevertheless, this is not the case since in our case the radius of convergence of the transseries is finite~\cite{Behtash:2019txb,Behtash:2019qtk}. 

Consider the truncated IR asympotic expansion of the Legendre mode $c_1$, i.e.,
\be
\label{eq:trunc_c1}
c_{1}(w) \sim \sum_{k=1}^{K} u_{1,k}^{(0)} w^{-k} + \mc O(w^{-K-1})\,,\hspace{1cm} 
\text{with}\,\, K \in {\mathbb N}.
\ee
where $K$ is the order of the truncation. The IR perturbative expansion~\eqref{eq:asympc_1} is asymptotically of Gevrais-1 class (see discussion in Sect.~IIID of Ref.~\cite{Behtash:2019txb}). Thus, the asymptotic form of the coefficients $u^{(0)}_{1,k}$ entering in Eq.~\eqref{eq:asympc_1} are
\be
\label{eq:asym_u0_gv1}
|u_{1,k}^{({\bf 0})}| \sim M(\hat{\theta}_0) S^{-k+\beta_1}\Gamma(k-\beta_1) \quad \mbox{ as } \quad k \rightarrow +\infty, 
\ee
where $M(\hat{\theta}_0)=M_0 \hat{\theta}_0^{\, \beta_1}$ ($M_0 \in {\mathbb R}$) is an overall factor that depends on the angle $\hat{\theta}_0$ defined in the Borel plane.  The function $K_{\rm op}(w)$ which optimizes the error between the asymptotic series and the exact solution can be calculated by evaluating the convergence rate as follows
\begin{eqnarray}
| u_{1,k}^{({\bf 0})}| w^{-k} > |u_{1,k+1}^{({\bf 0})}| w^{-k-1}  
&\quad \Rightarrow \quad &  k <  S w + \beta_1 \\
&\quad \Rightarrow \quad &  K_{\rm op}(w) =  \lfloor S w + \beta_1 \rfloor.
\end{eqnarray}
where $\lfloor \bullet \rfloor$ is the Floor function. Thus, the induced error $R_{K}(w)$ in the truncated expansion~\eqref{eq:trunc_c1} is
\begin{eqnarray}
R_{K_{\rm op}}(w) &\sim& M(\hat{\theta}_0) S^{-(K_{\rm op} - \beta_1+1)} \Gamma(K_{op}-\beta_1+1) \, w^{-(K_{\rm op}+1)} \nl
&\approx& M(\hat{\theta}_0)  \Gamma(S w+1) \, (S w)^{-S w} S^{-1} w^{-\beta_1-1}
\qquad (K_{\rm op} \approx S w + \beta_1) \nl
&\sim& \sqrt{2 \pi} M(\hat{\theta}_0)  \, (S w)^{-1/2} w^{-\beta_1} e^{-S w} \qquad \mbox{ as } \ w  \rightarrow + \infty\,,
\end{eqnarray}
where we used the Stirling's formula for the last line. Clearly, this estimate depends on the asymptotic behavior of $|u^{({\bf 0})}_{k}|$~\eqref{eq:asym_u0_gv1}. In that case if $k<K_{\rm op}$ the error is small when $|u^{({\bf 0})}_{k}|$ behaves like in Eq.~\eqref{eq:asym_u0_gv1} and depends explicitly on $w$. Thus, when adding higher orders $k>K_{\rm op}$ to the truncated expansion~\eqref{eq:trunc_c1} the mismatch between transseries and the exact solution increases. The situation gets more worrisome when solving the general dynamical system~\eqref{eq:ODE_tau}. In this general case, the leading order perturbative contribution to $c_{\ell}(\hat{\tau}) = O({\hat{\tau}}^{-\ell})$~\cite{Behtash:2019txb}. As a result, the disagreement between the multi-parameter transseries and the numerical solutions of the Legendre moments $c_{\ell}(\hat{\tau})$ is larger and expected if $\ell > K_{\rm op}$. This was precisely what some of us observed and extensively discussed in Refs.~\cite{Behtash:2019txb,Behtash:2019qtk}.

\section{IR Transseries solutions in the proper time $\tau$ variable}
\label{app:timetrans}

In Sects.~\ref{sec:l1case} and~\ref{sec:generalcase} we studied in detail the resurgent properties of the nonlinear ODEs~\eqref{eq:ODE_w} in terms of the $w$ variable. However, one can also analyze the original dynamical system in terms of the variable $\tau$ as it was done for the RTA Boltzmann equation~\cite{Behtash:2018moe,Behtash:2019txb} for the detail. For completeness we present this analysis in this section. 

The IR fixed points of the Legendre moments are $\bar{c}_{\ell}=0$ $\forall\,\ell>0$, namely $c_{\ell}(\tau) \rightarrow 0$ as $\tau \rightarrow + \infty$. From Eq.~\eqref{eq:consqnTtau}, one obtains the exact formal solution of $T(\tau)$, i.e., 
\be
\label{eq:asymptT_tau}
\besp
T(\tau) &= \sigma_T\,\left(\sigma_T \tau\right)^{-1/3}\, \exp \left[ -\frac{1}{30} \int \frac{d\tau}{\tau} \, c_{1}(\tau) \right] \\
&=: \frac{\sigma_T}{\hat{\tau}^{1/2}} \left(1 + \bar{T}(\hat{\tau}) \right),
\end{split}
\ee
where $\sigma_T \in {\mathbb R}^+$ is the integration constant with dimensions of energy and $\hat{\tau}:= (\sigma_T\tau)^{{2/3}}$~\footnote{
    The integration and the series expansions are commutative with each other \textit{up to the value of integration constant} determined from a given initial condition in general.
In this sense, the value of $\sigma_T$ in a transseries is individually determined for each fixed points from a given initial condition.}. Notice that that $\hat{\tau}$ is dimensionless and guarantees the scale invariance of $c_\ell$ in the $w$-coordinate. In addition, in the second line we expanded asymptotically the solution of $c_1$ while satisfying that $\bar{T}(\tau) \rightarrow 0$ as $\hat{\tau} \rightarrow + \infty$.
By substituting the solution~\eqref{eq:asymptT_tau} into Eq.~\eqref{eq:cldynsys}, one obtains
\be
\label{eq:ODE_c2} 
\besp
\frac{d c_{\ell}(\hat{\tau})}{d \hat{\tau}}  &= F_{\ell}({\bf c},\hat{\tau})\,,\\ 
F_{\ell}({\bf c},\hat{\tau})&=- \frac{3}{2 \hat{\tau}}  \left[ {\frak A}_{\ell} c_{\ell+1}(\hat{\tau}) + \left( \bar{\frak B}_{\ell} - \frac{2}{15} c_{1}(\hat{\tau}) \right)  c_{\ell}(\hat{\tau}) + {\frak C}_{\ell} c_{\ell-1}(\hat{\tau})   \right] \\
& -  \left( 1 + \bar{T}(\hat{\tau})\right) \left[  {\Lambda}(\ell)
  c_{\ell}(\hat{\tau})+  \hat{\kappa} \sum_{m,n=1}^{|m-n| \le \ell} \Omega_{\ell m n} c_{m}(\hat{\tau}) c_{n}(\hat{\tau})
    + \hat{\kappa}   
    \sum_{n = 1}^{+\infty} \frac{(2\ell - 1)( \ell + 1) }{3(4 n + 1) } c_{n}(\hat{\tau})^2   c_{\ell}(\hat{\tau})   \right] \,,
\end{split}
\ee
It is straightforward to find that $c_{\ell}(\hat{\tau}) = O(\hat{\tau}^{-\ell})$ by considering the asymptotic behavior of the solutions of Eqs.~\eqref{eq:ODE_tau}. For instance, one obtains that $c_1$  and $\bar{T}$ behave asymptotically as follows
\be
 c_{1}(\hat{\tau}) \sim -\frac{3 {\frak C}_{1}}{2 {\Lambda}(1) \hat{\tau}} = - \frac{4}{\Lambda(1) \hat{\tau}}, \\
\hspace{1cm} \bar{T}(\hat{\tau}) \sim - \frac{1}{20} \int \frac{d \hat{\tau}}{\hat{\tau}} c_{1}(\hat{\tau}) \sim - \frac{1}{5 \Lambda(1) \hat{\tau}}.
\ee
As we proceeded in Sects.~\ref{sec:l1case} and~\ref{sec:generalcase} we rewrite the nonlinear ODEs~\eqref{eq:ODE_c2} in a matrix form, i.e., 
\be
\label{eq:ODE_taur}
\frac{d{\bf c}}{d\hat{\tau}}={\bf F}({\bf c},\hat{\tau}) 
\ee
where 
\bs
\label{eq:matrODE_not}
\beal
{\bf c}(\hat{\tau}) &= (c_{1}(\hat{\tau}),\cdots, c_{L}(\hat{\tau}))^{\top}, \\
\label{eq:matrF_tau}
{\bf F}({\bf c},\hat{\tau}) &=  - \frac{1}{\hat{\tau}} \left[ {\frak X}({\bf c}) {\bf c}(\hat{\tau}) + {\bf \Gamma} \right] -  \left( 1 + \bar{T}(\hat{\tau})\right) \left[  \Lambda + {\frak Y}({\bf c}) + {\frak Z}({\bf c})    \right] {\bf c}(\hat{\tau}).
\end{align}
\es
The matrices entering into the previous expressions are defined in Eqs.~\eqref{eq:chimat} and~\eqref{eq:matODE}. In order to build the transmonomials we linearized the dynamical system~\eqref{eq:ODE_tau} around the IR fixed point, 
\be 
\label{eq:lindc_W}
\besp
\frac{d \delta {\bf c}(\hat{\tau})}{d \hat{\tau}}  &= \sum_{\ell=1}^{L} \frac{\pd {\bf F}({\bf c},\hat{\tau})}{\pd c_{\ell}}  \delta c_{\ell} (\hat{\tau}) \\
&= - \left[ \Lambda + \frac{1}{\hat{\tau}} \left( \bar{\frak X} - \frac{4\bar{\frak Y}}{\Lambda(1)}   - \frac{\Lambda}{5 \Lambda(1)} \right) \right]\delta {\bf c}(\hat{\tau}) - \frac{1}{\hat{\tau}} \delta {\frak Y}
\begin{pmatrix}
  c_{1}(\hat{\tau}) \\
  0 \\
  \vdots \\
  0
\end{pmatrix}
+ O({\bf c} \delta {\bf c}/\hat{\tau}, \delta {\bf c}/\hat{\tau}^2) \\
&= - \left[ \Lambda + \frac{1}{\hat{\tau}} \bar{\frak W}  \right] \delta {\bf c}(\hat{\tau}) + O({\bf c} \delta {\bf c}/\hat{\tau}, \delta {\bf c}/\hat{\tau}^2), \\
\end{split}
\ee
where 
\bs
\beal
\bar{\frak W} &:=  
\bar{\frak X}  - \frac{4}{\Lambda(1)} \left( 2 \bar{\frak Y} + \frac{\Lambda}{20} \right), \\
\delta {\frak Y}(\delta {\bf c}) = \hat{\kappa}
\begin{pmatrix}
  \sum_{m=1}^{2} \Omega_{1m1} \delta c_{m}(\hat{\tau}) &  0  &  \cdots & 0 \\
  \vdots   & \vdots  &  & \vdots  \\
  \sum_{m=\ell-1}^{\ell + 1} \Omega_{\ell m1} \delta c_{m}(\hat{\tau}) &  0  &  \cdots & 0 \\
  \vdots   & \vdots  &  & \vdots  \\
\sum_{m=L-1}^{L+1} \Omega_{Lm1} \delta c_{m}(\hat{\tau})    &  0  &  \cdots & 0 \\
\end{pmatrix}, &\qquad
\bar{\frak Y} = 
 \hat{\kappa} 
\begin{pmatrix}
  \Omega_{111}   & \Omega_{112} &  &  &    \\
  \Omega_{211} & \Omega _{212} & \Omega_{213} & & \\
     & \ddots & \ddots & \ddots & \\
  & &  \Omega_{L-11L-2} &  \Omega_{L-11L-1} & \Omega_{L-11L}  \\
  & &  & \Omega_{L1L-1} & \Omega_{L1L}  \\
\end{pmatrix} .
\end{align}
\es
In order to solve find the behavior of the fluctuations close to the IR fixed points it is needed to solve the eigenproblem for the matrix $\bar{\frak W}$. Following Costin's prescription~\cite{costin1998} we introduce the matrix $V$ as follows
\be
\delta \tilde{\bf c} := \tilde{V}(\hat{\tau}) \delta {\bf c}, \qquad \tilde{V}(\hat{\tau}) := {\bf 1}_{L} + V \hat{\tau}^{-1}.
\ee
Thus, the linearized equation Eq.~\eqref{eq:lindc_W} is modified as
\be
\besp
\frac{d \delta \tilde{\bf c}}{d \hat{\tau}} 
&= - \tilde{V}(\hat{\tau}) \left[ \Lambda + \frac{1}{\hat{\tau}} \bar{\frak W}  \right] \tilde{V}(\hat{\tau})^{-1} \delta \tilde{\bf c}(\hat{\tau})  -V \tilde{V}(\hat{\tau})^{-1} \hat{\tau}^{-2} \delta \tilde{\bf c} \nl
&= -  \left[ \Lambda + \frac{1}{\hat{\tau}} \left( \bar{\frak W} + [V,\Lambda]  \right)   \right]  \delta \tilde{\bf c}(\hat{\tau})  + O( \delta \tilde{\bf c}/\hat{\tau}^2), \label{eq:lindc2} 
\end{split}
\ee
where the commutator between two matrices is denoted as $[A,B]:=AB-BA$. One can take ${\rm diag.}(V)=(0,\cdots,0)$, and the other components can be chosen such that 
\be
\bar{\frak W} + [V,\Lambda] \ : \mapsto  \ \hat{\frak W} := {\rm diag.}({\frak w}_1,\cdots,{\frak w}_L).
\ee
Therefore, the solution of Eq.(\ref{eq:lindc2}) is given by
\be
\label{eq:sol_lin}
\delta \tilde{c}_{\ell}(\hat{\tau}) = \sigma_{\ell} \frac{e^{- \Lambda(\ell) \hat{\tau}}}{\hat{\tau}^{{\frak w}_{\ell}}} \quad \Rightarrow \quad \delta {c}_{\ell}(\hat{\tau}) = \sigma_{\ell} \frac{e^{- \Lambda(\ell) \hat{\tau}}}{\hat{\tau}^{{\frak w}_{\ell}}} + \mc O (e^{- \Lambda(\ell) \hat{\tau}}/\hat{\tau}^{{\frak w}_{\ell}+1}). 
\ee
From the solution of the linearized fluctuations, the transseries takes the following form
\be
\label{eq:trans}
\besp
c_{\ell}(\tau) &= \sum_{|{\bf n}|=0}^{+\infty} \sum_{k=0}^{+\infty} u^{(\bf n)}_{\ell,k} \Phi^{\bf n}_{k}, \\
\Phi^{\bf n}_{k}&:= \left( \prod_{j=1}^{L} \zeta_j^{n_{j}} \right) \hat{\tau}^{-k}, \qquad  \zeta_{j}:= \sigma_j \frac{e^{-S_j \hat{\tau}}}{\hat{\tau}^{\beta_{j}}}.
\end{split}
\ee
where ${\bf n}\in {\mathbb N}^L_{0}$, $\sigma_{i}\in {\mathbb C}$ is the integration constant,  and $L \in {\mathbb N}$ is the truncation order of $\ell$, namely  $1 \le \ell \le L$.

\section{On the existence of the Lyapunov functional}
\label{sec:Lyapfunc}

In this section we present a proof of the existence of a Lyapunov function for the dynamical system of ODEs~\eqref{eq:prepODE_c_w}. Roughly speaking, Lyapunov functions are positive definite functions which are monotonically decreasing along the trajectories of the flows in phase space. These functions are used to determine the stability properties of ODEs and PDEs. In the RG flow approach to dynamical systems the Lyapunov function plays an analogous role of the $c$ function in QFT and/or the dynamical effective potential in non-newtonian mechanics. For a non-autonomous dynamical system, the existence of the Lyapunov function allows us to identify the dynamical system as a RG flow from the global point of view. 

In our approach the existence of the Lyapunov function is inferred directly from the dynamical system of ODEs~\eqref{eq:prepODE_c_w}. First, we promote this non-autonomous dynamical system to an autonomous one of one dimension higher by introducing an ODE for $w$ in terms of a new flow time $\rho$ as follows
\bs 
\label{eq:RGaut}
\beal
\frac{d{\bf c}(\rho)}{d\log\rho}&=\beta({\bf c}(\rho),w(\rho))\equiv -\frac{\partial\mc V}{\partial c_i}\,,\\
\frac{dw(\rho)}{d\log\rho}&=\beta_w(w(\rho))\equiv-\frac{\partial\mc V}{\partial w}\,,
\end{align}
\es
where in the RHS of the previous expressions we introduce a positive definite differentiable function $\mc V$. It is straightforward to show that the function $\mc V$ decreases monotonically, i.e., 
\be
\label{eq:Vcond}
\besp
\frac{d\mc V({\bf c}(\rho),w(\rho))}{d\log\rho}&=\sum_{i=1}^L\,\frac{dc_i}{d\log\rho}.\frac{\partial\mc V({\bf c}(\rho),w(\rho))}{\partial\,c_i(\rho)}\,+\,\frac{dw(\rho)}{d\log\rho}.\frac{\partial\mc V({\bf c}(\rho),w(\rho))}{\partial\,w(\rho)}\,,\\
&=-\left[\sum_{i=1}^L\,|\beta_i({\bf c}(\rho),w(\rho))|^2\,+\,|\beta_w({\bf c}(\rho),w(\rho))|^2\right]\,\leq\, 0\,.
\end{split}
\ee
Therefore, $\mc V$ satisfies the properties required to be a candidate for the global Lyapunov function of the dynamical system~\eqref{eq:prepODE_c_w} (see Eq. B9 in Ref.~\cite{Behtash:2019txb}). Although we have proven its existence for the studied case here, it is in general extremely difficult to calculate the exact Lyapunov function (cf. ~\cite{Behtash:2017wqg}) and this is beyond the scope of this work.

\section{Existence of the forward attractor and higher nonlinearities of the collisional kernel}
\label{sec:invman}

In this section we outline the generic conditions that ensure the existence of a non-equilibrium forward attractor for systems undergoing Bjorken expansion described by relativistic kinetic theory within different approximations for the collisional kernel. Right now we have strong numerical evidence of this type of attractor for systems undergoing Bjorken expansion described microscopically within the relativistic kinetic theory framework and for different types of collisional kernels~\cite{Kurkela:2011ub,Kurkela:2015qoa,Kurkela:2018wud,Kurkela:2018vqr,Giacalone:2019ldn,Almaalol:2020rnu,Almaalol:2018jmz,Almaalol:2018ynx,Strickland:2019hff,Strickland:2018ayk,Berges:2013fga,Berges:2013eia,Blaizot:2019scw,Blaizot:2017ucy,Blaizot:2017lht,Berges:2020fwq,Mazeliauskas:2018yef,Tanji:2017suk,Blaizot:2019dut,Jaiswal:2019cju}. 

Let us assume that the distribution function can be expanded around the thermal equilibrium while the deviations from this state is written in terms of a set of Legendre moments like  in Eq.~\eqref{eq:ansatzf}. The conservation law is the same for any physical system regardless of the collisional kernel~\eqref{eq:consqnTtau}. On the other hand, the dynamical system of ODEs for the Legendre moments must satisfy certain symmetries given the constraints of the Bjorken flow. In this case, the dilation $\tau \rightarrow \gamma \tau$ (being $\gamma$ a constant) together with the conformal dimensionality of the Legendre moments and the temperature are $(\Delta_{\bf c},\Delta_T)=(0,1)$ restricts strongly the form of the differential equation. In addition, and given the results outlined in Fig.~\ref{fig:univatt}, one can assume that the small $\tau$ limit is dominated by singular terms which go like $\mc O(\tau^{-1})$ such that  the collision of kernel is $\mc O(\tau^0)$. Thus, the generic equation which satisfies these constraints is written generically as~\footnote{If one can considers a more general ansatz which encode information about the high energy tails like the one taken in Ref.~\cite{Blaizot:2019dut}, one gets a set of nonlinear ODEs which mathematically resembles our general ODE~\eqref{eq:generic}.}
\bs
\label{eq:generic}
\beal
& \frac{d {\bf c}({\tau})}{d {\tau}} = {\bf F}({\bf c},{\tau}), \\
&{\bf F}({\bf c},{\tau}) =  - \frac{1}{{\tau}} \left[ {\frak X}({\bf c}) {\bf c}({\tau}) + \Gamma \right] - {\bf G}({\bf c},T,\tau),
\end{align}
\es
where the terms entering into the matrix operator ${\frak X}({\bf c}) {\bf c}({\tau}) + \Gamma$ are given by Eqs.~\eqref{eq:chimat} and~\eqref{eq:gamvec} respectively, while ${\bf G}({\bf c})$ is a generic function determined uniquely by the collisional kernel. This function depends on the moments, proper time and temperature and its generic form must respect the symmetry restrictions mentioned above. Thus, the following general form of ${\bf G}({\bf c})$ is
\be
\label{eq:gencoll_moments}
   {\bf G}({\bf c}) = \sum_{h=0}^{H}  \sum_{n=1}^{N} \sum_{\ell_1 \ge \cdots \ge \ell_n =1}^{L} {\bf g}^{(\ell_1,\cdots,\ell_n)}_{h,n}   T^{h+1} \tau^h \prod_{n^\prime=1}^{n} c_{\ell_{n^\prime}},
\ee
where $H \in {\mathbb N}_0$ and $N \in {\mathbb N}$.
In addition, $g^{(\ell_1,\cdots,\ell_n)}_{\ell,h,n} \in {\mathbb R}$ is a dimensionless coefficient. \footnote{${\bf g}^{(\ell_1)}_{H,1}$ is the linearized version of the collisional kernel in the eigenmodes and it corresponds to a $L$-by-$L$ matrix of the form
\be
   {\bf g}^{(\ell_1)}_{H,1} =
   \begin{pmatrix}
     g^{(1)}_{1,H,1} & \cdots & g^{(L)}_{1,H,1} \\
     \vdots & \ddots & \vdots \\
     g^{(1)}_{L,H,1} & \cdots & g^{(L)}_{L,H,1} 
   \end{pmatrix}.
\ee
} The dynamical system of ODEs for the RTA Boltzmann~\eqref{eq:ODE_cl_tau_RTA} and FPE~\eqref{eq:cldynsys} cases are a particular case of the previous expression.

Our transasymptotic analysis discussed in this work and in Refs.~\cite{Behtash:2017wqg,Behtash:2018moe,Behtash:2019txb,Behtash:2019qtk} shows that the existence of the forward attractor is ensure if the following conditions are satisfied
\begin{enumerate}
\item If the fixed point equation for the long time limit given by
\be
\sum_{n=1}^{N} \sum_{\ell_1 \ge \cdots \ge \ell_n =1}^{L} {\bf g}^{(\ell_1,\cdots,\ell_n)}_{H,n} \prod_{n^\prime=1}^{n} \bar{c}_{\ell_{n^\prime}}=0
\ee
has a trivial solution, i.e. $\bar{\bf c}={\bf 0}$, then, the IR fixed point corresponds to the local thermal equilibrium state.
\item
  If the eigenvalues of the matrix ${\bf g}^{(\ell_1)}_{H,1}$, around $\bar{\bf c}={\bf 0}$ are all positive, then, the fixed point gives a (local) forward attractor.
\item  In case that there is a subset of either exactly vanishing or positive eigenvalues of the matrix ${\bf g}^{(\ell_1)}_{H,1}$, then there exists a sub- (local) forward attractor $\frak M$ in the phase space of dynamical variables $({\bf c},T,\tau)$-space.  In this case, ${\rm dim}[{\frak M}]=L+2-p$  where $p$ is the number of zeros or negative eigenvalues. In the the $({\bf c},w)$-space, ${\rm dim}[{\frak A}]=L+1-p$.
\item In addition to the previous conditions, if $H=1$, then, the transmonomials in the transseries are of the form~\eqref{eq:genIRtransm} like in the RTA and FPE cases respectively.
\end{enumerate}
On the other hand, the pullback attractor will be determined by the term $\mc O(\tau^{-1})$ in Eq.~\eqref{eq:generic}. Notice however that in the most general case it is not necessarily ensured that there is only one invariant flow (generally dubbed as 'attractor solution') that connects the UV and IR. A more careful analysis needs to be made by considering techniques based on Morse theory, cf.~\cite{Behtash:2019qtk}, and center manifolds~\cite{caraballo2017applied,kloeden} if there are vanishing eigenvalues for the collisional kernel. 

\end{widetext}

\bibliography{fokkerplanck}

\begin{thebibliography}{117}
\expandafter\ifx\csname natexlab\endcsname\relax\def\natexlab#1{#1}\fi
\expandafter\ifx\csname bibnamefont\endcsname\relax
  \def\bibnamefont#1{#1}\fi
\expandafter\ifx\csname bibfnamefont\endcsname\relax
  \def\bibfnamefont#1{#1}\fi
\expandafter\ifx\csname citenamefont\endcsname\relax
  \def\citenamefont#1{#1}\fi
\expandafter\ifx\csname url\endcsname\relax
  \def\url#1{\texttt{#1}}\fi
\expandafter\ifx\csname urlprefix\endcsname\relax\def\urlprefix{URL }\fi
\providecommand{\bibinfo}[2]{#2}
\providecommand{\eprint}[2][]{\url{#2}}

\bibitem[{\citenamefont{Chatrchyan et~al.}(2013)}]{Chatrchyan:2013nka}
\bibinfo{author}{\bibfnamefont{S.}~\bibnamefont{Chatrchyan}}
  \bibnamefont{et~al.} (\bibinfo{collaboration}{CMS}), \bibinfo{journal}{Phys.
  Lett. B} \textbf{\bibinfo{volume}{724}}, \bibinfo{pages}{213}
  (\bibinfo{year}{2013}), \eprint{1305.0609}.

\bibitem[{\citenamefont{Adare et~al.}(2013)}]{Adare:2013piz}
\bibinfo{author}{\bibfnamefont{A.}~\bibnamefont{Adare}} \bibnamefont{et~al.}
  (\bibinfo{collaboration}{PHENIX}), \bibinfo{journal}{Phys. Rev. Lett.}
  \textbf{\bibinfo{volume}{111}}, \bibinfo{pages}{212301}
  (\bibinfo{year}{2013}), \eprint{1303.1794}.

\bibitem[{\citenamefont{Aad et~al.}(2013)}]{Aad:2012gla}
\bibinfo{author}{\bibfnamefont{G.}~\bibnamefont{Aad}} \bibnamefont{et~al.}
  (\bibinfo{collaboration}{ATLAS}), \bibinfo{journal}{Phys. Rev. Lett.}
  \textbf{\bibinfo{volume}{110}}, \bibinfo{pages}{182302}
  (\bibinfo{year}{2013}), \eprint{1212.5198}.

\bibitem[{\citenamefont{Abelev et~al.}(2013)}]{Abelev:2012ola}
\bibinfo{author}{\bibfnamefont{B.}~\bibnamefont{Abelev}} \bibnamefont{et~al.}
  (\bibinfo{collaboration}{ALICE}), \bibinfo{journal}{Phys. Lett. B}
  \textbf{\bibinfo{volume}{719}}, \bibinfo{pages}{29} (\bibinfo{year}{2013}),
  \eprint{1212.2001}.

\bibitem[{\citenamefont{Aidala et~al.}(2019)}]{PHENIX:2018lia}
\bibinfo{author}{\bibfnamefont{C.}~\bibnamefont{Aidala}} \bibnamefont{et~al.}
  (\bibinfo{collaboration}{PHENIX}), \bibinfo{journal}{Nature Phys.}
  \textbf{\bibinfo{volume}{15}}, \bibinfo{pages}{214} (\bibinfo{year}{2019}),
  \eprint{1805.02973}.

\bibitem[{\citenamefont{Gale et~al.}(2013)\citenamefont{Gale, Jeon, and
  Schenke}}]{Gale:2013da}
\bibinfo{author}{\bibfnamefont{C.}~\bibnamefont{Gale}},
  \bibinfo{author}{\bibfnamefont{S.}~\bibnamefont{Jeon}}, \bibnamefont{and}
  \bibinfo{author}{\bibfnamefont{B.}~\bibnamefont{Schenke}},
  \bibinfo{journal}{Int. J. Mod. Phys. A} \textbf{\bibinfo{volume}{28}},
  \bibinfo{pages}{1340011} (\bibinfo{year}{2013}), \eprint{1301.5893}.

\bibitem[{\citenamefont{Heinz and Snellings}(2013)}]{Heinz:2013th}
\bibinfo{author}{\bibfnamefont{U.}~\bibnamefont{Heinz}} \bibnamefont{and}
  \bibinfo{author}{\bibfnamefont{R.}~\bibnamefont{Snellings}},
  \bibinfo{journal}{Ann. Rev. Nucl. Part. Sci.} \textbf{\bibinfo{volume}{63}},
  \bibinfo{pages}{123} (\bibinfo{year}{2013}), \eprint{1301.2826}.

\bibitem[{\citenamefont{Bluhm et~al.}(2017)\citenamefont{Bluhm, Hou, and
  Sch\"afer}}]{Bluhm:2017rnf}
\bibinfo{author}{\bibfnamefont{M.}~\bibnamefont{Bluhm}},
  \bibinfo{author}{\bibfnamefont{J.}~\bibnamefont{Hou}}, \bibnamefont{and}
  \bibinfo{author}{\bibfnamefont{T.}~\bibnamefont{Sch\"afer}},
  \bibinfo{journal}{Phys. Rev. Lett.} \textbf{\bibinfo{volume}{119}},
  \bibinfo{pages}{065302} (\bibinfo{year}{2017}), \eprint{1704.03720}.

\bibitem[{\citenamefont{Bluhm and Schaefer}(2016)}]{Bluhm:2015bzi}
\bibinfo{author}{\bibfnamefont{M.}~\bibnamefont{Bluhm}} \bibnamefont{and}
  \bibinfo{author}{\bibfnamefont{T.}~\bibnamefont{Schaefer}},
  \bibinfo{journal}{Phys. Rev. Lett.} \textbf{\bibinfo{volume}{116}},
  \bibinfo{pages}{115301} (\bibinfo{year}{2016}), \eprint{1512.00862}.

\bibitem[{\citenamefont{Bluhm and Sch\"afer}(2015)}]{Bluhm:2015raa}
\bibinfo{author}{\bibfnamefont{M.}~\bibnamefont{Bluhm}} \bibnamefont{and}
  \bibinfo{author}{\bibfnamefont{T.}~\bibnamefont{Sch\"afer}},
  \bibinfo{journal}{Phys. Rev. A} \textbf{\bibinfo{volume}{92}},
  \bibinfo{pages}{043602} (\bibinfo{year}{2015}), \eprint{1505.00846}.

\bibitem[{\citenamefont{Sch\"afer}(2016)}]{Schaefer:2016yzd}
\bibinfo{author}{\bibfnamefont{T.}~\bibnamefont{Sch\"afer}},
  \bibinfo{journal}{Phys. Rev. A} \textbf{\bibinfo{volume}{94}},
  \bibinfo{pages}{043644} (\bibinfo{year}{2016}), \eprint{1608.05083}.

\bibitem[{\citenamefont{Berges et~al.}(2020)\citenamefont{Berges, Heller,
  Mazeliauskas, and Venugopalan}}]{Berges:2020fwq}
\bibinfo{author}{\bibfnamefont{J.}~\bibnamefont{Berges}},
  \bibinfo{author}{\bibfnamefont{M.~P.} \bibnamefont{Heller}},
  \bibinfo{author}{\bibfnamefont{A.}~\bibnamefont{Mazeliauskas}},
  \bibnamefont{and}
  \bibinfo{author}{\bibfnamefont{R.}~\bibnamefont{Venugopalan}}
  (\bibinfo{year}{2020}), \eprint{2005.12299}.

\bibitem[{\citenamefont{Florkowski et~al.}(2018)\citenamefont{Florkowski,
  Heller, and Spalinski}}]{Florkowski:2017olj}
\bibinfo{author}{\bibfnamefont{W.}~\bibnamefont{Florkowski}},
  \bibinfo{author}{\bibfnamefont{M.~P.} \bibnamefont{Heller}},
  \bibnamefont{and}
  \bibinfo{author}{\bibfnamefont{M.}~\bibnamefont{Spalinski}},
  \bibinfo{journal}{Rept. Prog. Phys.} \textbf{\bibinfo{volume}{81}},
  \bibinfo{pages}{046001} (\bibinfo{year}{2018}), \eprint{1707.02282}.

\bibitem[{\citenamefont{Romatschke and Romatschke}(2019)}]{Romatschke:2017ejr}
\bibinfo{author}{\bibfnamefont{P.}~\bibnamefont{Romatschke}} \bibnamefont{and}
  \bibinfo{author}{\bibfnamefont{U.}~\bibnamefont{Romatschke}},
  \emph{\bibinfo{title}{{Relativistic Fluid Dynamics In and Out of
  Equilibrium}}}, Cambridge Monographs on Mathematical Physics
  (\bibinfo{publisher}{Cambridge University Press}, \bibinfo{year}{2019}), ISBN
  \bibinfo{isbn}{978-1-108-48368-1, 978-1-108-75002-8}, \eprint{1712.05815}.

\bibitem[{\citenamefont{Heller and Spalinski}(2015)}]{Heller:2015dha}
\bibinfo{author}{\bibfnamefont{M.~P.} \bibnamefont{Heller}} \bibnamefont{and}
  \bibinfo{author}{\bibfnamefont{M.}~\bibnamefont{Spalinski}},
  \bibinfo{journal}{Phys. Rev. Lett.} \textbf{\bibinfo{volume}{115}},
  \bibinfo{pages}{072501} (\bibinfo{year}{2015}), \eprint{1503.07514}.

\bibitem[{\citenamefont{Israel and Stewart}(1979)}]{Israel:1979wp}
\bibinfo{author}{\bibfnamefont{W.}~\bibnamefont{Israel}} \bibnamefont{and}
  \bibinfo{author}{\bibfnamefont{J.~M.} \bibnamefont{Stewart}},
  \bibinfo{journal}{Ann. Phys.} \textbf{\bibinfo{volume}{118}},
  \bibinfo{pages}{341} (\bibinfo{year}{1979}).

\bibitem[{\citenamefont{Aniceto et~al.}(2019)\citenamefont{Aniceto, Meiring,
  Jankowski, and Spali\'nski}}]{Aniceto:2018uik}
\bibinfo{author}{\bibfnamefont{I.}~\bibnamefont{Aniceto}},
  \bibinfo{author}{\bibfnamefont{B.}~\bibnamefont{Meiring}},
  \bibinfo{author}{\bibfnamefont{J.}~\bibnamefont{Jankowski}},
  \bibnamefont{and}
  \bibinfo{author}{\bibfnamefont{M.}~\bibnamefont{Spali\'nski}},
  \bibinfo{journal}{JHEP} \textbf{\bibinfo{volume}{02}}, \bibinfo{pages}{073}
  (\bibinfo{year}{2019}), \eprint{1810.07130}.

\bibitem[{\citenamefont{Spali\'nski}(2018{\natexlab{a}})}]{Spalinski:2018mqg}
\bibinfo{author}{\bibfnamefont{M.}~\bibnamefont{Spali\'nski}},
  \bibinfo{journal}{Phys. Lett. B} \textbf{\bibinfo{volume}{784}},
  \bibinfo{pages}{21} (\bibinfo{year}{2018}{\natexlab{a}}),
  \eprint{1805.11689}.

\bibitem[{\citenamefont{Spali\'nski}(2018{\natexlab{b}})}]{Spalinski:2017mel}
\bibinfo{author}{\bibfnamefont{M.}~\bibnamefont{Spali\'nski}},
  \bibinfo{journal}{Phys. Lett. B} \textbf{\bibinfo{volume}{776}},
  \bibinfo{pages}{468} (\bibinfo{year}{2018}{\natexlab{b}}),
  \eprint{1708.01921}.

\bibitem[{\citenamefont{Buchel et~al.}(2016)\citenamefont{Buchel, Heller, and
  Noronha}}]{Buchel:2016cbj}
\bibinfo{author}{\bibfnamefont{A.}~\bibnamefont{Buchel}},
  \bibinfo{author}{\bibfnamefont{M.~P.} \bibnamefont{Heller}},
  \bibnamefont{and} \bibinfo{author}{\bibfnamefont{J.}~\bibnamefont{Noronha}},
  \bibinfo{journal}{Phys. Rev. D} \textbf{\bibinfo{volume}{94}},
  \bibinfo{pages}{106011} (\bibinfo{year}{2016}), \eprint{1603.05344}.

\bibitem[{\citenamefont{Heller et~al.}(2018)\citenamefont{Heller, Kurkela,
  Spali\'nski, and Svensson}}]{Heller:2016rtz}
\bibinfo{author}{\bibfnamefont{M.~P.} \bibnamefont{Heller}},
  \bibinfo{author}{\bibfnamefont{A.}~\bibnamefont{Kurkela}},
  \bibinfo{author}{\bibfnamefont{M.}~\bibnamefont{Spali\'nski}},
  \bibnamefont{and} \bibinfo{author}{\bibfnamefont{V.}~\bibnamefont{Svensson}},
  \bibinfo{journal}{Phys. Rev. D} \textbf{\bibinfo{volume}{97}},
  \bibinfo{pages}{091503} (\bibinfo{year}{2018}), \eprint{1609.04803}.

\bibitem[{\citenamefont{Aniceto and Spali\'nski}(2016)}]{Aniceto:2015mto}
\bibinfo{author}{\bibfnamefont{I.}~\bibnamefont{Aniceto}} \bibnamefont{and}
  \bibinfo{author}{\bibfnamefont{M.}~\bibnamefont{Spali\'nski}},
  \bibinfo{journal}{Phys. Rev. D} \textbf{\bibinfo{volume}{93}},
  \bibinfo{pages}{085008} (\bibinfo{year}{2016}), \eprint{1511.06358}.

\bibitem[{\citenamefont{Basar and Dunne}(2015)}]{Basar:2015ava}
\bibinfo{author}{\bibfnamefont{G.}~\bibnamefont{Basar}} \bibnamefont{and}
  \bibinfo{author}{\bibfnamefont{G.~V.} \bibnamefont{Dunne}},
  \bibinfo{journal}{Phys. Rev. D} \textbf{\bibinfo{volume}{92}},
  \bibinfo{pages}{125011} (\bibinfo{year}{2015}), \eprint{1509.05046}.

\bibitem[{\citenamefont{Casalderrey-Solana
  et~al.}(2018)\citenamefont{Casalderrey-Solana, Gushterov, and
  Meiring}}]{Casalderrey-Solana:2017zyh}
\bibinfo{author}{\bibfnamefont{J.}~\bibnamefont{Casalderrey-Solana}},
  \bibinfo{author}{\bibfnamefont{N.~I.} \bibnamefont{Gushterov}},
  \bibnamefont{and} \bibinfo{author}{\bibfnamefont{B.}~\bibnamefont{Meiring}},
  \bibinfo{journal}{JHEP} \textbf{\bibinfo{volume}{04}}, \bibinfo{pages}{042}
  (\bibinfo{year}{2018}), \eprint{1712.02772}.

\bibitem[{\citenamefont{Romatschke}(2018{\natexlab{a}})}]{Romatschke:2017vte}
\bibinfo{author}{\bibfnamefont{P.}~\bibnamefont{Romatschke}},
  \bibinfo{journal}{Phys. Rev. Lett.} \textbf{\bibinfo{volume}{120}},
  \bibinfo{pages}{012301} (\bibinfo{year}{2018}{\natexlab{a}}),
  \eprint{1704.08699}.

\bibitem[{\citenamefont{Blaizot and Yan}(2020{\natexlab{a}})}]{Blaizot:2020gql}
\bibinfo{author}{\bibfnamefont{J.-P.} \bibnamefont{Blaizot}} \bibnamefont{and}
  \bibinfo{author}{\bibfnamefont{L.}~\bibnamefont{Yan}}
  (\bibinfo{year}{2020}{\natexlab{a}}), \eprint{2006.08815}.

\bibitem[{\citenamefont{Romatschke}(2017)}]{Romatschke:2016hle}
\bibinfo{author}{\bibfnamefont{P.}~\bibnamefont{Romatschke}},
  \bibinfo{journal}{Eur. Phys. J. C} \textbf{\bibinfo{volume}{77}},
  \bibinfo{pages}{21} (\bibinfo{year}{2017}), \eprint{1609.02820}.

\bibitem[{\citenamefont{Weller and Romatschke}(2017)}]{Weller:2017tsr}
\bibinfo{author}{\bibfnamefont{R.~D.} \bibnamefont{Weller}} \bibnamefont{and}
  \bibinfo{author}{\bibfnamefont{P.}~\bibnamefont{Romatschke}},
  \bibinfo{journal}{Phys. Lett. B} \textbf{\bibinfo{volume}{774}},
  \bibinfo{pages}{351} (\bibinfo{year}{2017}), \eprint{1701.07145}.

\bibitem[{\citenamefont{Behtash et~al.}(2018)\citenamefont{Behtash,
  Cruz-Camacho, and Martinez}}]{Behtash:2017wqg}
\bibinfo{author}{\bibfnamefont{A.}~\bibnamefont{Behtash}},
  \bibinfo{author}{\bibfnamefont{C.}~\bibnamefont{Cruz-Camacho}},
  \bibnamefont{and} \bibinfo{author}{\bibfnamefont{M.}~\bibnamefont{Martinez}},
  \bibinfo{journal}{Phys. Rev. D} \textbf{\bibinfo{volume}{97}},
  \bibinfo{pages}{044041} (\bibinfo{year}{2018}), \eprint{1711.01745}.

\bibitem[{\citenamefont{Behtash
  et~al.}(2019{\natexlab{a}})\citenamefont{Behtash, Cruz-Camacho, Kamata, and
  Martinez}}]{Behtash:2018moe}
\bibinfo{author}{\bibfnamefont{A.}~\bibnamefont{Behtash}},
  \bibinfo{author}{\bibfnamefont{C.}~\bibnamefont{Cruz-Camacho}},
  \bibinfo{author}{\bibfnamefont{S.}~\bibnamefont{Kamata}}, \bibnamefont{and}
  \bibinfo{author}{\bibfnamefont{M.}~\bibnamefont{Martinez}},
  \bibinfo{journal}{Phys. Lett. B} \textbf{\bibinfo{volume}{797}},
  \bibinfo{pages}{134914} (\bibinfo{year}{2019}{\natexlab{a}}),
  \eprint{1805.07881}.

\bibitem[{\citenamefont{Behtash
  et~al.}(2019{\natexlab{b}})\citenamefont{Behtash, Kamata, Martinez, and
  Shi}}]{Behtash:2019txb}
\bibinfo{author}{\bibfnamefont{A.}~\bibnamefont{Behtash}},
  \bibinfo{author}{\bibfnamefont{S.}~\bibnamefont{Kamata}},
  \bibinfo{author}{\bibfnamefont{M.}~\bibnamefont{Martinez}}, \bibnamefont{and}
  \bibinfo{author}{\bibfnamefont{H.}~\bibnamefont{Shi}},
  \bibinfo{journal}{Phys. Rev. D} \textbf{\bibinfo{volume}{99}},
  \bibinfo{pages}{116012} (\bibinfo{year}{2019}{\natexlab{b}}),
  \eprint{1901.08632}.

\bibitem[{\citenamefont{Behtash et~al.}(2020)\citenamefont{Behtash, Kamata,
  Martinez, and Shi}}]{Behtash:2019qtk}
\bibinfo{author}{\bibfnamefont{A.}~\bibnamefont{Behtash}},
  \bibinfo{author}{\bibfnamefont{S.}~\bibnamefont{Kamata}},
  \bibinfo{author}{\bibfnamefont{M.}~\bibnamefont{Martinez}}, \bibnamefont{and}
  \bibinfo{author}{\bibfnamefont{H.}~\bibnamefont{Shi}},
  \bibinfo{journal}{JHEP} \textbf{\bibinfo{volume}{07}}, \bibinfo{pages}{226}
  (\bibinfo{year}{2020}), \eprint{1911.06406}.

\bibitem[{\citenamefont{Bjorken}(1983)}]{Bjorken:1982qr}
\bibinfo{author}{\bibfnamefont{J.}~\bibnamefont{Bjorken}},
  \bibinfo{journal}{Phys. Rev. D} \textbf{\bibinfo{volume}{27}},
  \bibinfo{pages}{140} (\bibinfo{year}{1983}).

\bibitem[{\citenamefont{Heller et~al.}(2020)\citenamefont{Heller, Jefferson,
  Spali\'nski, and Svensson}}]{Heller:2020anv}
\bibinfo{author}{\bibfnamefont{M.~P.} \bibnamefont{Heller}},
  \bibinfo{author}{\bibfnamefont{R.}~\bibnamefont{Jefferson}},
  \bibinfo{author}{\bibfnamefont{M.}~\bibnamefont{Spali\'nski}},
  \bibnamefont{and} \bibinfo{author}{\bibfnamefont{V.}~\bibnamefont{Svensson}},
  \bibinfo{journal}{Phys. Rev. Lett.} \textbf{\bibinfo{volume}{125}},
  \bibinfo{pages}{132301} (\bibinfo{year}{2020}), \eprint{2003.07368}.

\bibitem[{\citenamefont{Gubser}(2010)}]{Gubser:2010ze}
\bibinfo{author}{\bibfnamefont{S.~S.} \bibnamefont{Gubser}},
  \bibinfo{journal}{Phys. Rev. D} \textbf{\bibinfo{volume}{82}},
  \bibinfo{pages}{085027} (\bibinfo{year}{2010}), \eprint{1006.0006}.

\bibitem[{\citenamefont{Gubser and Yarom}(2011)}]{Gubser:2010ui}
\bibinfo{author}{\bibfnamefont{S.~S.} \bibnamefont{Gubser}} \bibnamefont{and}
  \bibinfo{author}{\bibfnamefont{A.}~\bibnamefont{Yarom}},
  \bibinfo{journal}{Nucl. Phys. B} \textbf{\bibinfo{volume}{846}},
  \bibinfo{pages}{469} (\bibinfo{year}{2011}), \eprint{1012.1314}.

\bibitem[{\citenamefont{Denicol and Noronha}(2016)}]{Denicol:2016bjh}
\bibinfo{author}{\bibfnamefont{G.~S.} \bibnamefont{Denicol}} \bibnamefont{and}
  \bibinfo{author}{\bibfnamefont{J.}~\bibnamefont{Noronha}}
  (\bibinfo{year}{2016}), \eprint{1608.07869}.

\bibitem[{\citenamefont{Caraballo and Han}(2017)}]{caraballo2017applied}
\bibinfo{author}{\bibfnamefont{T.}~\bibnamefont{Caraballo}} \bibnamefont{and}
  \bibinfo{author}{\bibfnamefont{X.}~\bibnamefont{Han}},
  \emph{\bibinfo{title}{Applied Nonautonomous and Random Dynamical Systems:
  Applied Dynamical Systems}}, SpringerBriefs in Mathematics
  (\bibinfo{publisher}{Springer International Publishing},
  \bibinfo{year}{2017}).

\bibitem[{\citenamefont{Kloeden and Rasmussen}(2011)}]{kloeden}
\bibinfo{author}{\bibfnamefont{P.}~\bibnamefont{Kloeden}} \bibnamefont{and}
  \bibinfo{author}{\bibfnamefont{M.}~\bibnamefont{Rasmussen}},
  \emph{\bibinfo{title}{Nonautonomous Dynamical Systems}}, Mathematical surveys
  and monographs (\bibinfo{publisher}{American Mathematical Soc.},
  \bibinfo{year}{2011}).

\bibitem[{\citenamefont{Mueller}(2000{\natexlab{a}})}]{Mueller:1999pi}
\bibinfo{author}{\bibfnamefont{A.~H.} \bibnamefont{Mueller}},
  \bibinfo{journal}{Phys. Lett. B} \textbf{\bibinfo{volume}{475}},
  \bibinfo{pages}{220} (\bibinfo{year}{2000}{\natexlab{a}}),
  \eprint{hep-ph/9909388}.

\bibitem[{\citenamefont{Mueller}(2000{\natexlab{b}})}]{Mueller:1999fp}
\bibinfo{author}{\bibfnamefont{A.~H.} \bibnamefont{Mueller}},
  \bibinfo{journal}{Nucl. Phys. B} \textbf{\bibinfo{volume}{572}},
  \bibinfo{pages}{227} (\bibinfo{year}{2000}{\natexlab{b}}),
  \eprint{hep-ph/9906322}.

\bibitem[{\citenamefont{Bjoraker and Venugopalan}(2001)}]{Bjoraker:2000cf}
\bibinfo{author}{\bibfnamefont{J.}~\bibnamefont{Bjoraker}} \bibnamefont{and}
  \bibinfo{author}{\bibfnamefont{R.}~\bibnamefont{Venugopalan}},
  \bibinfo{journal}{Phys. Rev. C} \textbf{\bibinfo{volume}{63}},
  \bibinfo{pages}{024609} (\bibinfo{year}{2001}), \eprint{hep-ph/0008294}.

\bibitem[{\citenamefont{Hong and Teaney}(2010)}]{Hong:2010at}
\bibinfo{author}{\bibfnamefont{J.}~\bibnamefont{Hong}} \bibnamefont{and}
  \bibinfo{author}{\bibfnamefont{D.}~\bibnamefont{Teaney}},
  \bibinfo{journal}{Phys. Rev. C} \textbf{\bibinfo{volume}{82}},
  \bibinfo{pages}{044908} (\bibinfo{year}{2010}), \eprint{1003.0699}.

\bibitem[{\citenamefont{Mehtar-Tani}(2017)}]{Mehtar-Tani:2016bay}
\bibinfo{author}{\bibfnamefont{Y.}~\bibnamefont{Mehtar-Tani}},
  \bibinfo{journal}{Nucl. Phys. A} \textbf{\bibinfo{volume}{966}},
  \bibinfo{pages}{241} (\bibinfo{year}{2017}), \eprint{1611.01527}.

\bibitem[{\citenamefont{Blaizot and Tanji}(2019)}]{Blaizot:2019dut}
\bibinfo{author}{\bibfnamefont{J.-P.} \bibnamefont{Blaizot}} \bibnamefont{and}
  \bibinfo{author}{\bibfnamefont{N.}~\bibnamefont{Tanji}},
  \bibinfo{journal}{Nucl. Phys. A} \textbf{\bibinfo{volume}{992}},
  \bibinfo{pages}{121618} (\bibinfo{year}{2019}), \eprint{1904.08244}.

\bibitem[{\citenamefont{Blaizot et~al.}(2013)\citenamefont{Blaizot, Liao, and
  McLerran}}]{Blaizot:2013lga}
\bibinfo{author}{\bibfnamefont{J.-P.} \bibnamefont{Blaizot}},
  \bibinfo{author}{\bibfnamefont{J.}~\bibnamefont{Liao}}, \bibnamefont{and}
  \bibinfo{author}{\bibfnamefont{L.}~\bibnamefont{McLerran}},
  \bibinfo{journal}{Nucl. Phys. A} \textbf{\bibinfo{volume}{920}},
  \bibinfo{pages}{58} (\bibinfo{year}{2013}), \eprint{1305.2119}.

\bibitem[{\citenamefont{Blaizot et~al.}(2014)\citenamefont{Blaizot, Wu, and
  Yan}}]{Blaizot:2014jna}
\bibinfo{author}{\bibfnamefont{J.-P.} \bibnamefont{Blaizot}},
  \bibinfo{author}{\bibfnamefont{B.}~\bibnamefont{Wu}}, \bibnamefont{and}
  \bibinfo{author}{\bibfnamefont{L.}~\bibnamefont{Yan}},
  \bibinfo{journal}{Nucl. Phys. A} \textbf{\bibinfo{volume}{930}},
  \bibinfo{pages}{139} (\bibinfo{year}{2014}), \eprint{1402.5049}.

\bibitem[{\citenamefont{Tanji and Venugopalan}(2017)}]{Tanji:2017suk}
\bibinfo{author}{\bibfnamefont{N.}~\bibnamefont{Tanji}} \bibnamefont{and}
  \bibinfo{author}{\bibfnamefont{R.}~\bibnamefont{Venugopalan}},
  \bibinfo{journal}{Phys. Rev. D} \textbf{\bibinfo{volume}{95}},
  \bibinfo{pages}{094009} (\bibinfo{year}{2017}), \eprint{1703.01372}.

\bibitem[{\citenamefont{Moore and Teaney}(2005)}]{Moore:2004tg}
\bibinfo{author}{\bibfnamefont{G.~D.} \bibnamefont{Moore}} \bibnamefont{and}
  \bibinfo{author}{\bibfnamefont{D.}~\bibnamefont{Teaney}},
  \bibinfo{journal}{Phys. Rev. C} \textbf{\bibinfo{volume}{71}},
  \bibinfo{pages}{064904} (\bibinfo{year}{2005}), \eprint{hep-ph/0412346}.

\bibitem[{\citenamefont{Grad}(1949)}]{Grad}
\bibinfo{author}{\bibfnamefont{H.}~\bibnamefont{Grad}},
  \bibinfo{journal}{Commun.Pure Appl.Math.} \textbf{\bibinfo{volume}{2}},
  \bibinfo{pages}{331} (\bibinfo{year}{1949}).

\bibitem[{\citenamefont{Boyd}(1999)}]{boyd:1999}
\bibinfo{author}{\bibfnamefont{J.~P.} \bibnamefont{Boyd}},
  \bibinfo{journal}{Acta Applicandae Mathematica}
  \textbf{\bibinfo{volume}{56}}, \bibinfo{pages}{1} (\bibinfo{year}{1999}).

\bibitem[{\citenamefont{Costin}(2008)}]{costin2008asymptotics}
\bibinfo{author}{\bibfnamefont{O.}~\bibnamefont{Costin}},
  \emph{\bibinfo{title}{Asymptotics and Borel Summability}}, Monographs and
  Surveys in Pure and Applied Mathematics (\bibinfo{publisher}{CRC Press},
  \bibinfo{year}{2008}).

\bibitem[{\citenamefont{Costin}(1995)}]{costin1995}
\bibinfo{author}{\bibfnamefont{O.}~\bibnamefont{Costin}},
  \bibinfo{journal}{International Mathematics Research Notices}
  \textbf{\bibinfo{volume}{1995}}, \bibinfo{pages}{377} (\bibinfo{year}{1995}).

\bibitem[{\citenamefont{Costin}(1998)}]{costin1998}
\bibinfo{author}{\bibfnamefont{O.}~\bibnamefont{Costin}},
  \bibinfo{journal}{Duke Math. J.} \textbf{\bibinfo{volume}{93}},
  \bibinfo{pages}{289} (\bibinfo{year}{1998}).

\bibitem[{\citenamefont{Gukov}(2016)}]{Gukov:2015qea}
\bibinfo{author}{\bibfnamefont{S.}~\bibnamefont{Gukov}},
  \bibinfo{journal}{JHEP} \textbf{\bibinfo{volume}{01}}, \bibinfo{pages}{020}
  (\bibinfo{year}{2016}), \eprint{1503.01474}.

\bibitem[{\citenamefont{Gukov}(2017)}]{Gukov:2016tnp}
\bibinfo{author}{\bibfnamefont{S.}~\bibnamefont{Gukov}},
  \bibinfo{journal}{Nucl. Phys. B} \textbf{\bibinfo{volume}{919}},
  \bibinfo{pages}{583} (\bibinfo{year}{2017}), \eprint{1608.06638}.

\bibitem[{\citenamefont{Kurkela et~al.}(2020)\citenamefont{Kurkela, van~der
  Schee, Wiedemann, and Wu}}]{Kurkela:2019set}
\bibinfo{author}{\bibfnamefont{A.}~\bibnamefont{Kurkela}},
  \bibinfo{author}{\bibfnamefont{W.}~\bibnamefont{van~der Schee}},
  \bibinfo{author}{\bibfnamefont{U.~A.} \bibnamefont{Wiedemann}},
  \bibnamefont{and} \bibinfo{author}{\bibfnamefont{B.}~\bibnamefont{Wu}},
  \bibinfo{journal}{Phys. Rev. Lett.} \textbf{\bibinfo{volume}{124}},
  \bibinfo{pages}{102301} (\bibinfo{year}{2020}), \eprint{1907.08101}.

\bibitem[{\citenamefont{Jaiswal et~al.}(2019)\citenamefont{Jaiswal,
  Chattopadhyay, Jaiswal, Pal, and Heinz}}]{Jaiswal:2019cju}
\bibinfo{author}{\bibfnamefont{S.}~\bibnamefont{Jaiswal}},
  \bibinfo{author}{\bibfnamefont{C.}~\bibnamefont{Chattopadhyay}},
  \bibinfo{author}{\bibfnamefont{A.}~\bibnamefont{Jaiswal}},
  \bibinfo{author}{\bibfnamefont{S.}~\bibnamefont{Pal}}, \bibnamefont{and}
  \bibinfo{author}{\bibfnamefont{U.}~\bibnamefont{Heinz}},
  \bibinfo{journal}{Phys. Rev. C} \textbf{\bibinfo{volume}{100}},
  \bibinfo{pages}{034901} (\bibinfo{year}{2019}), \eprint{1907.07965}.

\bibitem[{\citenamefont{Blaizot and Yan}(2020{\natexlab{b}})}]{Blaizot:2019scw}
\bibinfo{author}{\bibfnamefont{J.-P.} \bibnamefont{Blaizot}} \bibnamefont{and}
  \bibinfo{author}{\bibfnamefont{L.}~\bibnamefont{Yan}},
  \bibinfo{journal}{Annals Phys.} \textbf{\bibinfo{volume}{412}},
  \bibinfo{pages}{167993} (\bibinfo{year}{2020}{\natexlab{b}}),
  \eprint{1904.08677}.

\bibitem[{\citenamefont{Arnold et~al.}(2003)\citenamefont{Arnold, Moore, and
  Yaffe}}]{Arnold:2002zm}
\bibinfo{author}{\bibfnamefont{P.~B.} \bibnamefont{Arnold}},
  \bibinfo{author}{\bibfnamefont{G.~D.} \bibnamefont{Moore}}, \bibnamefont{and}
  \bibinfo{author}{\bibfnamefont{L.~G.} \bibnamefont{Yaffe}},
  \bibinfo{journal}{JHEP} \textbf{\bibinfo{volume}{01}}, \bibinfo{pages}{030}
  (\bibinfo{year}{2003}), \eprint{hep-ph/0209353}.

\bibitem[{\citenamefont{Lifshitz and Pitaevskii}(1981)}]{Lifshitz81}
\bibinfo{author}{\bibfnamefont{E.}~\bibnamefont{Lifshitz}} \bibnamefont{and}
  \bibinfo{author}{\bibfnamefont{L.}~\bibnamefont{Pitaevskii}},
  \emph{\bibinfo{title}{Physical Kinetics}} (\bibinfo{publisher}{Pergamon
  Press, Oxford}, \bibinfo{year}{1981}).

\bibitem[{\citenamefont{M.~Le~Bellac}(2000)}]{lebellac}
\bibinfo{author}{\bibfnamefont{M.}~\bibnamefont{M.~Le~Bellac}},
  \emph{\bibinfo{title}{Thermal Field Theory}} (\bibinfo{publisher}{Cambridge
  University Press, Cambridge}, \bibinfo{year}{2000}).

\bibitem[{\citenamefont{Blaizot and Yan}(2017)}]{Blaizot:2017lht}
\bibinfo{author}{\bibfnamefont{J.-P.} \bibnamefont{Blaizot}} \bibnamefont{and}
  \bibinfo{author}{\bibfnamefont{L.}~\bibnamefont{Yan}},
  \bibinfo{journal}{JHEP} \textbf{\bibinfo{volume}{11}}, \bibinfo{pages}{161}
  (\bibinfo{year}{2017}), \eprint{1703.10694}.

\bibitem[{\citenamefont{Baym et~al.}(1990)\citenamefont{Baym, Monien, Pethick,
  and Ravenhall}}]{Baym:1990uj}
\bibinfo{author}{\bibfnamefont{G.}~\bibnamefont{Baym}},
  \bibinfo{author}{\bibfnamefont{H.}~\bibnamefont{Monien}},
  \bibinfo{author}{\bibfnamefont{C.}~\bibnamefont{Pethick}}, \bibnamefont{and}
  \bibinfo{author}{\bibfnamefont{D.}~\bibnamefont{Ravenhall}},
  \bibinfo{journal}{Phys. Rev. Lett.} \textbf{\bibinfo{volume}{64}},
  \bibinfo{pages}{1867} (\bibinfo{year}{1990}).

\bibitem[{\citenamefont{Heiselberg}(1994)}]{Heiselberg:1994vy}
\bibinfo{author}{\bibfnamefont{H.}~\bibnamefont{Heiselberg}},
  \bibinfo{journal}{Phys. Rev. D} \textbf{\bibinfo{volume}{49}},
  \bibinfo{pages}{4739} (\bibinfo{year}{1994}), \eprint{hep-ph/9401309}.

\bibitem[{\citenamefont{Abraao~York et~al.}(2014)\citenamefont{Abraao~York,
  Kurkela, Lu, and Moore}}]{York:2014wja}
\bibinfo{author}{\bibfnamefont{M.~C.} \bibnamefont{Abraao~York}},
  \bibinfo{author}{\bibfnamefont{A.}~\bibnamefont{Kurkela}},
  \bibinfo{author}{\bibfnamefont{E.}~\bibnamefont{Lu}}, \bibnamefont{and}
  \bibinfo{author}{\bibfnamefont{G.~D.} \bibnamefont{Moore}},
  \bibinfo{journal}{Phys. Rev. D} \textbf{\bibinfo{volume}{89}},
  \bibinfo{pages}{074036} (\bibinfo{year}{2014}), \eprint{1401.3751}.

\bibitem[{\citenamefont{Bazow et~al.}(2016{\natexlab{a}})\citenamefont{Bazow,
  Denicol, Heinz, Martinez, and Noronha}}]{Bazow:2015dha}
\bibinfo{author}{\bibfnamefont{D.}~\bibnamefont{Bazow}},
  \bibinfo{author}{\bibfnamefont{G.}~\bibnamefont{Denicol}},
  \bibinfo{author}{\bibfnamefont{U.}~\bibnamefont{Heinz}},
  \bibinfo{author}{\bibfnamefont{M.}~\bibnamefont{Martinez}}, \bibnamefont{and}
  \bibinfo{author}{\bibfnamefont{J.}~\bibnamefont{Noronha}},
  \bibinfo{journal}{Phys. Rev. Lett.} \textbf{\bibinfo{volume}{116}},
  \bibinfo{pages}{022301} (\bibinfo{year}{2016}{\natexlab{a}}),
  \eprint{1507.07834}.

\bibitem[{\citenamefont{Bazow et~al.}(2016{\natexlab{b}})\citenamefont{Bazow,
  Denicol, Heinz, Martinez, and Noronha}}]{Bazow:2016oky}
\bibinfo{author}{\bibfnamefont{D.}~\bibnamefont{Bazow}},
  \bibinfo{author}{\bibfnamefont{G.}~\bibnamefont{Denicol}},
  \bibinfo{author}{\bibfnamefont{U.}~\bibnamefont{Heinz}},
  \bibinfo{author}{\bibfnamefont{M.}~\bibnamefont{Martinez}}, \bibnamefont{and}
  \bibinfo{author}{\bibfnamefont{J.}~\bibnamefont{Noronha}},
  \bibinfo{journal}{Phys. Rev. D} \textbf{\bibinfo{volume}{94}},
  \bibinfo{pages}{125006} (\bibinfo{year}{2016}{\natexlab{b}}),
  \eprint{1607.05245}.

\bibitem[{\citenamefont{Krook and Wu}(1976)}]{krookwu}
\bibinfo{author}{\bibfnamefont{M.}~\bibnamefont{Krook}} \bibnamefont{and}
  \bibinfo{author}{\bibfnamefont{T.~T.} \bibnamefont{Wu}},
  \bibinfo{journal}{Phys. Rev. Lett.} \textbf{\bibinfo{volume}{36}},
  \bibinfo{pages}{1107} (\bibinfo{year}{1976}).

\bibitem[{\citenamefont{Bobylev}(1976)}]{bobylev1976fourier}
\bibinfo{author}{\bibfnamefont{A.}~\bibnamefont{Bobylev}},
  \bibinfo{journal}{Sov. Phys. Dokl} \textbf{\bibinfo{volume}{20}},
  \bibinfo{pages}{820} (\bibinfo{year}{1976}).

\bibitem[{\citenamefont{Molnar et~al.}(2016)\citenamefont{Molnar, Niemi, and
  Rischke}}]{Molnar:2016vvu}
\bibinfo{author}{\bibfnamefont{E.}~\bibnamefont{Molnar}},
  \bibinfo{author}{\bibfnamefont{H.}~\bibnamefont{Niemi}}, \bibnamefont{and}
  \bibinfo{author}{\bibfnamefont{D.}~\bibnamefont{Rischke}},
  \bibinfo{journal}{Phys. Rev. D} \textbf{\bibinfo{volume}{93}},
  \bibinfo{pages}{114025} (\bibinfo{year}{2016}), \eprint{1602.00573}.

\bibitem[{\citenamefont{Martinez et~al.}(2017)\citenamefont{Martinez, McNelis,
  and Heinz}}]{Martinez:2017ibh}
\bibinfo{author}{\bibfnamefont{M.}~\bibnamefont{Martinez}},
  \bibinfo{author}{\bibfnamefont{M.}~\bibnamefont{McNelis}}, \bibnamefont{and}
  \bibinfo{author}{\bibfnamefont{U.}~\bibnamefont{Heinz}},
  \bibinfo{journal}{Phys. Rev. C} \textbf{\bibinfo{volume}{95}},
  \bibinfo{pages}{054907} (\bibinfo{year}{2017}), \eprint{1703.10955}.

\bibitem[{\citenamefont{Martinez and Strickland}(2009)}]{Martinez:2009mf}
\bibinfo{author}{\bibfnamefont{M.}~\bibnamefont{Martinez}} \bibnamefont{and}
  \bibinfo{author}{\bibfnamefont{M.}~\bibnamefont{Strickland}},
  \bibinfo{journal}{Phys. Rev. C} \textbf{\bibinfo{volume}{79}},
  \bibinfo{pages}{044903} (\bibinfo{year}{2009}), \eprint{0902.3834}.

\bibitem[{\citenamefont{Denicol et~al.}(2010)\citenamefont{Denicol, Koide, and
  Rischke}}]{Denicol:2010xn}
\bibinfo{author}{\bibfnamefont{G.}~\bibnamefont{Denicol}},
  \bibinfo{author}{\bibfnamefont{T.}~\bibnamefont{Koide}}, \bibnamefont{and}
  \bibinfo{author}{\bibfnamefont{D.}~\bibnamefont{Rischke}},
  \bibinfo{journal}{Phys. Rev. Lett.} \textbf{\bibinfo{volume}{105}},
  \bibinfo{pages}{162501} (\bibinfo{year}{2010}), \eprint{1004.5013}.

\bibitem[{\citenamefont{Florkowski
  et~al.}(2013{\natexlab{a}})\citenamefont{Florkowski, Ryblewski, and
  Strickland}}]{Florkowski:2013lza}
\bibinfo{author}{\bibfnamefont{W.}~\bibnamefont{Florkowski}},
  \bibinfo{author}{\bibfnamefont{R.}~\bibnamefont{Ryblewski}},
  \bibnamefont{and}
  \bibinfo{author}{\bibfnamefont{M.}~\bibnamefont{Strickland}},
  \bibinfo{journal}{Nucl. Phys. A} \textbf{\bibinfo{volume}{916}},
  \bibinfo{pages}{249} (\bibinfo{year}{2013}{\natexlab{a}}),
  \eprint{1304.0665}.

\bibitem[{\citenamefont{Florkowski
  et~al.}(2013{\natexlab{b}})\citenamefont{Florkowski, Ryblewski, and
  Strickland}}]{Florkowski:2013lya}
\bibinfo{author}{\bibfnamefont{W.}~\bibnamefont{Florkowski}},
  \bibinfo{author}{\bibfnamefont{R.}~\bibnamefont{Ryblewski}},
  \bibnamefont{and}
  \bibinfo{author}{\bibfnamefont{M.}~\bibnamefont{Strickland}},
  \bibinfo{journal}{Phys. Rev. C} \textbf{\bibinfo{volume}{88}},
  \bibinfo{pages}{024903} (\bibinfo{year}{2013}{\natexlab{b}}),
  \eprint{1305.7234}.

\bibitem[{\citenamefont{Denicol et~al.}(2011)\citenamefont{Denicol, Noronha,
  Niemi, and Rischke}}]{Denicol:2011fa}
\bibinfo{author}{\bibfnamefont{G.~S.} \bibnamefont{Denicol}},
  \bibinfo{author}{\bibfnamefont{J.}~\bibnamefont{Noronha}},
  \bibinfo{author}{\bibfnamefont{H.}~\bibnamefont{Niemi}}, \bibnamefont{and}
  \bibinfo{author}{\bibfnamefont{D.~H.} \bibnamefont{Rischke}},
  \bibinfo{journal}{Phys. Rev. D} \textbf{\bibinfo{volume}{83}},
  \bibinfo{pages}{074019} (\bibinfo{year}{2011}), \eprint{1102.4780}.

\bibitem[{\citenamefont{Teaney and Yan}(2014)}]{Teaney:2013gca}
\bibinfo{author}{\bibfnamefont{D.}~\bibnamefont{Teaney}} \bibnamefont{and}
  \bibinfo{author}{\bibfnamefont{L.}~\bibnamefont{Yan}},
  \bibinfo{journal}{Phys. Rev. C} \textbf{\bibinfo{volume}{89}},
  \bibinfo{pages}{014901} (\bibinfo{year}{2014}), \eprint{1304.3753}.

\bibitem[{\citenamefont{Chattopadhyay et~al.}(2015)\citenamefont{Chattopadhyay,
  Jaiswal, Pal, and Ryblewski}}]{Chattopadhyay:2014lya}
\bibinfo{author}{\bibfnamefont{C.}~\bibnamefont{Chattopadhyay}},
  \bibinfo{author}{\bibfnamefont{A.}~\bibnamefont{Jaiswal}},
  \bibinfo{author}{\bibfnamefont{S.}~\bibnamefont{Pal}}, \bibnamefont{and}
  \bibinfo{author}{\bibfnamefont{R.}~\bibnamefont{Ryblewski}},
  \bibinfo{journal}{Phys. Rev. C} \textbf{\bibinfo{volume}{91}},
  \bibinfo{pages}{024917} (\bibinfo{year}{2015}), \eprint{1411.2363}.

\bibitem[{\citenamefont{Bhalerao et~al.}(2014)\citenamefont{Bhalerao, Jaiswal,
  Pal, and Sreekanth}}]{Bhalerao:2013pza}
\bibinfo{author}{\bibfnamefont{R.~S.} \bibnamefont{Bhalerao}},
  \bibinfo{author}{\bibfnamefont{A.}~\bibnamefont{Jaiswal}},
  \bibinfo{author}{\bibfnamefont{S.}~\bibnamefont{Pal}}, \bibnamefont{and}
  \bibinfo{author}{\bibfnamefont{V.}~\bibnamefont{Sreekanth}},
  \bibinfo{journal}{Phys. Rev. C} \textbf{\bibinfo{volume}{89}},
  \bibinfo{pages}{054903} (\bibinfo{year}{2014}), \eprint{1312.1864}.

\bibitem[{\citenamefont{Jaiswal}(2013{\natexlab{a}})}]{Jaiswal:2013vta}
\bibinfo{author}{\bibfnamefont{A.}~\bibnamefont{Jaiswal}},
  \bibinfo{journal}{Phys. Rev. C} \textbf{\bibinfo{volume}{88}},
  \bibinfo{pages}{021903} (\bibinfo{year}{2013}{\natexlab{a}}),
  \eprint{1305.3480}.

\bibitem[{\citenamefont{Jaiswal}(2013{\natexlab{b}})}]{Jaiswal:2013npa}
\bibinfo{author}{\bibfnamefont{A.}~\bibnamefont{Jaiswal}},
  \bibinfo{journal}{Phys. Rev. C} \textbf{\bibinfo{volume}{87}},
  \bibinfo{pages}{051901} (\bibinfo{year}{2013}{\natexlab{b}}),
  \eprint{1302.6311}.

\bibitem[{\citenamefont{Romatschke}(2012)}]{Romatschke:2011qp}
\bibinfo{author}{\bibfnamefont{P.}~\bibnamefont{Romatschke}},
  \bibinfo{journal}{Phys. Rev. D} \textbf{\bibinfo{volume}{85}},
  \bibinfo{pages}{065012} (\bibinfo{year}{2012}), \eprint{1108.5561}.

\bibitem[{\citenamefont{Yan}(2012)}]{Yan:2012jb}
\bibinfo{author}{\bibfnamefont{L.}~\bibnamefont{Yan}}, \bibinfo{journal}{J.
  Phys. Conf. Ser.} \textbf{\bibinfo{volume}{389}}, \bibinfo{pages}{012014}
  (\bibinfo{year}{2012}), \eprint{1208.3011}.

\bibitem[{\citenamefont{Denicol et~al.}(2012)\citenamefont{Denicol, Moln\'ar,
  Niemi, and Rischke}}]{Denicol:2012es}
\bibinfo{author}{\bibfnamefont{G.}~\bibnamefont{Denicol}},
  \bibinfo{author}{\bibfnamefont{E.}~\bibnamefont{Moln\'ar}},
  \bibinfo{author}{\bibfnamefont{H.}~\bibnamefont{Niemi}}, \bibnamefont{and}
  \bibinfo{author}{\bibfnamefont{D.}~\bibnamefont{Rischke}},
  \bibinfo{journal}{Eur. Phys. J. A} \textbf{\bibinfo{volume}{48}},
  \bibinfo{pages}{170} (\bibinfo{year}{2012}), \eprint{1206.1554}.

\bibitem[{\citenamefont{Baier et~al.}(2008)\citenamefont{Baier, Romatschke,
  Son, Starinets, and Stephanov}}]{Baier:2007ix}
\bibinfo{author}{\bibfnamefont{R.}~\bibnamefont{Baier}},
  \bibinfo{author}{\bibfnamefont{P.}~\bibnamefont{Romatschke}},
  \bibinfo{author}{\bibfnamefont{D.~T.} \bibnamefont{Son}},
  \bibinfo{author}{\bibfnamefont{A.~O.} \bibnamefont{Starinets}},
  \bibnamefont{and} \bibinfo{author}{\bibfnamefont{M.~A.}
  \bibnamefont{Stephanov}}, \bibinfo{journal}{JHEP}
  \textbf{\bibinfo{volume}{04}}, \bibinfo{pages}{100} (\bibinfo{year}{2008}),
  \eprint{0712.2451}.

\bibitem[{\citenamefont{Costin and Costin}(2001)}]{Costin2001}
\bibinfo{author}{\bibfnamefont{O.}~\bibnamefont{Costin}} \bibnamefont{and}
  \bibinfo{author}{\bibfnamefont{R.}~\bibnamefont{Costin}},
  \bibinfo{journal}{Inventiones mathematicae} \textbf{\bibinfo{volume}{145}},
  \bibinfo{pages}{425} (\bibinfo{year}{2001}).

\bibitem[{\citenamefont{Lang}(2010)}]{lang2010complex}
\bibinfo{author}{\bibfnamefont{S.}~\bibnamefont{Lang}},
  \emph{\bibinfo{title}{Complex Analysis}}, Graduate Texts in Mathematics
  (\bibinfo{publisher}{Springer New York}, \bibinfo{year}{2010}).

\bibitem[{\citenamefont{Heiselberg and Levy}(1999)}]{Heiselberg:1998es}
\bibinfo{author}{\bibfnamefont{H.}~\bibnamefont{Heiselberg}} \bibnamefont{and}
  \bibinfo{author}{\bibfnamefont{A.-M.} \bibnamefont{Levy}},
  \bibinfo{journal}{Phys. Rev. C} \textbf{\bibinfo{volume}{59}},
  \bibinfo{pages}{2716} (\bibinfo{year}{1999}), \eprint{nucl-th/9812034}.

\bibitem[{\citenamefont{Kurkela
  et~al.}(2019{\natexlab{a}})\citenamefont{Kurkela, Wiedemann, and
  Wu}}]{Kurkela:2019kip}
\bibinfo{author}{\bibfnamefont{A.}~\bibnamefont{Kurkela}},
  \bibinfo{author}{\bibfnamefont{U.~A.} \bibnamefont{Wiedemann}},
  \bibnamefont{and} \bibinfo{author}{\bibfnamefont{B.}~\bibnamefont{Wu}},
  \bibinfo{journal}{Eur. Phys. J. C} \textbf{\bibinfo{volume}{79}},
  \bibinfo{pages}{965} (\bibinfo{year}{2019}{\natexlab{a}}),
  \eprint{1905.05139}.

\bibitem[{\citenamefont{Kurkela
  et~al.}(2019{\natexlab{b}})\citenamefont{Kurkela, Wiedemann, and
  Wu}}]{Kurkela:2018qeb}
\bibinfo{author}{\bibfnamefont{A.}~\bibnamefont{Kurkela}},
  \bibinfo{author}{\bibfnamefont{U.~A.} \bibnamefont{Wiedemann}},
  \bibnamefont{and} \bibinfo{author}{\bibfnamefont{B.}~\bibnamefont{Wu}},
  \bibinfo{journal}{Eur. Phys. J. C} \textbf{\bibinfo{volume}{79}},
  \bibinfo{pages}{759} (\bibinfo{year}{2019}{\natexlab{b}}),
  \eprint{1805.04081}.

\bibitem[{\citenamefont{Borghini and Gombeaud}(2011)}]{Borghini:2010hy}
\bibinfo{author}{\bibfnamefont{N.}~\bibnamefont{Borghini}} \bibnamefont{and}
  \bibinfo{author}{\bibfnamefont{C.}~\bibnamefont{Gombeaud}},
  \bibinfo{journal}{Eur. Phys. J. C} \textbf{\bibinfo{volume}{71}},
  \bibinfo{pages}{1612} (\bibinfo{year}{2011}), \eprint{1012.0899}.

\bibitem[{\citenamefont{Romatschke}(2018{\natexlab{b}})}]{Romatschke:2018wgi}
\bibinfo{author}{\bibfnamefont{P.}~\bibnamefont{Romatschke}},
  \bibinfo{journal}{Eur. Phys. J. C} \textbf{\bibinfo{volume}{78}},
  \bibinfo{pages}{636} (\bibinfo{year}{2018}{\natexlab{b}}),
  \eprint{1802.06804}.

\bibitem[{\citenamefont{Blaizot and Yan}(2018)}]{Blaizot:2017ucy}
\bibinfo{author}{\bibfnamefont{J.-P.} \bibnamefont{Blaizot}} \bibnamefont{and}
  \bibinfo{author}{\bibfnamefont{L.}~\bibnamefont{Yan}},
  \bibinfo{journal}{Phys. Lett. B} \textbf{\bibinfo{volume}{780}},
  \bibinfo{pages}{283} (\bibinfo{year}{2018}), \eprint{1712.03856}.

\bibitem[{\citenamefont{Kamata et~al.}(2020)\citenamefont{Kamata, Martinez,
  Plaschke, Ochsenfeld, and Schlichting}}]{Kamata:2020mka}
\bibinfo{author}{\bibfnamefont{S.}~\bibnamefont{Kamata}},
  \bibinfo{author}{\bibfnamefont{M.}~\bibnamefont{Martinez}},
  \bibinfo{author}{\bibfnamefont{P.}~\bibnamefont{Plaschke}},
  \bibinfo{author}{\bibfnamefont{S.}~\bibnamefont{Ochsenfeld}},
  \bibnamefont{and}
  \bibinfo{author}{\bibfnamefont{S.}~\bibnamefont{Schlichting}},
  \bibinfo{journal}{Phys. Rev. D} \textbf{\bibinfo{volume}{102}},
  \bibinfo{pages}{056003} (\bibinfo{year}{2020}), \eprint{2004.06751}.

\bibitem[{\citenamefont{Giacalone et~al.}(2019)\citenamefont{Giacalone,
  Mazeliauskas, and Schlichting}}]{Giacalone:2019ldn}
\bibinfo{author}{\bibfnamefont{G.}~\bibnamefont{Giacalone}},
  \bibinfo{author}{\bibfnamefont{A.}~\bibnamefont{Mazeliauskas}},
  \bibnamefont{and}
  \bibinfo{author}{\bibfnamefont{S.}~\bibnamefont{Schlichting}},
  \bibinfo{journal}{Phys. Rev. Lett.} \textbf{\bibinfo{volume}{123}},
  \bibinfo{pages}{262301} (\bibinfo{year}{2019}), \eprint{1908.02866}.

\bibitem[{\citenamefont{Schlichting and Teaney}(2019)}]{Schlichting:2019abc}
\bibinfo{author}{\bibfnamefont{S.}~\bibnamefont{Schlichting}} \bibnamefont{and}
  \bibinfo{author}{\bibfnamefont{D.}~\bibnamefont{Teaney}},
  \bibinfo{journal}{Ann. Rev. Nucl. Part. Sci.} \textbf{\bibinfo{volume}{69}},
  \bibinfo{pages}{447} (\bibinfo{year}{2019}), \eprint{1908.02113}.

\bibitem[{\citenamefont{Dusling et~al.}(2010)\citenamefont{Dusling, Moore, and
  Teaney}}]{Dusling:2009df}
\bibinfo{author}{\bibfnamefont{K.}~\bibnamefont{Dusling}},
  \bibinfo{author}{\bibfnamefont{G.~D.} \bibnamefont{Moore}}, \bibnamefont{and}
  \bibinfo{author}{\bibfnamefont{D.}~\bibnamefont{Teaney}},
  \bibinfo{journal}{Phys. Rev. C} \textbf{\bibinfo{volume}{81}},
  \bibinfo{pages}{034907} (\bibinfo{year}{2010}), \eprint{0909.0754}.

\bibitem[{\citenamefont{Kurkela
  et~al.}(2019{\natexlab{c}})\citenamefont{Kurkela, Mazeliauskas, Paquet,
  Schlichting, and Teaney}}]{Kurkela:2018wud}
\bibinfo{author}{\bibfnamefont{A.}~\bibnamefont{Kurkela}},
  \bibinfo{author}{\bibfnamefont{A.}~\bibnamefont{Mazeliauskas}},
  \bibinfo{author}{\bibfnamefont{J.-F.} \bibnamefont{Paquet}},
  \bibinfo{author}{\bibfnamefont{S.}~\bibnamefont{Schlichting}},
  \bibnamefont{and} \bibinfo{author}{\bibfnamefont{D.}~\bibnamefont{Teaney}},
  \bibinfo{journal}{Phys. Rev. Lett.} \textbf{\bibinfo{volume}{122}},
  \bibinfo{pages}{122302} (\bibinfo{year}{2019}{\natexlab{c}}),
  \eprint{1805.01604}.

\bibitem[{\citenamefont{Kurkela
  et~al.}(2019{\natexlab{d}})\citenamefont{Kurkela, Mazeliauskas, Paquet,
  Schlichting, and Teaney}}]{Kurkela:2018vqr}
\bibinfo{author}{\bibfnamefont{A.}~\bibnamefont{Kurkela}},
  \bibinfo{author}{\bibfnamefont{A.}~\bibnamefont{Mazeliauskas}},
  \bibinfo{author}{\bibfnamefont{J.-F.} \bibnamefont{Paquet}},
  \bibinfo{author}{\bibfnamefont{S.}~\bibnamefont{Schlichting}},
  \bibnamefont{and} \bibinfo{author}{\bibfnamefont{D.}~\bibnamefont{Teaney}},
  \bibinfo{journal}{Phys. Rev. C} \textbf{\bibinfo{volume}{99}},
  \bibinfo{pages}{034910} (\bibinfo{year}{2019}{\natexlab{d}}),
  \eprint{1805.00961}.

\bibitem[{\citenamefont{Bazow et~al.}(2015)\citenamefont{Bazow, Heinz, and
  Martinez}}]{Bazow:2015cha}
\bibinfo{author}{\bibfnamefont{D.}~\bibnamefont{Bazow}},
  \bibinfo{author}{\bibfnamefont{U.~W.} \bibnamefont{Heinz}}, \bibnamefont{and}
  \bibinfo{author}{\bibfnamefont{M.}~\bibnamefont{Martinez}},
  \bibinfo{journal}{Phys. Rev. C} \textbf{\bibinfo{volume}{91}},
  \bibinfo{pages}{064903} (\bibinfo{year}{2015}), \eprint{1503.07443}.

\bibitem[{\citenamefont{Denicol
  et~al.}(2014{\natexlab{a}})\citenamefont{Denicol, Heinz, Martinez, Noronha,
  and Strickland}}]{Denicol:2014tha}
\bibinfo{author}{\bibfnamefont{G.~S.} \bibnamefont{Denicol}},
  \bibinfo{author}{\bibfnamefont{U.~W.} \bibnamefont{Heinz}},
  \bibinfo{author}{\bibfnamefont{M.}~\bibnamefont{Martinez}},
  \bibinfo{author}{\bibfnamefont{J.}~\bibnamefont{Noronha}}, \bibnamefont{and}
  \bibinfo{author}{\bibfnamefont{M.}~\bibnamefont{Strickland}},
  \bibinfo{journal}{Phys. Rev. D} \textbf{\bibinfo{volume}{90}},
  \bibinfo{pages}{125026} (\bibinfo{year}{2014}{\natexlab{a}}),
  \eprint{1408.7048}.

\bibitem[{\citenamefont{Denicol
  et~al.}(2014{\natexlab{b}})\citenamefont{Denicol, Heinz, Martinez, Noronha,
  and Strickland}}]{Denicol:2014xca}
\bibinfo{author}{\bibfnamefont{G.~S.} \bibnamefont{Denicol}},
  \bibinfo{author}{\bibfnamefont{U.~W.} \bibnamefont{Heinz}},
  \bibinfo{author}{\bibfnamefont{M.}~\bibnamefont{Martinez}},
  \bibinfo{author}{\bibfnamefont{J.}~\bibnamefont{Noronha}}, \bibnamefont{and}
  \bibinfo{author}{\bibfnamefont{M.}~\bibnamefont{Strickland}},
  \bibinfo{journal}{Phys. Rev. Lett.} \textbf{\bibinfo{volume}{113}},
  \bibinfo{pages}{202301} (\bibinfo{year}{2014}{\natexlab{b}}),
  \eprint{1408.5646}.

\bibitem[{\citenamefont{Tinti et~al.}(2016)\citenamefont{Tinti, Ryblewski,
  Florkowski, and Strickland}}]{Tinti:2015xra}
\bibinfo{author}{\bibfnamefont{L.}~\bibnamefont{Tinti}},
  \bibinfo{author}{\bibfnamefont{R.}~\bibnamefont{Ryblewski}},
  \bibinfo{author}{\bibfnamefont{W.}~\bibnamefont{Florkowski}},
  \bibnamefont{and}
  \bibinfo{author}{\bibfnamefont{M.}~\bibnamefont{Strickland}},
  \bibinfo{journal}{Nucl. Phys. A} \textbf{\bibinfo{volume}{946}},
  \bibinfo{pages}{29} (\bibinfo{year}{2016}), \eprint{1505.06456}.

\bibitem[{\citenamefont{Denicol
  et~al.}(2014{\natexlab{c}})\citenamefont{Denicol, Florkowski, Ryblewski, and
  Strickland}}]{Denicol:2014mca}
\bibinfo{author}{\bibfnamefont{G.~S.} \bibnamefont{Denicol}},
  \bibinfo{author}{\bibfnamefont{W.}~\bibnamefont{Florkowski}},
  \bibinfo{author}{\bibfnamefont{R.}~\bibnamefont{Ryblewski}},
  \bibnamefont{and}
  \bibinfo{author}{\bibfnamefont{M.}~\bibnamefont{Strickland}},
  \bibinfo{journal}{Phys. Rev. C} \textbf{\bibinfo{volume}{90}},
  \bibinfo{pages}{044905} (\bibinfo{year}{2014}{\natexlab{c}}),
  \eprint{1407.4767}.

\bibitem[{\citenamefont{Florkowski
  et~al.}(2014{\natexlab{a}})\citenamefont{Florkowski, Ryblewski, Strickland,
  and Tinti}}]{Florkowski:2014bba}
\bibinfo{author}{\bibfnamefont{W.}~\bibnamefont{Florkowski}},
  \bibinfo{author}{\bibfnamefont{R.}~\bibnamefont{Ryblewski}},
  \bibinfo{author}{\bibfnamefont{M.}~\bibnamefont{Strickland}},
  \bibnamefont{and} \bibinfo{author}{\bibfnamefont{L.}~\bibnamefont{Tinti}},
  \bibinfo{journal}{Phys. Rev. C} \textbf{\bibinfo{volume}{89}},
  \bibinfo{pages}{054909} (\bibinfo{year}{2014}{\natexlab{a}}),
  \eprint{1403.1223}.

\bibitem[{\citenamefont{Florkowski
  et~al.}(2014{\natexlab{b}})\citenamefont{Florkowski, Maksymiuk, Ryblewski,
  and Strickland}}]{Florkowski:2014sfa}
\bibinfo{author}{\bibfnamefont{W.}~\bibnamefont{Florkowski}},
  \bibinfo{author}{\bibfnamefont{E.}~\bibnamefont{Maksymiuk}},
  \bibinfo{author}{\bibfnamefont{R.}~\bibnamefont{Ryblewski}},
  \bibnamefont{and}
  \bibinfo{author}{\bibfnamefont{M.}~\bibnamefont{Strickland}},
  \bibinfo{journal}{Phys. Rev. C} \textbf{\bibinfo{volume}{89}},
  \bibinfo{pages}{054908} (\bibinfo{year}{2014}{\natexlab{b}}),
  \eprint{1402.7348}.

\bibitem[{\citenamefont{Kurkela and Moore}(2011)}]{Kurkela:2011ub}
\bibinfo{author}{\bibfnamefont{A.}~\bibnamefont{Kurkela}} \bibnamefont{and}
  \bibinfo{author}{\bibfnamefont{G.~D.} \bibnamefont{Moore}},
  \bibinfo{journal}{JHEP} \textbf{\bibinfo{volume}{1111}}, \bibinfo{pages}{120}
  (\bibinfo{year}{2011}), \eprint{1108.4684}.

\bibitem[{\citenamefont{Kurkela and Zhu}(2015)}]{Kurkela:2015qoa}
\bibinfo{author}{\bibfnamefont{A.}~\bibnamefont{Kurkela}} \bibnamefont{and}
  \bibinfo{author}{\bibfnamefont{Y.}~\bibnamefont{Zhu}},
  \bibinfo{journal}{Phys. Rev. Lett.} \textbf{\bibinfo{volume}{115}},
  \bibinfo{pages}{182301} (\bibinfo{year}{2015}), \eprint{1506.06647}.

\bibitem[{\citenamefont{Almaalol et~al.}(2020)\citenamefont{Almaalol, Kurkela,
  and Strickland}}]{Almaalol:2020rnu}
\bibinfo{author}{\bibfnamefont{D.}~\bibnamefont{Almaalol}},
  \bibinfo{author}{\bibfnamefont{A.}~\bibnamefont{Kurkela}}, \bibnamefont{and}
  \bibinfo{author}{\bibfnamefont{M.}~\bibnamefont{Strickland}}
  (\bibinfo{year}{2020}), \eprint{2004.05195}.

\bibitem[{\citenamefont{Almaalol et~al.}(2019)\citenamefont{Almaalol,
  Alqahtani, and Strickland}}]{Almaalol:2018jmz}
\bibinfo{author}{\bibfnamefont{D.}~\bibnamefont{Almaalol}},
  \bibinfo{author}{\bibfnamefont{M.}~\bibnamefont{Alqahtani}},
  \bibnamefont{and}
  \bibinfo{author}{\bibfnamefont{M.}~\bibnamefont{Strickland}},
  \bibinfo{journal}{Phys. Rev. C} \textbf{\bibinfo{volume}{99}},
  \bibinfo{pages}{014903} (\bibinfo{year}{2019}), \eprint{1808.07038}.

\bibitem[{\citenamefont{Almaalol and Strickland}(2018)}]{Almaalol:2018ynx}
\bibinfo{author}{\bibfnamefont{D.}~\bibnamefont{Almaalol}} \bibnamefont{and}
  \bibinfo{author}{\bibfnamefont{M.}~\bibnamefont{Strickland}},
  \bibinfo{journal}{Phys. Rev. C} \textbf{\bibinfo{volume}{97}},
  \bibinfo{pages}{044911} (\bibinfo{year}{2018}), \eprint{1801.10173}.

\bibitem[{\citenamefont{Strickland and Tantary}(2019)}]{Strickland:2019hff}
\bibinfo{author}{\bibfnamefont{M.}~\bibnamefont{Strickland}} \bibnamefont{and}
  \bibinfo{author}{\bibfnamefont{U.}~\bibnamefont{Tantary}},
  \bibinfo{journal}{JHEP} \textbf{\bibinfo{volume}{10}}, \bibinfo{pages}{069}
  (\bibinfo{year}{2019}), \eprint{1903.03145}.

\bibitem[{\citenamefont{Strickland}(2018)}]{Strickland:2018ayk}
\bibinfo{author}{\bibfnamefont{M.}~\bibnamefont{Strickland}},
  \bibinfo{journal}{JHEP} \textbf{\bibinfo{volume}{12}}, \bibinfo{pages}{128}
  (\bibinfo{year}{2018}), \eprint{1809.01200}.

\bibitem[{\citenamefont{Berges et~al.}(2013{\natexlab{a}})\citenamefont{Berges,
  Boguslavski, Schlichting, and Venugopalan}}]{Berges:2013fga}
\bibinfo{author}{\bibfnamefont{J.}~\bibnamefont{Berges}},
  \bibinfo{author}{\bibfnamefont{K.}~\bibnamefont{Boguslavski}},
  \bibinfo{author}{\bibfnamefont{S.}~\bibnamefont{Schlichting}},
  \bibnamefont{and}
  \bibinfo{author}{\bibfnamefont{R.}~\bibnamefont{Venugopalan}}
  (\bibinfo{year}{2013}{\natexlab{a}}), \eprint{1311.3005}.

\bibitem[{\citenamefont{Berges et~al.}(2013{\natexlab{b}})\citenamefont{Berges,
  Boguslavski, Schlichting, and Venugopalan}}]{Berges:2013eia}
\bibinfo{author}{\bibfnamefont{J.}~\bibnamefont{Berges}},
  \bibinfo{author}{\bibfnamefont{K.}~\bibnamefont{Boguslavski}},
  \bibinfo{author}{\bibfnamefont{S.}~\bibnamefont{Schlichting}},
  \bibnamefont{and}
  \bibinfo{author}{\bibfnamefont{R.}~\bibnamefont{Venugopalan}}
  (\bibinfo{year}{2013}{\natexlab{b}}), \eprint{1303.5650}.

\bibitem[{\citenamefont{Mazeliauskas and Berges}(2019)}]{Mazeliauskas:2018yef}
\bibinfo{author}{\bibfnamefont{A.}~\bibnamefont{Mazeliauskas}}
  \bibnamefont{and} \bibinfo{author}{\bibfnamefont{J.}~\bibnamefont{Berges}},
  \bibinfo{journal}{Phys. Rev. Lett.} \textbf{\bibinfo{volume}{122}},
  \bibinfo{pages}{122301} (\bibinfo{year}{2019}), \eprint{1810.10554}.

\end{thebibliography}

\end{document}